\documentclass[conference,compsoc]{IEEEtran}
\usepackage[utf8]{inputenc}
\usepackage[T1]{fontenc}
\usepackage{microtype}
\usepackage{amsmath,amssymb}
\usepackage{booktabs}
\usepackage{longtable}

\usepackage{placeins}
\usepackage{tabularx}
\usepackage{graphicx}
\usepackage{adjustbox}
\usepackage{tikz}
\usetikzlibrary{positioning,arrows.meta}
\usepackage{xcolor}
\usepackage{url}
\usepackage{enumitem}
\usepackage[hidelinks]{hyperref}
\setlist{leftmargin=*,itemsep=1pt,topsep=2pt}
\usepackage{wasysym}
\newcommand{\halfcirc}{\LEFTcircle}
\newenvironment{progblock}{\par\medskip\noindent\begin{minipage}{\columnwidth}\ttfamily\footnotesize\raggedright\setlength{\parindent}{0pt}}{\end{minipage}\par\medskip\noindent}
\makeatletter\let\@afterindenttrue\@afterindentfalse\makeatother
\begin{document}
\title{SoK: Timestamps Are Not a Time Model:\\ Rethinking Temporal Reasoning in Host-Based Intrusion Detection}
\author{\IEEEauthorblockN{Rene Heinzl and Markus Nissl}
\IEEEauthorblockA{datAInsights GmbH\\
\{rene.heinzl, markus.nissl\}@data-insights.ai}}
\maketitle

\begin{abstract}
Host-based intrusion detectors order events, window audit streams and sequence attack stages by their timestamps. Every conclusion they draw therefore rests on assumptions about clocks, delivery, entity identity and the integrity of the evidence, and most of these assumptions go unstated. This Systematization of Knowledge (SoK) makes them explicit. Drawing on temporal databases, stream processing and runtime verification, we distill five components of a time model, together with a checklist of evidence assumptions and four levels of guarantee, and walk through a running campaign with them. A documented search yields fifteen families of detection and investigation. For papers selected by screening role and period, plus named systems and landmarks, we check the stated assumptions in the full texts, and we assess eleven temporal knowledge representation and reasoning (KRR) formalisms and a learned baseline on nine attack patterns. Some early alert correlators stated clock and delivery assumptions, and forensics measures clock error. Current provenance-based detectors state no clock bound. Four audited papers describe input-ordering conditions or conventions, but none specifies conditions for conclusion validity across a stated class of delivery failures. In the selected provenance-based systems, campaign composition uses system-specific procedures or objectives, learned state, causal graph queries or language models, and beyond CEP-based query systems, we found no temporal rule language or stream reasoner from KRR applied to host audit data. In 28 of 38 audited provenance-era detector papers, capture or stored evidence is trusted or its manipulation is placed out of scope, and none says which conclusions survive deleted, delayed or forged records. Five open challenges follow from these gaps.
\end{abstract}

\section{Introduction}
Host auditing and provenance mechanisms such as Linux Audit, Event Tracing for Windows with its kernel providers, and CamFlow record what processes do on a host: they create processes, read and write files, map memory and open sockets. Such records are commonly assembled into whole-system provenance graphs, in which system entities are nodes and timestamped, directed edges stand for recorded interactions and potential information-flow dependencies~\cite{hanetal2018}. Timestamps matter at several points when a multi-stage intrusion is detected and reconstructed. Learned detectors encode edge times and cut the stream into windows, rule-based systems order attack stages, and investigators follow edges forward and backward in time.

Yet a timestamp is not a time model. By itself it does not tell us whether it marks the moment an event happened or the moment it was observed, how far the clocks of two hosts may drift apart, whether records can arrive late, out of order or not at all, or what an attacker could have altered. A system that leaves these questions open still answers them implicitly in its output, and every campaign claim built on that output inherits the answers.

In practice, four conditions bring these questions to the surface:

\begin{enumerate}
\item \textbf{Long horizons.} Advanced persistent threats (APTs) stage activity over days or weeks, causing evidence to fall into different analysis windows.
\item \textbf{Living off the land.} Attackers abuse legitimate binaries such as PowerShell, Windows Management Instrumentation (WMI) and certutil, producing graph structures that closely resemble benign administration.
\item \textbf{Time manipulation and skew.} Timestomping rewrites file time attributes~\cite{mitreattckt1070006}, log tampering can delete or alter records, and clock skew between hosts distorts cross-host order. These are distinct problems that leave distinct evidence (Section 2).
\item \textbf{Mimicry.} Attackers shape their activity to look benign, a problem known for system-call detectors~\cite{wagnerandsoto2002} and shown for provenance-graph detectors~\cite{goyaletal2023}.
\end{enumerate}

The field has developed along two largely separate lines. On the symbolic side, work began with attack specification languages~\cite{ilgunetal1995,eckmannetal2002}, alert correlation~\cite{ningetal2002,valeuretal2004} and chronicle recognition~\cite{morinanddebar2003}. Provenance-based attack investigation grew out of backtracking~\cite{kingandchen2003}, gained execution partitioning and practical provenance tracing~\cite{leeetal2013a,maetal2016}, and eventually produced rule-based provenance detectors such as SLEUTH and HOLMES~\cite{hossainetal2017,milajerdietal2019a}. Conclusions on this side are derived explicitly from rules, which experts often write and maintain by hand. The learned side started with anomaly detection over system-call sequences~\cite{forrestetal1996}, system logs~\cite{duetal2017} and streaming provenance graphs~\cite{manzooretal2016}, and today systems such as KAIROS~\cite{chengetal2024} and FLASH~\cite{rehmanetal2024} scale to raw telemetry. Their output is a statistical score without a derivation behind it, and re-evaluations find that reported benchmark accuracy overstates detection quality once evaluation conditions become more realistic~\cite{bilotetal2025}.

Meanwhile, knowledge representation and reasoning (KRR) has built formalisms and engines for temporal order, incomplete evidence and rule-based derivation, mostly for applications outside security. They include runtime monitors for first-order temporal logic, complex event recognition, stream reasoning, temporal rule languages, probabilistic and annotated logics, argumentation, temporal ontologies and causal models. A recent survey of neuro-symbolic cybersecurity covers 127 publications, but none of the symbolic foundations it describes has a temporal operator~\cite{hakimetal2025}. A recent example combining learning with temporal KRR is the proof of concept of Eckhoff et al., which composes security information and event management (SIEM) alerts against attack patterns in temporal answer set programming (ASP)~\cite{eckhoffetal2025}. We know of no survey that relates the breadth of temporal KRR to the reasoning problems of host-based intrusion detection.

\smallskip\noindent \textbf{Main argument.} We argue that the missing time model is one instance of a broader problem. A campaign claim depends on assumptions about its evidence: when the events occurred, how they are ordered across hosts, which entities they concern and whether the relevant records are complete. Current systems mostly leave these assumptions unstated, and where they state anything, it is often a strong trust assumption, such as an uncompromised kernel and unmodified audit logs. Formal reasoning can establish conclusions relative to such assumptions. Whether the assumptions hold is a separate question, and answering it takes evidence about capture, identity and the trusted computing base.

\smallskip\noindent \textbf{Relation to existing surveys.} The closest prior systematization is Inam et al., who organize system auditing and provenance analysis along the pipeline of capture, reduction, detection and investigation~\cite{inametal2023}. We instead ask what that pipeline assumes about time and evidence, and what temporal KRR offers for stating those assumptions and reasoning under them. Other surveys each cover one side of the problem (Table 1). Zipperle et al. survey provenance-based intrusion detection~\cite{zipperleetal2022}. Salah et al. cover symbolic multi-step alert correlation before the provenance and deep-learning era~\cite{salahetal2013}. The neuro-symbolic survey covers the integration of learning and logic, but its symbolic side is atemporal~\cite{hakimetal2025}. Stream reasoning surveys cover temporal KRR over data streams without a security focus~\cite{dellaglioetal2017}. Recent SoKs cover adjacent steps. Ammar et al. systematize the integrity and delivery of execution evidence at the auditing layer~\cite{ammaretal2025}, Breitinger et al. systematize timeline-based event reconstruction in digital forensics, including the interpretation and reliability of timestamps~\cite{breitingeretal2025}, and Büchel et al. find an apparent performance plateau for natural-language processing (NLP) and large language model (LLM) extractors of attack techniques from threat reports~\cite{bucheletal2025}, a limit for the knowledge acquisition that symbolic families depend on (Section 7.6). Bilot et al. complement these surveys with an empirical re-evaluation of learned detectors~\cite{bilotetal2025}.

\begin{table*}[!tp]
\caption{Scope of this survey compared with related surveys.}
\label{tab:1}
\centering \footnotesize
\begin{tabular}{>{\raggedright\arraybackslash}p{0.132\textwidth} >{\raggedright\arraybackslash}p{0.226\textwidth} >{\raggedright\arraybackslash}p{0.113\textwidth} >{\raggedright\arraybackslash}p{0.132\textwidth} >{\raggedright\arraybackslash}p{0.132\textwidth} >{\raggedright\arraybackslash}p{0.113\textwidth}}
\toprule
\textbf{Survey} & \textbf{Focus} & \textbf{Learned detectors} & \textbf{Symbolic security systems} & \textbf{Temporal KRR formalisms} & \textbf{Joint comparison of detectors and temporal KRR} \\
\midrule
Zipperle et al. 2022 & Provenance-based intrusion detection & Yes & Yes & No & No \\
Inam et al. 2023 & System auditing and provenance analysis & Partly & Yes & No & No \\
Salah et al. 2013 & Alert correlation & No & Yes & Partly & No \\
Hakim et al. 2025 & Neuro-symbolic AI for cybersecurity & Yes & Partly & No & No \\
Ammar et al. 2025 & Integrity, attestation and auditing of program execution & No & No & No & No \\
Breitinger et al. 2025 & Timeline-based event reconstruction in digital forensics & No & Partly & No & No \\
Büchel et al. 2025 & Technique extraction from threat reports & Yes, NLP and LLM extractors & No & No & No; compares extractors empirically \\
Dell'Aglio et al. 2017 & Stream reasoning & No & No & Yes & No \\
Bilot et al. 2025 & Empirical re-evaluation of learned provenance detectors & Yes & No & No & No; re-evaluates detectors empirically \\
This survey & Temporal reasoning over host audit data & Yes & Yes & Yes, ten domains & Yes \\
\bottomrule
\end{tabular}
\end{table*}
\smallskip\noindent \textbf{Research questions.} Three questions guide this systematization:

\begin{itemize}
\item \textbf{RQ1.} Which notions of time and which evidence assumptions do learned and symbolic systems for host audit data rely on, and which of them do they state?
\item \textbf{RQ2.} What does the state of the art in KRR offer for stating a time model and reasoning under it, with which guarantees and at which cost?
\item \textbf{RQ3.} What must a campaign claim make explicit about time and evidence for its conclusion to be checkable, and which of these requirements do current systems document?
\end{itemize}

\smallskip\noindent \textbf{Contributions.}

\begin{enumerate}
\item A framework of five time-model components drawn from temporal databases, stream processing and runtime verification, four notions of time, a checklist of time and evidence assumptions, and four levels of guarantee, which together make the assumptions behind a temporal claim explicit (Section 2).
\item A running example that applies the checklist to one campaign and separates observed facts, derived hypotheses and correspondence assumptions step by step (Section 3).
\item A documented selection of the literature and a survey of fifteen paradigm families, from learned and hybrid pipelines and host behavior models to symbolic, investigation and knowledge-based approaches, with the time and evidence assumptions of a rule-based sample of papers checked against their full texts (Section 4, Appendix A).
\item A comparison of the families on fourteen dimensions that record capability, evidence status and preconditions separately, a systematization of six integration patterns, and an analysis of where campaign composition happens (Section 5, Appendix C).
\item An analysis of five design constraints and theoretical limits (Section 6).
\item A problem-centered review of temporal KRR for six security reasoning problems, and a catalog of nine precisely specified attack-reasoning patterns against which eleven symbolic formalisms and a learned baseline are assessed, with schematic formalizations behind the symbolic rows (Section 7, Appendices B and D).
\item Five open challenges, each with a research question and a way to measure progress against a baseline (Section 8).
\end{enumerate}

\smallskip\noindent \textbf{Scope.} We cover reasoning over host audit data and provenance graphs, together with the alert correlation and threat-intelligence work that feeds it. Network-flow classification, malware classification and cloud-log analytics are out of scope. All absence findings are bounded by those searches (Section 9). As a systematization of knowledge, this paper introduces no new experiments, and its findings are derived from the literature and from the formalizations in Appendix D.

\smallskip\noindent \textbf{Method.} We searched OpenAlex on 23 and 24 October 2026 with nine documented queries over 21 security venues and twelve over 39 KRR venues, from 2000 onward, each also without venue restriction. The 4,869 security and 2,550 KRR records were screened against stated inclusion rules with AI assistance under the authors' direction and a written reason for every decision, and 1,003 and 1,009 of them were retained. Snowballing and a citation chase added lines of work that the queries' vocabulary missed. From the retained security records we derived the families of Section 4, audited the assumptions of a rule-based sample of papers in full text (Section 4.3), rated the families (Section 5) and assessed the formalisms on encoded patterns (Section 7.7). Appendix A documents every step.

\section{Time, evidence and guarantees}
To interpret what the surveyed systems output, one needs to know what a time model fixes, what is assumed about the evidence, and at which of four levels a system gives a guarantee. We close the section with a checklist of six questions, which the rest of the paper applies to the running example, to the paradigm families and to the formalisms.

\smallskip\noindent \textbf{Time model.} A timestamp is a value, and a time model says what that value means and what may be inferred from it. We do not propose a new model. Fields that reason over timestamped data have each made part of it explicit, and we collect those parts into five components that serve as our criteria for comparison. (i) A time domain, such as the integers at a stated resolution or the rationals, with a convention for equal timestamps, as temporal logics and temporal databases fix~\cite{koymans1990,jensenandsnodgrass1999}. (ii) For each source s, the notion of time its timestamps carry (Table 2) and a clock assumption, for example that the clock of s deviates from a reference clock by at most $\delta$(s), including the rounding of its timestamps, as monitoring under imprecise timestamps assumes~\cite{basinetal2014}. (iii) A delivery assumption, such as a bound $\lambda$ on how late a record can arrive after its event time, as heartbeats and watermarks express in stream processing~\cite{srivastavaandwidom2004,akidauetal2015}, or none. (iv) An order basis: order may come from timestamps within one source, from timestamps across sources, or from recorded causal dependencies~\cite{lamport1978}. Under these clock-error bounds alone, a timestamp comparison guarantees that one event precedes another when their timestamps differ by more than the sum of the relevant bounds. (v) A verdict policy, stating when a query result becomes final and which results are provisional until then, as watermarks and three-valued monitoring do~\cite{akidauetal2015,baueretal2011}. A bounded past-time query becomes final once the delivery assumption rules out any further record in its time range, whereas queries that depend on future events need their own completion condition. The four notions in Table 2 are the possible meanings in (ii), not a partition of timestamps: event and observation time interpret a timestamp, file metadata time comprises several attributes with their own update rules, and causal order is a relation rather than a time value.

\begin{table*}[!tp]
\caption{Four notions used to interpret timestamps and order.}
\label{tab:2}
\centering \footnotesize
\begin{tabular}{>{\raggedright\arraybackslash}p{0.122\textwidth} >{\raggedright\arraybackslash}p{0.282\textwidth} >{\raggedright\arraybackslash}p{0.207\textwidth} >{\raggedright\arraybackslash}p{0.254\textwidth}}
\toprule
\textbf{Notion} & \textbf{What it is} & \textbf{Typical source} & \textbf{What can go wrong} \\
\midrule
Event time & The time at which the sensor says an action occurred, on the sensor's clock & Kernel audit record timestamp & Host clock skew; records lost in races between occurrence and commitment~\cite{paccagnellaetal2020b} \\
Observation time & When a record reaches the analysis pipeline & Collector or SIEM ingest timestamp & Delay and reordering across hosts \\
File metadata time & Time attributes stored with a file object & Modification, access, change and birth times & Timestomping rewrites them~\cite{mitreattckt1070006} \\
Causal order & The order implied by information flow, such as a fork before the child's first action & Provenance edges & Missing edges when records are lost or deleted \\
\bottomrule
\end{tabular}
\end{table*}
Event time is assigned by the capture layer, such as Linux Audit, provenance-aware storage~\cite{muniswamyreddyetal2006}, Hi-Fi~\cite{pohlyetal2012}, SPADE~\cite{gehaniandtariq2012}, CamFlow~\cite{pasquieretal2017} or eAudit~\cite{sekaretal2024}, so what event time means is fixed by that layer, and gaps in the causal dependencies it captures are a problem of their own~\cite{kimmetal2026}. Causal order is likewise limited to the program-order and communication events that this layer records. Logical clocks have been used to consolidate log files from different sources~\cite{gomezetal2006}. Timestamps are imprecise in ways that must be kept apart: records tie at the resolution of their timestamps, each timestamp carries the error of the clock that assigned it, and intervals emitted by a learned component have uncertain endpoints. An interval cannot decide order at a finer resolution than its uncertainty. Clock error is not negligible in practice: a 2007 study of about 8,000 web servers found only about 74\% within 10 s of a reference clock~\cite{buchholzandtjaden2007}, and digital forensics estimates skew and drift by correlating timestamps with reference sources~\cite{schatzetal2006} or tests hypotheses about past clock adjustments against the evidence~\cite{willassen2008}. Within one host, timestamps are also a weak proxy for causality, since asynchronous tasks and delayed user interaction separate cause and effect in time~\cite{leeetal2013a}.

Each failure in Table 2 calls for a different remedy. When a trusted audit mechanism captures the operation that changes a file's timestamps, timestomping changes the file attributes but not that independently recorded audit event, so a mismatch between the two is a contradiction relative to a stated invariant, such as that the modification time equals the time of the last recorded write, although benign operations that preserve timestamps, clock error and identity mistakes must be excluded before the mismatch counts as evidence of manipulation. Forensic practice applies such checks to the timestamps that NTFS keeps for each file~\cite{cho2013,palmbachandbreitinger2020,vaninietal2025}, tests whether a clock was correct against time anchors, events whose time is known independently~\cite{vaninietal2024}, orders events by implicit timing information where timestamps are missing or unreliable~\cite{dreieretal2024}, and detects manipulation of the system clock itself~\cite{jangetal2026}. A late record need not carry a wrong event time, and out-of-order monitoring handles it~\cite{basinetal2020}. A fixed offset between hosts leaves the order on each host intact, provided each clock is monotonic and ties are handled explicitly, while a forged sequence of event times can remain internally consistent, so checks of timestamp consistency alone do not establish authenticity.

\smallskip\noindent \textbf{Evidence assumptions.} Beyond time, a reasoning system states or silently assumes which records are complete, which timestamps are trusted and how entities are identified across processes and hosts. Where detectors state a threat model, it is typically that the kernel and audit framework are trusted and the audit data are not tampered with (Section 4.3): the attacker chooses its actions but cannot corrupt the committed records. Such statements protect the integrity of specified components or records but say nothing about missing or delayed records, which trusted components can still lose through capacity limits or capture races. This trusts the record, not the recorded behavior: a trusted kernel faithfully records an attacker who changes file attributes through ordinary system calls.

\smallskip\noindent \textbf{Attacker model for evidence.} Claims about robustness are meaningful only relative to what the attacker can do. Shaping actions to resemble benign activity, as in mimicry, is a capability over behavior. Over the evidence we distinguish four capabilities, which overlap and are not ordered by strength: (i) changing file metadata only, as by timestomping, (ii) deleting or delaying individual records, as by exploiting race conditions in the audit framework~\cite{paccagnellaetal2020b}, (iii) controlling one logger or host, and (iv) controlling all sensors. Tamper-evident auditing detects manipulation of committed records under a trusted-execution assumption~\cite{paccagnellaetal2020a}, which covers part of the second and third capability. Against the fourth, records alone cannot establish that an unobserved event occurred. Clocks can also be shifted without controlling a host, by attacking time synchronization~\cite{malhotraetal2016}, which secure synchronization makes harder~\cite{deutschetal2018,finkenzelleretal2024}.

\smallskip\noindent \textbf{Four levels of guarantee.} A statement that a detector is formally correct can mean four different things (Table 3). We separate them because the surveyed systems give guarantees at different levels, and a guarantee at one level does not carry over to another. Table 6 rates the evaluation and certificate levels for each family, in its columns on correct evaluation of a declarative specification and on independent checkability, with a specification as their precondition, while the running example (Section 3) and Section 6 show that campaign claims depend most on the fourth.

\begin{table*}[!tp]
\caption{Four levels of guarantee.}
\label{tab:3}
\centering \footnotesize
\begin{tabular}{>{\raggedright\arraybackslash}p{0.122\textwidth} >{\raggedright\arraybackslash}p{0.338\textwidth} >{\raggedright\arraybackslash}p{0.414\textwidth}}
\toprule
\textbf{Level} & \textbf{Question} & \textbf{Example} \\
\midrule
Specification & Is it defined unambiguously what counts as a match or a campaign? & A chronicle, an Event Calculus specification in RTEC, a program in the stream reasoning language LARS \\
Evaluation & Is the algorithm correct, and possibly complete, with respect to that definition? & VeriMon is formally verified for metric first-order temporal logic (MFOTL)~\cite{schneideretal2019,zinggetal2022}; for a fixed query in the unary-filter fragment of complex event logic (CEL), the compiled automaton has proven semantics and constant update cost per event, with results enumerated separately~\cite{grezetal2019} \\
Certificate & Can a single output be checked independently? & WhyMon's proof trees, checked by a formally verified checker~\cite{limaetal2024} \\
Correspondence & Under which assumptions about sensors, extraction and attacker does the output describe actual behavior? & Only partial guarantees exist: tamper-evident auditing shows that committed records were not altered, under a trusted-execution assumption~\cite{paccagnellaetal2020a}; the inspected systems state some trust assumptions (Table C1), but none gives a contract per conclusion covering capture, extraction, identity, timestamp meaning and attacker influence \\
\bottomrule
\end{tabular}
\end{table*}
A precise semantics or optimization objective is a specification, but it does not show that an algorithm computes it, which is a separate, evaluation-level result. Even a verified algorithm does not let a user check an individual output, which needs either an independent re-evaluation or a certificate that a separate checker can verify without rerunning the full algorithm. None of the three says whether the output describes what happened on the host, which is correspondence. Two consequences follow for temporal evidence. First, evaluation-level theorems hold only under their input assumptions: manipulated timestamps or a changed delivery order can violate them, and then the theorem does not apply. Second, a certificate is bound to the data version it was checked against: under monotone semantics, a late record that only adds true facts preserves a positive derivation, but late, corrected or retracted records can invalidate claims that use negation, completeness, counting or maximality, and claims whose premises they change. Formal reasoning therefore guarantees correctness relative to its input but says nothing about whether that input is true or complete. Manipulated evidence attacks exactly this, and no evaluation-level guarantee repairs it. Song et al. already reasoned about a host-based detector in this way: they verified its specifications against a security policy in the theorem prover ACL2, under assumptions the detector itself does not check~\cite{songetal2004}, a design-time guarantee about specifications rather than a certificate for an individual alert.

\smallskip\noindent \textbf{A checklist for time and evidence.} A temporal claim about host activity depends on the answers to six questions. Q1 to Q4 ask for the components of the time model that a system description can state: Q1 the meaning of each timestamp, Q2 the basis and scope of order, Q3 the clock assumption and the resolution of timestamps, and Q4 delivery. The time domain and the verdict policy are chosen by a system and its deployment. The audit of Section 4.3 does not code them, but Section 7 compares the constructs that formalisms provide for them, under the fixed conventions of Appendix D. Q5 and Q6 ask for the evidence assumptions. For formal comparison, Section 7.7 adds conventions on equal timestamps, interval endpoints and strict order.

\begin{itemize}
\item \textbf{Q1. Notion of time.} Does each timestamp give event time or observation time, and are the two kept apart?
\item \textbf{Q2. Order.} What establishes order (timestamps, sequence numbers or causal edges), and is order used within one window, across windows, or across hosts?
\item \textbf{Q3. Clocks and precision.} What is assumed about the offset and drift between the clocks that produced the timestamps, and about the precision of each timestamp: its granularity, ties between records, and, for intervals extracted by a learned component, the uncertainty of the extracted endpoints?
\item \textbf{Q4. Delivery.} What is assumed about late, reordered and lost records, and what does the paper state about the validity or completeness of its outputs when those assumptions fail?
\item \textbf{Q5. Identity.} How are entities, such as processes, accounts and credentials, identified over time and across hosts?
\item \textbf{Q6. Attacker capability.} Which of the four capabilities above may the attacker have over the evidence?
\end{itemize}

Section 3 illustrates these questions on one campaign and identifies the assumptions it needs, Section 4.3 records what the sampled systems state about them, and Section 7 asks which formalisms can express the answers.

\section{A running example: from audit records to a campaign claim}
This section applies the checklist to one concrete attack, to show which parts of a campaign claim are observed and which rest on assumptions. An attacker reads a credential on host A, a different process later uses it, the credential authenticates a session on host B, and data later leaves B. Table 4 lists, for each step, what the audit records show, what a detector would conclude from them, and what must additionally hold for that conclusion to be true. Steps 1 to 3 depend on identity (Q5), step 3 on the clock offset between A and B (Q3), which we assume to be at most $\delta$, rounding included, every step on the authenticity of the records it uses and on what the attacker can change in them (Q6), and any claim that rests on an absence or an exact count on complete delivery of the relevant records (Q4).

\begin{table*}[!tp]
\caption{The running example: observed facts, derived hypotheses and correspondence assumptions.}
\label{tab:4}
\centering \scriptsize
\begin{tabular}{>{\raggedright\arraybackslash}p{0.085\textwidth} >{\raggedright\arraybackslash}p{0.160\textwidth} >{\raggedright\arraybackslash}p{0.132\textwidth} >{\raggedright\arraybackslash}p{0.320\textwidth} >{\raggedright\arraybackslash}p{0.169\textwidth}}
\toprule
\textbf{Step} & \textbf{Observed fact} & \textbf{Derived hypothesis} & \textbf{Correspondence assumption} & \textbf{If evidence is missing} \\
\midrule
1. Credential read on A & Read(p, f)@t1, with t1 on A's clock & p read a store of credential c; that p obtained c needs evidence that the read returned it & f's path or ontology class identifies a store of c; process identifiers are not reused within the horizon & The fact and the hypothesis are not derived; this says nothing about whether a read occurred \\
2. Use by another process & Flow edges from f, or from p's memory, to process q, with event times & q holds c & The recorded flow carried c itself rather than other data; the edges of the witness path were captured and are authentic. A provenance path shows dependency, not the identity of a credential & The link between steps 1 and 3 is not derivable \\
3. Authentication on B & Logon(u, from address a)@t3, with t3 on B's clock & u authenticated with c, after step 1 & u's credential is c, which the logon record alone does not show; a identifies host A; the relative offset of A's and B's timestamps, rounding included, is at most $\delta$, so step 1 precedes step 3 if t3 $-$ t1 > $\delta$ & The hypothesis is not derived \\
4. Exfiltration from B & Outbound transfer by process r on B at t4, with its volume & r's outbound transfer exfiltrated data from A & r descends from u's session; the transfer carries data from A, which its volume alone does not show & The hypothesis is not derived \\
\bottomrule
\end{tabular}
\end{table*}
Composing the four hypotheses into the claim "a credential taken on A was used to reach B and exfiltrate data from A" needs one more step: a rule that joins the read and the logon on the shared credential, links the logon to the session and to process r on B, and orders the steps in time. The symbolic formalisms of Section 7 that support joins across entities can state such a rule (pattern AP3 in Table 9), and a reasoner can then derive the claim correctly from the facts it was given. Whether the claim describes what happened is a different question. It depends on the assumptions in the fourth column of Table 4 holding together, on complete records wherever the claim relies on an absence, an exact count or an upper-bound count, and on a bound on what the attacker can change (Section 2). None of the systems we inspected states these assumptions together.

The example supports three observations. The first concerns where a claim becomes uncertain. The audit records establish each observed fact, but the claim as a whole rests on three attributions that the records do not establish: that the credential read on A is the one used later (identity, steps 1 to 3), that the read on A happened before the logon on B (cross-host order, step 3), and that the data leaving B originate from A (content, step 4). Each of these is an assumption about the correspondence between records and events, not an observation.

The second observation concerns missing evidence. If the logon record of step 3 is absent and no other evidence or rule establishes the logon, a reasoner cannot derive the step, but this does not show that the logon did not take place. In logical terms, a knowledge base K that does not entail q (K $\nvDash$ q) is not the same as K entailing $\neg$q (K $\models$ $\neg$q), and the predicate "a logon was recorded" differs from "a logon occurred". A system that reads non-derivation as non-occurrence adds a completeness assumption, which should be stated for the predicates and intervals concerned, as pattern AP6 does in Section 7.7. A system that instead reports such a claim as undetermined needs a semantics for that verdict, for example stated completeness of a source over an interval, a three-valued reading or abduction over the possible explanations.

The third observation concerns what an output must contain to be checked. Because the claim depends on its derivation and on the assumptions above, a candidate evidence interface for independent checking is a structured evidence object: it records the derivation, the identifiers of the records used, the rules that mapped records to facts and the correspondence assumptions on which the conclusion rests. None of the surveyed systems produces such an object.

\section{Paradigm families and the time models they assume}
This section organizes the surveyed literature and records what it assumes about time and evidence (RQ1). We organize the literature into fifteen families. L1 to L4 center on learned or hybrid pipelines, L5 covers host behavior models, including statistical or learned models, authored execution policies and models derived from program code, and S1 to S10 cover the other reasoning, investigation and knowledge-based approaches. A family is defined by how it represents attack knowledge and how it reasons, so that systems within a family share most of the properties rated in Section 5, with exceptions noted in Appendix C. Figure 1 places the families by the pipeline stage they serve, Sections 4.1 and 4.2 describe them, and Section 4.3 checks their time and evidence assumptions against the full texts of a rule-based sample.

\begin{figure*}[t]
\centering
\begin{tikzpicture}[
  stage/.style={draw, rounded corners=2pt, fill=gray!8, minimum height=0.8cm, text width=3.2cm, align=center, font=\small\bfseries},
  fam/.style={draw, rounded corners=1pt, fill=white, text width=3.2cm, align=left, execute at begin node={\hyphenpenalty=10000 \exhyphenpenalty=10000}, font=\scriptsize, inner sep=3pt},
  decl/.style={fam, fill=blue!7, draw=blue!60!black},
  proc/.style={fam, fill=orange!10, draw=orange!70!black},
  arr/.style={-{Latex[length=2mm]}, thick, gray!60!black},
  node distance=0.18cm and 0.9cm]
\node[stage] (k) {Knowledge};
\node[stage, right=of k] (r) {Recognition};
\node[stage, right=of r] (c) {Composition};
\node[stage, right=of c] (i) {Investigation and explanation};
\node[fam, below=of k] (s9) {S9 Threat-intelligence\\knowledge};
\node[fam, below=of s9] (s8) {S8 Rule and specification learning};
\node[fam, below=of s8] (s6) {S6 Attack graphs};
\node[fam, below=of r] (l1) {L1 Dynamic graph learning};
\node[fam, below=of l1] (l2) {L2 Semantic-structural fusion};
\node[decl, below=of l2] (s2) {S2 Event-pattern correlation};
\node[decl, below=of s2] (s3) {S3 Temporal-logic monitoring\\[1pt]\textit{certified explanations}};
\node[fam, below=of s3] (s5) {S5 Provenance and knowledge-graph queries};
\node[fam, below=of s5] (l5) {L5 Host behavior models};
\node[decl, below=of c] (s1) {S1 Alert correlation, plan recognition};
\node[proc, below=of s1] (s4) {S4 Rule-based provenance analysis\\[1pt]\textit{system-specific procedure}};
\node[fam, below=of s4] (l4) {L4 Knowledge-guided learning};
\node[proc, below=of i] (l3) {L3 LLM investigation agents\\[1pt]\textit{LLM composes the campaign}};
\node[fam, below=of l3] (s7) {S7 Argumentation};
\node[fam, below=of s7] (s10) {S10 Provenance tracing and attack investigation};
\draw[arr] (k) -- (r); \draw[arr] (r) -- (c); \draw[arr] (c) -- (i);
\node[font=\scriptsize, anchor=north west, text width=15.2cm] at ([yshift=-0.25cm]s6.west |- l5.south) {Shaded blue: families whose recognition or composition rests on a declarative temporal specification. Shaded orange: families in which campaigns are composed by a system-specific procedure or by an LLM agent.};
\end{tikzpicture}
\caption{The fifteen families, placed by the pipeline stage they mainly serve: knowledge, recognition, composition, and investigation and explanation. Blue marks the families whose recognition or composition rests on a declarative temporal specification, of which temporal-logic monitoring can also certify its explanations. Orange marks the families in which campaigns are composed by a system-specific procedure or an LLM agent. Section 5.2 compares how selected systems compose campaigns.}
\label{fig:families}
\end{figure*}

\subsection{Learned and hybrid families}
\noindent \textbf{L1 Dynamic graph learning.} Learned detectors over streaming provenance graphs model benign behavior and score deviations from it. Unicorn summarizes the graph with sketches~\cite{hanetal2020}. SIGL and ATLAS apply graph and sequence learning to software installation and to attack investigation~\cite{hanetal2021,alsaheeletal2021}. KAIROS learns how the graph's structure evolves, scores each event and reconstructs compact attack summaries~\cite{chengetal2024}. R-CAID links nodes to precomputed root causes, which improves resistance to the mimicry attacks evaluated under its attacker model~\cite{goyaletal2024}. NODLINK keeps alerts fine-grained by solving a Steiner tree problem online~\cite{lietal2024}. TFLAG applies deviation-aware self-supervised learning on temporal provenance graphs~\cite{jiangetal2025}. Other detectors embed causal paths~\cite{wangetal2020}, recommend likely interactions~\cite{zengetal2022}, learn node-level or masked graph representations~\cite{wangetal2022b,jiaetal2024}, process edge streams in temporal order~\cite{kingetal2023} or attribute detections to attack-related nodes~\cite{jiangetal2025b}, and related models detect lateral movement over authentication graphs~\cite{kingandhuang2022,khouryetal2024}. Heterogeneous log graphs~\cite{liuetal2019}, spatial-temporal encodings of provenance graphs~\cite{sangetal2025} and timing anomalies in kernel activity~\cite{landaueretal2025} extend the same idea.

\smallskip\noindent \textbf{L2 Semantic-structural fusion.} Fusion models combine a language model of node attributes or command lines with a model of graph structure. FLASH pairs a Word2Vec encoder with a graph neural network (GNN)~\cite{rehmanetal2024}, EagleEye embeds command lines with a Transformer over long event sequences~\cite{gyseletal2024}, and CrossGuard fuses a graph-attention branch with a Transformer over provenance paths~\cite{yoonetal2025}. CrossGuard's 15-minute windows are shorter than the 45 to 75 minutes between attack steps in its industrial testbed, so multi-stage correlation is left to an unspecified SIEM step.

\smallskip\noindent \textbf{L3 LLM investigation agents.} Here a large language model composes detections into an attack story. OCR-APT detects anomalous subgraphs with GNNs and lets an LLM reconstruct the story~\cite{alyetal2025}, and ProvSEEK orchestrates several agents over provenance queries and hypotheses~\cite{mukherjeeandkantarcioglu2025}. ANCHOR adds a retrieval-augmented baseline of normal behavior and a confidence-gated cache that tracks attacks across windows~\cite{zhengetal2026}, and ProvAgent coordinates agents in a similar way~\cite{yanetal2026}. LLM agents also correlate events across sources~\cite{hmimouetal2025}, adapt detection policies~\cite{karunanayakeetal2025} and reconstruct forensic timelines by retrieval-augmented generation~\cite{loumachietal2025}.

\smallskip\noindent \textbf{L4 Knowledge-guided learning.} In this family, symbolic knowledge constrains what is learned. KnowGraph injects weighted first-order rules into GNN-based graph anomaly detection~\cite{zhouetal2024}. CAPTAIN keeps a rule-based tag-propagation detector but learns its node, edge and threshold parameters by gradient descent~\cite{wangetal2025b}. Eckhoff et al. label clusters of alert embeddings with kill-chain phases by ASP optimization, preferring labels for which the timestamp order of related alerts matches the order of the phases. In an experiment, they label 1,900 alerts in six clusters~\cite{eckhoffetal2025}. APT-KGL learns over heterogeneous provenance graphs augmented with threat knowledge~\cite{chenetal2022}, and other systems guide sequence learning by attack intent~\cite{yueetal2024}. Apart from such ordering conditions over timestamps, the symbolic components in this family carry no temporal operators.

\smallskip\noindent \textbf{L5 Host behavior models.} The first learned line of host-based detection modeled the normal behavior of users and programs from audit data. Denning's model profiled subjects statistically over audit records~\cite{denning1986}, and Lee and Stolfo mined rules from audit and connection records~\cite{leeandstolfo1998}. Forrest et al. treated a process as normal if the short sequences of system calls it makes were seen in training~\cite{forrestetal1996,hofmeyretal1998}, and later work compared data models~\cite{warrenderetal1999,liaoandvemuri2002} and used automata~\cite{sekaretal2001}, models derived from program code~\cite{wagneranddean2001}, call arguments~\cite{kruegeletal2003,maggietal2010}, execution graphs~\cite{gaoetal2004}, dataflow~\cite{bhatkaretal2006}, access models~\cite{lanzietal2010} and long execution paths~\cite{shuetal2015}. Specification-based monitoring constrains the same events with written policies~\cite{koetal1997}. Mimicry attacks against these models~\cite{wagnerandsoto2002} anticipate those against provenance detectors. Learned models of system-log sequences~\cite{duetal2017,guoetal2021} and of the context of security events~\cite{vanedeetal2022} carry the idea to logs and alerts. The family models order within one process or log, mostly without timestamps, although DEEPCASE also bounds the elapsed time between an event and its context. Later learned work includes log-sequence models such as LogBERT alongside the graph-based provenance models of L1 and L2.

\subsection{Symbolic families}
\noindent \textbf{S1 Alert correlation and plan recognition.} Correlation approaches describe attack steps by prerequisites and consequences and chain alerts whose consequences enable later prerequisites~\cite{templetonandlevitt2001,ningetal2002,cuppensandortalo2000}. Early work fused alerts by the similarity of their attributes~\cite{valdesandskinner2001}, defined a formal data model for correlation~\cite{morinetal2002}, ranked alerts by mission impact~\cite{porrasetal2002}, built scenarios probabilistically or from attack models~\cite{dainandcunningham2002,cheungetal2003}, clustered alarms by root cause~\cite{julisch2003} and integrated these steps in one pipeline~\cite{valeuretal2004}. Other approaches aggregate alerts~\cite{debarandwespi2001}, correlate them across sensors~\cite{cuppensandmiege2002}, analyze statistical dependencies and causal networks~\cite{qinandlee2003,qinandlee2004}, learn attack strategies~\cite{ningandxu2003}, use attack graphs to hypothesize missed steps~\cite{wangetal2006}, and link alerts to OS-level dependencies within and across hosts~\cite{kingetal2005,zhaietal2006}, an early link between alert correlation and provenance. Between 2007 and 2016, Markov models projected attack sequences~\cite{favaetal2008}, alerts were matched to attack graphs~\cite{roschkeetal2011} and temporal attack models~\cite{albaneseetal2011}, and episodes were mined in real time~\cite{ramakietal2015}. HERCULE correlates multiple logs into one graph for attack reconstruction~\cite{peietal2016}, and Salah et al. survey the line~\cite{salahetal2013}. Recent systems correlate the stages of advanced persistent threats~\cite{ghafiretal2018,zimbaetal2020} and lateral movement~\cite{tianetal2019}. Plan recognition abduces the attacker's plan, including unobserved steps~\cite{geibandgoldman2001}. Eckhoff et al. check attack patterns in linear temporal logic on finite traces (LTLf), translated from threat reports by an LLM, against sequences of symbolic SIEM alerts with the temporal ASP solver telingo, as a proof of concept on short traces that the authors do not expect to scale to 100,000 alerts~\cite{eckhoffetal2025}.

\smallskip\noindent \textbf{S2 Event-pattern correlation.} Attacks are recognized here as patterns of events with order and timing constraints. Misuse detection had temporal attack languages early, from pattern matching over audit trails~\cite{kumarandspafford1994} and state transition analysis~\cite{ilgunetal1995} to the expert system P-BEST~\cite{lindqvistandporras1999}, the distributed monitors of EMERALD~\cite{porrasandneumann1997}, the network variant NetSTAT~\cite{vignaandkemmerer1999} and STATL~\cite{eckmannetal2002}, with faster matching later~\cite{meieretal2005}, and declarative signatures were given a formal semantics~\cite{pouzolandducasse2001}. Chronicle recognition matches event patterns with metric constraints over alert streams~\cite{morinanddebar2003}. Complex event processing (CEP) engines match sequences with time windows at high rates~\cite{wuetal2006}. Sequence matching over late and out-of-order events has a line of work with correctness guarantees~\cite{liuetal2009,chandramoulietal2010}, extended to distributed sensor streams~\cite{mutschlerandphilippsen2013} and to log analytics~\cite{chandramoulietal2018}, and pattern matching over events with imprecise timestamps has a formal treatment~\cite{zhangetal2010}. CEL gives CEP a denotational, compositional semantics and compiles fixed queries with unary filters to complex event automata with constant update cost per event, with result enumeration counted separately~\cite{grezetal2019}. The SIEM correlation languages used today, such as Sigma correlation rules and the sequences of Elastic's Event Query Language (EQL), have published specifications that define their syntax and describe in prose temporal conditions such as counts, ordered sequences and time spans~\cite{sigmahq2025,elastic2026}. None of them has a formal semantics, that is, a mathematical definition of the event sequences a rule matches against which implementations could be checked. These rules are also evadable: an analysis of public Sigma rules found many that small changes to commands or process attributes evade~\cite{uetzetal2024}.

\smallskip\noindent \textbf{S3 Temporal-logic monitoring.} Monitors check streams or logs of events against formulas of a temporal logic. A temporal-logic framework for intrusion detection~\cite{naldurgetal2004} and ORCHIDS, which matches temporal-logic signatures against event streams~\cite{olivainandgoubaultlarrecq2005}, applied this idea to security early, as did log auditing by model checking~\cite{rogerandgoubaultlarrecq2001} and temporal-logic models of forensic investigation~\cite{rekhisandboudriga2005}. Policy auditing over incomplete logs reports violations for the parts of a policy that the log determines~\cite{gargetal2011,chowdhuryetal2014}, and monitoring of incomplete or disagreeing logs marks what the logs do not determine~\cite{basinetal2012}. Monitoring with imprecise timestamps identifies the formula classes of metric temporal logic (MTL)~\cite{koymans1990} whose verdicts a standard monitor still computes correctly when each timestamp carries a bounded error~\cite{basinetal2014}. Runtime monitors for MFOTL, such as MonPoly, check policies over logs online~\cite{basinetal2015}. Separately, R2U2 combines propositional LTL and MTL observers with Bayesian diagnosis to monitor security threats for unmanned aircraft~\cite{schumannetal2015}. VeriMon is a formally verified MFOTL monitor~\cite{schneideretal2019}, since extended with a recursive let construct whose recursive occurrences lie strictly in the past~\cite{zinggetal2022}, and WhyMon explains each verdict with a proof tree certified by a formally verified checker~\cite{limaetal2024}. Temporal probabilistic logic has been used to allocate monitoring resources~\cite{jajodiaetal2016}. Outside these security applications, the monitors have been applied mainly to compliance checking.

\smallskip\noindent \textbf{S4 Rule-based provenance analysis.} Expert-written rules over provenance graphs detect and reconstruct attacks. They build on the tracing and reduction techniques of S10. SLEUTH propagates trust and confidentiality tags along provenance edges in real time~\cite{hossainetal2017}. HOLMES maps events to ATT\&CK-style tactics and links them into a high-level scenario graph that follows kill-chain stages~\cite{milajerdietal2019a}. RapSheet builds tactical provenance graphs from endpoint alerts and scores them by kill-chain order~\cite{hassanetal2020}. KnowHow lifts system identifiers to natural-language terms, matches events to attack techniques from threat reports, and discards technique sequences that violate the temporal order of tactics and APT lifecycle stages~\cite{mengetal2026}. These stage orders are design choices of the individual systems: ATT\&CK tactics are not meant to be read as a linear sequence~\cite{cisa2023}.

\smallskip\noindent \textbf{S5 Provenance and knowledge-graph queries.} Query languages let analysts or automated tools search stored audit data and live event streams. AIQL and SAQL give analysts declarative, time-aware queries over stored audit data and live streams~\cite{gaoetal2018a,gaoetal2018b}. POIROT aligns query graphs extracted from threat intelligence with provenance graphs~\cite{milajerdietal2019b}, and ThreatRaptor synthesizes threat-hunting queries from report text~\cite{gaoetal2021}. KRYSTAL represents audit data as a Resource Description Framework (RDF) knowledge graph and combines tag propagation, attack signatures and graph queries over it~\cite{kurniawanetal2022}. Extractor turns threat reports into provenance-graph queries~\cite{satvatetal2021}.

\smallskip\noindent \textbf{S6 Attack graphs.} Attack graphs model the sequences of exploits by which an attacker could move through a network. They were first generated by model checking~\cite{sheyneretal2002} and by scalable graph search~\cite{ammannetal2002,ingolsetal2006}, and their distances have been used to correlate intrusion alerts~\cite{noeletal2004}. MulVAL derives them as Datalog proofs~\cite{ouetal2005} and scales to large networks~\cite{ouetal2006}, and Bayesian attack graphs quantify and update the likelihood of attack paths~\cite{poolsappasitetal2012}. Later work defines probabilistic security metrics~\cite{wangetal2008}, models modern attacks and countermeasures~\cite{ingolsetal2009}, identifies zero-day attack paths from system-call dependencies~\cite{sunetal2018} and generates attack graphs from topology and attack patterns or with LLMs~\cite{ozdemirsonmezetal2022,zhangetal2025}. The underlying attack graphs are atemporal models of what an attacker could do, although some systems use them to correlate observed alerts or to reconstruct likely attack paths from system-call evidence~\cite{noeletal2004,sunetal2018}, without a time model for that evidence.

\smallskip\noindent \textbf{S7 Argumentation.} Argumentation weighs conflicting evidence through arguments that attack and defeat each other. A defeasible logic programming model raised the accuracy of attributing attacks to their culprits from 37\% to 62\% on a capture-the-flag dataset by pruning candidate culprits~\cite{nunesetal2016}, and an argumentation-based reasoner supports forensic attribution~\cite{karafilietal2020}.

\smallskip\noindent \textbf{S8 Rule and specification learning and synthesis.} Detection rules and specifications can also be learned from data or synthesized from text. Inductive logic programming learned valid-behavior specifications from the system-call traces of single programs early on~\cite{ko2000}, signatures were generated from execution traces~\cite{nandaandchiueh2008,lietal2009}, and correlation rules from attack trees~\cite{godefroyetal2015}. Invariants mined from sensor data outperformed engineer-designed ones on the SWaT testbed~\cite{fengetal2019}, and ensembles of temporal-logic formulas can be learned for anomaly detection in cyber-physical systems~\cite{indrietal2022}. LLMs translate natural language into temporal logic~\cite{cosleretal2023,eckhoffetal2025}, infer specifications for Datalog-based static analysis~\cite{lietal2025}, and translate threat reports into detection rules: SIGMERGE generates Sigma rules, four of which were accepted into the official repository~\cite{caietal2026}. YARA rules are learned by biclustering~\cite{raffetal2020} and cloud detection rules extracted with LLMs~\cite{schwartzetal2025}.

\smallskip\noindent \textbf{S9 Threat-intelligence knowledge.} Ontologies and knowledge graphs give threat intelligence a formal structure. Ontologies for intrusion detection date back two decades~\cite{undercofferetal2003,martinandplaza2003,heetal2004} and were extended to diagnosis and data semantics~\cite{coppolinoetal2009,balduccinietal2015}. Today ATT\&CK, D3FEND~\cite{kaloroumakisandsmith2021}, Structured Threat Information Expression (STIX)~\cite{oasis2021} and the Unified Cyber Ontology~\cite{uco2026} serve mainly as shared vocabularies. Temporal knowledge is emerging at technique level: ChronoCTI mines BEFORE and CONCURRENT relations between techniques~\cite{rahmanetal2024}, MITRE's Attack Flow encodes sequences of adversary behaviors~\cite{mitreattackflow}, threat actions and indicators are extracted from reports~\cite{husarietal2017,liaoetal2016}, technique knowledge graphs align reports with ATT\&CK~\cite{lietal2022}, and knowledge graphs, increasingly built with LLMs, support attribution and analysis~\cite{renetal2023,huetal2024,chengetal2025}.

\smallskip\noindent \textbf{S10 Provenance tracing and attack investigation.} Investigation reconstructs how an alert or a compromised object came about by following dependencies between processes, files and sockets backward and forward from it. Backtracking over operating-system dependencies~\cite{kingandchen2003} and the replay of past execution against vulnerability predicates~\cite{joshietal2005} were followed by intrusion recovery~\cite{goeletal2005}, tracing of process ancestry~\cite{purohitetal2004} and reconstruction in a virtual testbed~\cite{arnesetal2006}. Later work partitions executions into units~\cite{leeetal2013a,maetal2016,maetal2017}, reduces logs~\cite{leeetal2013b,xuetal2016,tangetal2018}, prioritizes or triages dependencies~\cite{liuetal2018,hassanetal2019,fangetal2022}, propagates tags~\cite{hossainetal2020}, tracks information flow on demand~\cite{jietal2017}, infers ramifications over temporal dependence networks~\cite{yangetal2018} and reconstructs missing behaviors~\cite{yangetal2022}, and TARDIS rolls back website snapshots to locate a compromise~\cite{paikasturietal2020}. Partitioning and reduction decide which events and timestamps survive, so they belong to the evidence assumptions of Section 2. Formal digital forensics reasons about the reconstruction itself, with finite-state machines~\cite{gladyshevandpatel2004}, intensional logic~\cite{mokhovetal2010} and models of anti-forensic attacks~\cite{rekhisandboudriga2012,rekhisandboudriga2012b,nealeetal2026}, and live analysis from the hypervisor addresses a compromised operating system~\cite{martignonietal2010}.

\subsection{What the systems assume about time and evidence}
This section answers RQ1. We sampled systems by a rule that can be reproduced from the artifact. The population is the set of security records retained by screening (Appendix A), each assigned to the family of its approach. For each screening role that describes detection or investigation and each of three periods, 2000 to 2006, 2007 to 2016 and 2017 to 2026, we took the three retained system papers with the highest citation rate under the rule in Appendix A.4, replacing surveys and position papers by the next in rank. The roles that pair L1 with L2, S2 with S3 and S4 with S5 were ranked jointly, and each selected paper was assigned to a family when it was coded. We added every system that Sections 4.1 and 4.2 name, an author-curated list of landmark systems, the first or most influential system of a mechanism within a family, and, as a check of the findings, every retained record that describes a detection, investigation or reasoning system and concerns clocks, timestamps, out-of-order or incomplete data, or the manipulation of evidence. Five digital-forensics papers on time and event reconstruction entered through the landmark list and are reported separately, because they are not detection systems. Of the 175 selected papers, we checked 160 in their full text, by keyword search per question followed by reading the matched passages and the threat-model sections. For two more, only an abstract or a description in another paper was available, from which we coded only positive statements on Q1 and Q2, and thirteen could not be obtained. The unit of the audit is a publication: several publications on the same system can contribute separate entries, so the counts and fractions below describe inspected papers, not distinct implementations or adoption across the field. Unless noted otherwise, counts refer to these audited papers. Table 5 summarizes the result per family, Table C1 lists every system, and the evidence ledger in the artifact gives the passage behind each entry and lists the papers we could not obtain.

\begin{table*}[!tp]
\caption{Time and evidence assumptions stated in the inspected full-text papers, by family.}
\label{tab:5}
\centering \scriptsize
\setlength{\tabcolsep}{3pt}\renewcommand{\arraystretch}{0.92}
\begin{adjustbox}{max width=\textwidth, max totalheight=0.88\textheight}
\begin{tabular}{>{\raggedright\arraybackslash}p{0.132\textwidth} >{\raggedright\arraybackslash}p{0.156\textwidth} >{\raggedright\arraybackslash}p{0.168\textwidth} >{\raggedright\arraybackslash}p{0.156\textwidth} >{\raggedright\arraybackslash}p{0.132\textwidth} >{\raggedright\arraybackslash}p{0.156\textwidth} >{\raggedright\arraybackslash}p{0.156\textwidth}}
\toprule
\smallskip\noindent \textbf{Family (full-text papers)} & \textbf{Q1 Notion of time} & \textbf{Q2 Scope of order} & \textbf{Q3 Clocks and precision} & \textbf{Q4 Delivery} & \textbf{Q5 Identity} & \textbf{Q6 Trust in the evidence} \\
\midrule
L1 Dynamic graph learning (18) & Event time as recorded, in edges, windows or batches & Within windows or batches; across them by learned state or a cache & None stated & Partial arrival order in Unicorn; chronological ingestion in KAIROS; observation-order streaming in EdgeTorrent; no validity account under violations & Hashes or node identifiers; hosts and users for lateral movement & Trusted capture, audit or logs in 11 of 18 \\
L2 Semantic-structural fusion (2) & Event time as recorded & Within windows or sequences & None stated & Missing events can change classification in EagleEye; no validity account & None stated & Trusted or tamper-resistant logs in 2 of 2 \\
L3 LLM investigation agents (7) & Event time as text & Across windows through agent memory or a cache & OCR-APT coarsens timestamps to seconds; no clock-offset bound & None stated & Process or node identifiers for lookup & Trusted capture and logs in 4 of 7; anti-forensics excluded in 1 \\
L4 Knowledge-guided learning (3) & Event time as recorded & Across windows through persistent tags & None stated & None stated & None stated & Trusted kernel and auditing in 2 of 3 \\
L5 Host behavior models (14) & None; order of system calls, records or log keys; audit timestamps in Denning & Within one process trace or log & None stated & None stated; no dropped records observed in Kruegel et al. & Processes and users & None stated \\
S1 Alert correlation, plan recognition (29) & Interval timestamps of alerts; event time distinguished from report and receipt time in M2D2 and Valeur et al. & Prerequisite before consequence; partial orders; across hosts through packets; by windows or time-series causality & Synchronized clocks required, drift tolerated or a maximum gap assumed in 6 of 29; no numeric bound & Late reports or out-of-order arrival allowed in 3; unobserved steps hypothesized or missing observations accommodated in 9, named as a limitation in 2 & Alert attributes; objects and packets; process identifiers & Spoofed sources considered in 4; injected alerts and delayed steps in Wang et al. 2006 \\
S2 Event-pattern correlation (8) & Event timestamps and timers; time points of chronicles & Within a scenario, signature or chronicle; across hosts after merging trails in NetSTAT & Clock gap tied to the chronicle time resolution in Morin and Debar; clock accuracy in Kumar and Spafford; no numeric bound & Delayed alerts integrated under that clock condition in Morin and Debar & User and process identifiers & Attacker varying time gaps noted in Morin and Debar \\
S3 Temporal-logic monitoring (9) & Order in a temporal logic; integer event times; ticks of one clock & Within a formula; across sources after merging by timestamp & Limited clock precision acknowledged in MonPoly; no bound & Incomplete logs as three-valued structures in 2; ordered input assumed in MonPoly & User identifiers; first-order data values & Spoofed sensor and GPS data diagnosed in R2U2 \\
S4 Rule-based provenance analysis (4) & Event time as recorded & Across windows through scenario graphs; partial order of stages in KnowHow & Millisecond resolution in SLEUTH & Detection despite missing events in one HOLMES scenario; no validity account & Process identifiers in SLEUTH & Trusted kernel, auditing or logs in 4 of 4 \\
S5 Provenance and knowledge-graph queries (6) & Event time as a queryable attribute; order of actions from text in Extractor & Within the time window of a query; along causal paths without time in KRYSTAL & NTP synchronization without a residual bound in AIQL & Timestamps assumed to increase monotonically per host in AIQL & Entity identifiers qualified by host in AIQL and SAQL; URIs in KRYSTAL & Trusted monitoring in 4 of 6 \\
S6 Attack graphs (12) & None; atemporal, except alert order in Noel et al. and timestamped objects in Sun et al. & Not applicable; event paths by attack-graph distance in Noel et al. & Not applicable to atemporal models; none stated by the timestamp-using exceptions & Missed detections bridged by graph distance in Noel et al. & Hosts and vulnerabilities as facts & None stated \\
S7 Argumentation (2) & Attack timestamps as data & Not applicable & None stated & Missing evidence abduced in Karafili et al. & Not applicable & Planted or forged evidence considered in 2 of 2 \\
S8 Rule and specification learning (9) & Time steps of sensor signals; none for generated rules or signatures & Within one learned formula, rule or signature & None stated & None stated & Files by inode and generation number in Ko & Threat reports assumed untampered; clean training traces in Ko \\
S9 Threat-intelligence knowledge (12) & Relative order of techniques from text, or none; alert timestamps in Coppolino et al. & Within one report or ontology; temporal proximity in Coppolino et al. & Not applicable, except none stated by Coppolino et al. & Not applicable to report-only resources; no delivery assumption located for the event-consuming ontology-based systems & Entities merged by similarity in CTINexus & None stated \\
S10 Provenance tracing and attack investigation (20) & Event timestamps; judged unreliable as a proxy for causality in BEEP & Dependencies traced backward and forward in temporal order; execution units & Unsynchronized distributed events noted in ViSe; clock-adjustment hypotheses in Rekhis and Boudriga & Out-of-order processing in ProTracer; incomplete logs in MORSE; missing behaviors reconstructed in RATScope & Process, file and socket identifiers; execution units & Kernel, audit or logs trusted in 8 of 20; compromised OS assumed in Martignoni et al.; anti-forensics modeled in 3 \\
\bottomrule
\end{tabular}
\end{adjustbox}
\end{table*}
\smallskip\noindent \textbf{Event time and observation time (Q1, Q2).} Two alert correlators built on the IDMEF alert format separate the two: M2D2 distinguishes when an event was detected from when an analyzer reported it, and Valeur et al. prefer the detection time, fall back to the report time and finally to the time of receipt, noting that these heuristics can introduce errors into the correlation~\cite{morinetal2002,valeuretal2004}. Debar and Wespi let alerts arrive out of time order as long as their timestamps come from one time frame, and Wang et al. trust neither the order of timestamps nor the order of arrival~\cite{debarandwespi2001,wangetal2006}. The provenance-era detectors (L1 to L4, S4, S5) use the event time recorded by the capture layer and take it as given, whereas host behavior models (L5) use the order of system calls within a process, mostly without timestamps. Investigation systems (S10) trace dependencies in timestamp order, although BEEP observes that asynchronous tasks and delays in user interaction make timestamps ineffective as a proxy for causality~\cite{leeetal2013a}. Order across windows is carried by learned state, caches, agent memory or scenario graphs, whose temporal semantics none of the systems states, with the exceptions of KnowHow's partial order over attack stages and the prerequisite-before-consequence condition of Ning et al.

\smallskip\noindent \textbf{Clocks (Q3).} None of the sampled systems states a numeric bound on the clock offset that remains after synchronization, although every cross-host order inferred from physical timestamps alone depends on one, as step 3 of the running example shows. Alert correlation and chronicles state the condition qualitatively: M2D2 matches events within a constant close to the maximum gap between synchronized clocks, Valdes and Skinner set a minimum time similarity to tolerate clock drift and late reports, Morin and Debar require the gap between sensor clocks to be coherent with the time resolution of the chronicles, and Debar and Wespi, Valeur et al. and Wang et al. require or assume synchronized sensor clocks~\cite{valdesandskinner2001,morinanddebar2003}. In the provenance era, AIQL synchronizes clocks by NTP and checks them against the server without a bound, and HERCULE and Naldurg et al. merge logs by timestamp without one~\cite{peietal2016,naldurgetal2004}. Digital forensics measures and models what the detectors leave unstated: Buchholz and Tjaden found only about 74\% of about 8,000 observed web servers within 10 s of a reference clock, Schatz et al. estimate skew and drift by correlating timestamps with reference sources and name deliberate clock tampering, and Willassen formalizes hypotheses about clock adjustments and tests them against the evidence~\cite{buchholzandtjaden2007,schatzetal2006,willassen2008}, while timeline reconstruction still assumes a correct clock on the investigated system~\cite{hargreavesandpatterson2012}. King et al. avoid the clock assumption for links between hosts by following recorded packets~\cite{kingetal2005}, as any order established by a trusted send and receive pair or by logical clocks does.

\smallskip\noindent \textbf{Late, reordered and missing records (Q4).} Whether the delivery question applies depends on a paper's input rather than its family. We count papers whose evaluated pipeline consumes host event records or hypotheses derived from them, and the ledger records the input layer of each paper. Of the papers of families L1 to L5, S4 and S5, 52 qualify. The two exceptions are Extractor, which translates threat reports into queries, and Karunanayake et al., who generate detection policies for IoT intrusion detection. Four of the 52 describe an input-ordering condition or convention. Unicorn exploits that all incoming edges of a vertex arrive before any outgoing ones and describes the additional updates needed when this partial order does not hold~\cite{hanetal2020}, KAIROS ingests edges chronologically as they appear~\cite{chengetal2024}, EdgeTorrent streams edges in the order in which they were observed~\cite{kingetal2023}, and AIQL relies on the timestamps of the collected data increasing monotonically per host when it partitions storage~\cite{gaoetal2018a}. Two papers report observations about missing records: Kruegel et al. found that no audit records were dropped in their evaluation~\cite{kruegeletal2003}, and HOLMES still flagged one attack scenario whose records lacked relevant events~\cite{milajerdietal2019a}. EagleEye notes that one or two missing behavior events can change how a pattern is classified~\cite{gyseletal2024}. None of the 52, however, specifies conditions under which its detection conclusions remain valid across a stated class of delayed or missing records. Alert correlators go further: they allow late reports and out-of-order arrival~\cite{valdesandskinner2001,debarandwespi2001}, and 9 of the 29 hypothesize unobserved attack steps or accommodate missing observations, from Ning et al.'s hypotheses validated against raw audit data to plan recognition over unobserved actions and CAML's policy for missing information~\cite{ningetal2004,geibandgoldman2001,cheungetal2003}, while two others name missing observations only as a limitation, and M-Correlator, which tolerates an incomplete fact base rather than missing observations, is not counted. Chronicles integrate alerts delayed by sensor processing or transport under the clock condition above~\cite{morinanddebar2003}. In investigation, ProTracer processes events concurrently and out of order across applications, MORSE reports incomplete logs in the DARPA TC data, and RATScope reconstructs behaviors missing from the records~\cite{maetal2016,hossainetal2020,yangetal2022}. Policy auditing represents incomplete logs as three-valued structures and decides only what the log determines~\cite{gargetal2011,chowdhuryetal2014}, while MonPoly assumes an ordered input~\cite{basinetal2015}. Unicorn also discusses race conditions in capture mechanisms, which can lose or misrecord events.

\smallskip\noindent \textbf{Identity (Q5).} Several systems name the identifiers they use for entities: NODLINK hashes the process identifier, thread identifier and command line, AIQL and SAQL qualify entity identifiers by host, SLEUTH uses process identifiers, and HERCULE relates log entries by process identifier and name, and R-CAID notes that identifiers differ between executions. Only Ko addresses identifier reuse, identifying files by inode and generation number~\cite{ko2000}, and none states how process identity is maintained under reuse or how corresponding entities, such as an account used on two hosts, are linked across hosts.

\smallskip\noindent \textbf{Trust in the evidence (Q6).} Where the provenance-era detectors state an attacker capability over the evidence, it is a strong trust assumption. Of the 38 eligible papers of families L1 to L4, S4 and S5, 28 state that the kernel, the audit framework or the stored logs are trusted or tamper-proof, rely on the guarantees of the capture system, or place the manipulation of evidence out of scope~\cite{gaoetal2018a,milajerdietal2019a,hanetal2020,chengetal2024,jiaetal2024,zengetal2022}. The question here concerns the integrity of host evidence, not of reports or training data. POIROT adds a separate secure log server, and R-CAID declares tamper-evident logging out of scope. The trust fraction concerns the provenance-era cohort L1 to L4, S4 and S5, with the input-based exclusions stated above, while the 14 L5 papers form a separate host-behavior cohort, none of which states an integrity assumption. The delivery cohort contains 52 papers, the provenance-era trust cohort 38. Eight of the twenty audited investigation papers state the trust assumption, and this family also contains the exceptions: Martignoni et al. assume a compromised operating system and analyze from the hypervisor, Purohit et al. consider a root attacker who tampers with a user-level tracing daemon, and formal forensics models anti-forensic attacks and reasons about which artifacts were tampered with~\cite{martignonietal2010,purohitetal2004,rekhisandboudriga2012,nealeetal2026}. Outside host detection, the argumentation systems take planted or forged evidence as input~\cite{nunesetal2016,karafilietal2020}, Wang et al. consider an attacker who delays steps or injects bogus alerts, Dain and Cunningham and Valdes and Skinner allow for forged source addresses, and R2U2 diagnoses spoofed sensor and GPS readings~\cite{schumannetal2015}. Guarantees exist at the logging layer, where secure and forward-secure logging~\cite{schneierandkelsey1999,maandtsudik2009,yavuzandning2009,bowersetal2014,karandeetal2017}, trustworthy whole-system provenance~\cite{batesetal2015}, tamper-evident auditing~\cite{paccagnellaetal2020a,zhaoetal2025}, the analyzed log sealing of journald~\cite{dorreandottenhues2025} and secure event ordering~\cite{correiaetal2022} address parts of the problem, although even secure logging can lose events through races between occurrence and commitment~\cite{paccagnellaetal2020b}. None of the detectors uses such guarantees in its reasoning, and none states which of its conclusions would survive the deletion, delay or forgery of records.

\smallskip\noindent \textbf{Formal treatments outside the detectors.} What the detectors leave unstated has formal treatments in the theory behind families S2 and S3: sequence matching and monitoring over late and out-of-order events~\cite{liuetal2009,basinetal2020}, pattern matching and monitoring with imprecise timestamps~\cite{zhangetal2010,basinetal2014}, and verdicts that mark what incomplete or disagreeing logs do not determine~\cite{basinetal2012}. Of the sampled systems, only the policy auditors give one of these problems, incomplete logs, a formal treatment, and none uses the treatments of late or imprecise data.

\smallskip\noindent \textbf{Summary.} The time and evidence assumptions of the checklist were stated, at least qualitatively, by a minority of the alert correlators and by chronicle recognition, almost all from 2000 to 2006, and they are measured and modeled in digital forensics and have formal treatments in monitoring and event processing. The provenance-era detectors document some timestamp, synchronization and identity conventions alongside trust in the capture layer. We located four input-ordering conditions or conventions, but no residual clock-error bound and no conditions under which detection conclusions remain valid across a stated class of delivery failures. Our data do not show why, but one possibility is that evaluation on recorded datasets replayed offline leaves capture and delivery properties implicit, which would require checking the dataset and replay conventions of each evaluation.

\section{Comparative assessment}
Section 4.3 recorded what individual systems state about time and evidence. This section asks what the families can do, from the capabilities that the reasoning problems of Section 7 require (Table 6) to where campaign composition happens (Section 5.2). The fourteen dimensions of Table 6 fall into four groups: the guarantees of Section 2 (correct evaluation of a declarative specification, the form of explanation and its independent checkability), temporal expressiveness (order and metric time, long horizons, recursion and joins, counting), reasoning over imperfect evidence (uncertainty; defaults, incompleteness and conflicts; semantic context) and operation (demonstrated raw-stream operation, late or out-of-order data, adversarial manipulation, knowledge update). Unlike Table 5, Table 6 rates capability at family level, from the representative systems and the cited work, and Appendix C defines each dimension and gives the evidence per family.

\smallskip\noindent \textbf{Rubric.} Symbols record capability only. A cell is assessed within the representative's stated task, language fragment and input model, and restrictions that define that task do not by themselves lower the rating: ++ means that the representative meets the full criterion of the dimension, defined in Appendix C, within that scope, + that it supports a named part of the criterion and leaves another part unsupported, $-$ a documented limitation, and $--$ absent by design. The criteria in Appendix C specify the assessed property and the requirement for full support, while intermediate ratings are qualitative judgments of partial support, explained in the evidence for each family. A cell records the strongest documented capability of a named representative within its stated scope. Correct-evaluation and late-data cells may rest on the cited language and evaluator theory even when the sampled security systems do not adopt that result, whereas raw-stream operation and adversarial resistance concern the evaluated systems. Appendix C identifies the supporting representative and scope, and Table 5 records separately what the sampled security papers adopt or assume. How widely it is shared is recorded separately, and a superscript \textsuperscript{1} marks a capability documented for only one representative system or approach, with publications on the same approach counted once. Limitations of other representatives are listed with the evidence in Appendix C.2. Three further values record the state of evidence rather than a level: dep (depends on the surrounding system, such as what an extractor supplies), n/s (not studied in the surveyed work) and n/a (not applicable), which are not points on an ordinal scale. Robustness is split into two columns, late or out-of-order data and adversarial manipulation, the latter judged against the attacker capabilities of Section 2. For the less obvious dimensions, ++ requires a checker separate from the producing system for independent checkability, metric temporal operators or constraints for order and metric time, no cut-off on how far back evidence can be related for long horizons, a probabilistic or annotated semantics for uncertainty, a formal treatment of out-of-order input for late data, and, for adversarial manipulation, a stated guarantee that the named conclusions remain valid for all manipulations a stated attacker model permits, while empirical resistance to the evaluated manipulations earns +. Appendix C defines each dimension and names the evidence and its source for every family.

\begin{table*}[!tp]
\caption{Families compared on fourteen dimensions.}
\label{tab:6}
\centering \scriptsize
\setlength{\tabcolsep}{3pt}
\begin{tabular}{>{\raggedright\arraybackslash}p{0.160\textwidth} c c c c c c c c c c c c c c }
\toprule
Family & \rotatebox{90}{\parbox{3.1cm}{\raggedright Correct evaluation of a declarative attack specification}} & \rotatebox{90}{\parbox{3.1cm}{\raggedright Explanation form}} & \rotatebox{90}{\parbox{3.1cm}{\raggedright Independently checkable}} & \rotatebox{90}{\parbox{3.1cm}{\raggedright Order and metric time}} & \rotatebox{90}{\parbox{3.1cm}{\raggedright Long horizons}} & \rotatebox{90}{\parbox{3.1cm}{\raggedright Recursion and joins}} & \rotatebox{90}{\parbox{3.1cm}{\raggedright Counting}} & \rotatebox{90}{\parbox{3.1cm}{\raggedright Uncertainty}} & \rotatebox{90}{\parbox{3.1cm}{\raggedright Defaults, incompleteness, conflicts}} & \rotatebox{90}{\parbox{3.1cm}{\raggedright Semantic context}} & \rotatebox{90}{\parbox{3.1cm}{\raggedright Demonstrated raw-stream operation}} & \rotatebox{90}{\parbox{3.1cm}{\raggedright Late or out-of-order data}} & \rotatebox{90}{\parbox{3.1cm}{\raggedright Adversarial manipulation}} & \rotatebox{90}{\parbox{3.1cm}{\raggedright Knowledge update}} \\
\midrule
L1 Dynamic graph learning & $--$ & Scores, summary graphs & $-$ & + & $-$ & $-$ & + & $-$ & $--$ & ++\textsuperscript{1} & ++ & n/s & +\textsuperscript{1} & $--$ \\
L2 Semantic-structural fusion & $--$ & Scores, attention weights & $-$ & + & $-$ & $-$ & + & $-$ & $--$ & ++ & ++ & n/s & +\textsuperscript{1} & $--$ \\
L3 LLM investigation agents & $-$ & Narratives & $-$ & n/s & + & $-$ & $-$ & n/s & $-$ & ++ & n/s & n/s & n/s & + \\
L4 Knowledge-guided learning & + & Rules with weights & n/s & +\textsuperscript{1} & + & + & + & ++\textsuperscript{1} & +\textsuperscript{1} & ++ & ++\textsuperscript{1} & n/s & +\textsuperscript{1} & + \\
L5 Host behavior models & $--$ & Anomalies, policy violations & $-$ & + & $-$ & $-$ & + & $-$ & $--$ & ++ & ++ & n/s & $-$ & + \\
S1 Alert correlation, plan recognition & + & Derivations, scenario graphs & + & ++ & + & + & n/s & ++ & + & dep & + & + & + & ++ \\
S2 Event-pattern correlation & ++ & Matches & $-$ & ++ & + & + & ++ & ++ & $-$ & dep & ++ & ++ & $-$ & ++ \\
S3 Temporal-logic monitoring & ++ & Verdicts, proof trees & ++\textsuperscript{1} & ++ & ++ & ++\textsuperscript{1} & ++ & ++ & + & dep & + & ++ & +\textsuperscript{1} & ++ \\
S4 Rule-based provenance analysis & + & Scenario graphs & $-$ & + & ++ & ++ & n/s & n/s & $-$ & + & ++ & n/s & +\textsuperscript{1} & + \\
S5 Provenance and knowledge-graph queries & + & Matches & $-$ & ++ & + & ++ & ++ & n/s & n/s & + & ++\textsuperscript{1} & n/s & n/s & ++ \\
S6 Attack graphs & ++ & Derivations & + & $-$ & n/a & ++ & n/s & ++ & n/s & dep & n/s & n/s & n/s & ++ \\
S7 Argumentation & ++ & Arguments & + & $-$ & n/a & + & n/s & n/s & ++ & dep & n/s & n/a & n/s & ++ \\
S8 Rule and specification learning & + & Rules & $-$ & + & n/s & + & n/s & n/s & n/s & + & + & n/s & n/s & ++ \\
S9 Threat-intelligence knowledge & dep & Vocabulary, ontology inferences & n/s & + & n/s & + & n/s & + & n/s & ++ & n/s & n/s & n/s & + \\
S10 Provenance tracing and attack investigation & $-$ & Dependency graphs & + & + & ++ & ++ & n/s & ++\textsuperscript{1} & $-$ & + & + & +\textsuperscript{1} & $-$ & + \\
\bottomrule
\end{tabular}
\end{table*}
No family dominates across the fourteen dimensions. Raw-stream operation is demonstrated in several learned and rule-based families, and semantic attributes are used both by learned encoders and by knowledge-based approaches, while explicit specifications, order constructs, recursion, joins and declarative knowledge update are concentrated in the symbolic families. Uncertainty has an explicit probabilistic or annotated semantics in alert correlation, event-pattern correlation, here for imprecise event times, temporal-logic monitoring, attack graphs, provenance tracing and knowledge-guided learning, and a partial combination mechanism in ontology-based diagnosis. The rating does not assess input calibration, which Section 7.3 examines for the outputs of learned provenance detectors. Defaults and incompleteness are treated mainly in argumentation and monitoring, and resistance to manipulated evidence is documented only partially and under bounded attacker models, which matches the evidence assumptions recorded in Table 5. For learned families, $--$ in the first column means that no declarative attack specification is evaluated, not that their computation is ill-defined. The ratings also rest on each family's own evaluations. Detector comparisons across families exist~\cite{lietal2024,bilotetal2025}, but the temporal reasoning formalisms have not been compared on common audit tasks (T4). Like all codings in this paper, the Table 6 ratings were assigned once (Section 9).

\subsection{Integration patterns in hybrid systems}
Table 6 rates families, but many recent systems combine learned and symbolic components, and what such a system can guarantee depends on how its components are connected. Table 7 classifies systems by integration pattern. The six patterns are not exclusive, since a system can instantiate more than one: symbolic reasoning over extracted facts (P1), learned detection (P2), and four ways of combining learned and symbolic stages. Wherever a pattern has a symbolic component, guarantees at the specification and evaluation levels cover only that component, and correspondence to real behavior depends on its inputs.

\begin{table*}[!tp]
\caption{Integration patterns observed in the literature.}
\label{tab:7}
\centering \scriptsize
\begin{tabular}{>{\raggedright\arraybackslash}p{0.132\textwidth} >{\raggedright\arraybackslash}p{0.132\textwidth} >{\raggedright\arraybackslash}p{0.169\textwidth} >{\raggedright\arraybackslash}p{0.207\textwidth} >{\raggedright\arraybackslash}p{0.226\textwidth}}
\toprule
\textbf{Pattern} & \textbf{Data flow} & \textbf{Representative systems} & \textbf{What is formally specified} & \textbf{Main limitation} \\
\midrule
P1 Symbolic over extracted facts & Expert rules or queries over extracted audit, configuration or vulnerability facts & SLEUTH, HOLMES, POIROT, KRYSTAL, MulVAL & The rules or queries; evaluation correctness is rarely stated & Rule coverage and the accuracy of extracted facts limit the conclusions; evasion has been demonstrated for particular rule sets \\
P2 Learned detection & Events to embeddings to scores & Unicorn, KAIROS, TFLAG, R-CAID, NODLINK, FLASH, EagleEye, CrossGuard & Training and inference procedures; no declarative campaign definition & Windows, expressiveness bounds, mimicry \\
P3 Learned detection, then LLM investigation & Scores select evidence; an LLM assembles the campaign & OCR-APT, ANCHOR, ProvSEEK, ProvAgent & No declarative campaign semantics identified & Composition without specification; bounded context \\
P4 Knowledge into learning & Rules shape training, learning tunes rule parameters, or logic labels embeddings & KnowGraph, CAPTAIN, Eckhoff et al. (alert embeddings) & Rule structure; parameters and weights are learned & Rules mostly atemporal; Eckhoff et al. add timestamp-order preferences as weak constraints; differentiable temporal logics exist outside security (Section 7.3) \\
P5 Learning into symbolic artifacts & Learned models draft rules, queries or specifications & SIGMERGE, ThreatRaptor, IRIS, invariant mining, LTLf patterns from reports (Eckhoff et al.) & The generated artifact, but not its agreement with the analyst's intent & Mostly single rules or one pattern template; validation ad hoc \\
P6 Recognition, then symbolic temporal composition & Recognized alerts or techniques are composed by ordering rules or temporal formulas & HOLMES, RapSheet, KnowHow; Eckhoff et al. (LTLf in telingo) & System-specific composition procedures in the first three; a temporal logic with known semantics in Eckhoff et al. & Recognition is rule- or signature-based; the general formalism is used only in a proof of concept over short symbolic alert traces \\
\bottomrule
\end{tabular}
\end{table*}
Pattern P6 comes closest to campaign-level temporal reasoning. In the provenance-based instances, the recognized units come from rules or from matching against threat reports, and the composition is defined by a procedure specific to the system. The one instance that uses a general temporal logic operates on symbolic SIEM alerts rather than host provenance, and its authors expect it not to scale to realistic alert volumes~\cite{eckhoffetal2025}. Patterns P3 and P6 thus differ in exactly the step that turns recognized units into a campaign claim, and Section 5.2 examines that step across systems.

\subsection{Where campaign composition happens}
In the running example of Section 3, the campaign claim arises in a final step that joins hypotheses across hosts and orders them in time. Table 8 locates this step in representative systems, from the alert correlators of the early 2000s to current provenance detectors: what each system recognizes, how it composes the recognized units across windows or hosts, and whether that composition has a declarative specification.

\begin{table*}[!tp]
\caption{Campaign composition in representative systems.}
\label{tab:8}
\centering \scriptsize
\begin{tabular}{>{\raggedright\arraybackslash}p{0.160\textwidth} >{\raggedright\arraybackslash}p{0.207\textwidth} >{\raggedright\arraybackslash}p{0.254\textwidth} >{\raggedright\arraybackslash}p{0.244\textwidth}}
\toprule
\textbf{System} & \textbf{Recognition} & \textbf{Composition across events, windows or hosts} & \textbf{Declarative specification of composition} \\
\midrule
Ning et al., CCS 2002 & Intrusion detection alerts & Prerequisite and consequence chaining & Yes, logical prerequisites and consequences \\
Morin and Debar, RAID 2003 & Intrusion detection alerts & Chronicle recognition with metric constraints & Yes, chronicles \\
Albanese et al., ESORICS 2011 & Incoming alerts associated with vulnerability exploits & Matching and indexing of attack occurrences against probabilistic temporal attack graphs, with possible continuations & Yes, an explicit temporal and probabilistic attack-occurrence model over alerts, a task distinct from learned provenance detection \\
HOLMES, S\&P 2019 & Rules map events to tactics & High-level scenario graph ordered by kill-chain stage & No; a system-specific construction procedure \\
RapSheet, S\&P 2020 & Endpoint alerts & Tactical provenance graph scored by temporal order & No; a scoring function \\
Unicorn, NDSS 2020 & Learned graph sketches & Evolutionary model over the sketch stream & No \\
KRYSTAL, Computers \& Security 2022 & Tag-propagation alerts, Sigma rules and graph patterns as SPARQL queries & Backward and forward chaining along causal property paths, with iterative alert scoring & Partly; SPARQL graph patterns and path queries, without temporal or metric constraints \\
KAIROS, S\&P 2024 & Learned event scores & Queue of anomalous windows merged into summary graphs & No \\
NODLINK, NDSS 2024 & Learned anomaly scores & Steiner tree over alerted nodes & No; a specified optimization objective, but no declarative definition of the campaign claim \\
CrossGuard, CPSIoTSec 2025 & Learned window classification & Delegated to an unspecified SIEM step & No \\
OCR-APT, CCS 2025 & GNN subgraph anomalies & LLM iteration over anomalous subgraphs & No \\
Eckhoff et al., Neurosymbolic AI 2025 & Signature-based SIEM alerts mapped to techniques & LTLf attack patterns checked by telingo over the alert trace & Yes, temporal equilibrium logic on finite traces; proof of concept \\
ANCHOR, preprint 2026 & Relation-aware anomaly detection & LLM with a confidence-gated attack-tracking cache & No \\
KnowHow, NDSS 2026 & Natural-language matching to techniques & Ordering by tactic and APT lifecycle stage & No; system-specific ordering constraints \\
\bottomrule
\end{tabular}
\end{table*}
The selected examples contrast two designs. The early alert correlators define composition declaratively, by prerequisites and consequences or by chronicles~\cite{ningetal2002,morinanddebar2003}. Formal temporal composition continued in alert-based systems during 2007 to 2016 (Albanese et al.), so what changed with the later provenance pipelines is the input layer and the way composition is specified, while temporal reasoning remained part of security research. In the selected provenance-based systems, composition is implemented by a construction procedure (HOLMES), a scoring function (RapSheet), an optimization objective (NODLINK), causal graph queries (KRYSTAL), learned state (Unicorn, KAIROS), system-specific ordering constraints (KnowHow) or an LLM (OCR-APT, ANCHOR), and CrossGuard leaves composition to an unspecified SIEM step. KRYSTAL states attack patterns declaratively as SPARQL graph patterns, without temporal or metric constraints or correspondence assumptions. The other procedures and objectives are precise, but none of the provenance-based systems defines the campaign predicate that its output is meant to establish, or shows that its computation establishes it, so a compact graph or a high score does not, on its own, certify the intended campaign claim. This is the gap behind RQ3 and challenge C1.

\subsection{Takeaways}
\begin{itemize}
\item \textbf{T1.} Declarative campaign composition is documented in alert correlation and in one recent proof of concept over SIEM alerts, whereas the selected provenance-based systems of Table 8 compose campaigns through system-specific procedures or objectives, learned state, causal graph queries without temporal constraints or LLMs.
\item \textbf{T2.} Combining learned detection with temporal reasoning requires an explicit mapping from scored outputs to typed hypotheses and a semantics for composing their uncertainty. No surveyed system composes the uncertain outputs of learned provenance detectors in a temporal formalism with a stated semantics for their confidences.
\item \textbf{T3.} The robustness evaluations we identified in the sampled host-detector cohort concern manipulated behavior, such as mimicry against learned detectors, and we found none evaluating deletion, delay or forgery of the input records. Most provenance-era detectors trust the capture components or the committed records without saying which conclusions survive late, lost or forged records, while formal forensics and argumentation reason about manipulated evidence outside that cohort (Section 4.3).
\item \textbf{T4.} Detectors have been compared across families, and temporal reasoners with each other, but we found no comparison of the surveyed temporal reasoning formalisms on common host-audit tasks with aligned pattern semantics and explicit clock and delivery assumptions.
\end{itemize}

\section{Design constraints and limits}
Section 5 showed that no family combines all the capabilities of Table 6. This section examines five considerations relevant to combining these capabilities: throughput against specified reasoning, structure against semantics, causal depth against information loss, expressiveness against decidability, and evaluation-level guarantees against an adversary who influences the inputs.

\subsection{Throughput versus specified reasoning}
Learned detectors produce scores, not consequences of stated rules, and their reported accuracy depends heavily on benchmark properties~\cite{bilotetal2025,liuetal2025}. A sound symbolic evaluator derives a conclusion only if the rules and facts entail it under its stated semantics, but expressive rule languages are costly over raw events (Section 6.4). Rule-based provenance systems such as SLEUTH and CAPTAIN show that simple tag-propagation rules do run in real time~\cite{hossainetal2017,wangetal2025b}, and for a fixed query in a formally specified fragment, complex event automata update in constant time per event, with result enumeration counted separately~\cite{grezetal2019}. The expressiveness of the rule language is a major determinant of worst-case cost, alongside implementation, retained state and output size, rather than the symbolic approach as such.

\subsection{Structure versus semantics}
Living-off-the-land attacks use administrative binaries whose graph neighborhoods closely resemble legitimate maintenance. Fusion models that embed command lines and attributes alongside graph context address this directly~\cite{rehmanetal2024,gyseletal2024}. Symbolic engines can express precise conditions over arguments, but only after a parser has turned unstructured strings into typed relations. If extraction misses or mislabels an argument, a reasoner cannot in general recover its true value without redundant evidence or domain constraints, and abductive alternatives remain hypotheses. Every system that reasons over extracted predicates thus depends directly on the quality of extraction.

\subsection{Multi-hop causality and expressiveness}
Campaigns can span many causal hops from entry to exfiltration. Where the receptive field grows rapidly with depth, message-passing GNNs compress information from many distant nodes into fixed-size vectors, a bottleneck that loses long-range signals~\cite{alonandyahav2021}, and deep GNNs tend toward indistinguishable node representations~\cite{lietal2018}. Logical characterizations make these limits precise, but only for specific architectures and tasks. Among node classifiers definable in first-order logic with two variables and counting, aggregate-combine GNNs express exactly those definable in graded modal logic, while adding a global readout captures more~\cite{barceloetal2020}. Mean- or softmax-style aggregators also lose neighbor multiplicities that sum aggregation keeps~\cite{xuetal2019}. For temporal message passing, global and local mechanisms have Weisfeiler--Leman characterizations and incomparable expressive power on arbitrary temporal graphs~\cite{waegaandrawson2025}.

These results bound a backbone on a stated input representation and task. They do not automatically extend to a whole detector, whose features, preprocessing, windowing and graph construction can encode a join, such as linking the process that reads a credential to the one that later uses it, as an edge or an attribute. A claim that a detector cannot express a pattern needs a pair of input graphs that its architecture cannot distinguish although the pattern holds in only one. Expressiveness, learnability and empirical generalization are separate questions, and none of the surveyed detectors states which pattern class it can distinguish.

\subsection{Expressiveness versus decidability}
For detection, the cost that matters is that of monitoring a trace or answering queries over a growing set of facts, not that of design-time tasks such as satisfiability. On this measure several formalisms have tractable cores, although the bounds below measure different resources (Appendix B). Propositional past-time temporal logic can be monitored with memory independent of trace length~\cite{havelundandrosu2002}, fixed queries in the unary-filter fragment of complex event logic have constant update cost per event~\cite{grezetal2019}, Datalog query evaluation is P-complete in data complexity~\cite{dantsinetal2001}, DL-Lite query answering is first-order rewritable~\cite{calvaneseetal2007}, regular path queries are evaluable in nondeterministic logarithmic space~\cite{barcelo2013}, and metric temporal rule languages have fragments in NL and TC\textsuperscript{0}~\cite{waegaetal2020}.

The features that security reasoning asks for beyond these cores are expensive in the worst case. Existential rules can entail an unobserved entity such as an attacker session, but for the existential extension of DatalogMTL, reasoning is undecidable under the natural semantics and decidable only under restricted semantics and program classes~\cite{lanzingeretal2023}, and selecting such an entity as an explanation additionally needs an abductive semantics. Defaults under stable-model semantics make model existence NP-complete already for finite propositional normal programs~\cite{dantsinetal2001}. Skeptical acceptance under preferred argumentation semantics, one way to adjudicate conflicting evidence, is $\Pi_2^P$-complete~\cite{dvorakanddunne2018}. Exact inference in probabilistic logic programs, one way to give uncertain composition (pattern AP7 in Section 7.7) a semantics, is \#P-hard~\cite{deraedtetal2007}. For the original Halpern--Pearl definition of actual causation, deciding causality is NP-complete for binary models and $\Sigma_2^P$-complete in general~\cite{eiterandlukasiewicz2002}, and later definitions have different bounds, all concerning counterfactual root causes rather than the backward dependency tracing that provenance systems call root-cause analysis. Consistency of qualitative interval constraints (AP8b) is NP-complete in Allen's algebra and polynomial in its ORD-Horn subclass~\cite{vilainandkautz1986,nebelandburckert1995}. Table B1 in Appendix B gives these bounds for each formalism.

Cost alone is not a reason to avoid symbolic reasoning, since the detection-time cores of several formalisms are tractable (T5). What these cores leave out is largely what R3 and R4 in Section 7 ask for, and for these features the literature gives worst-case results but no evaluation on audit workloads (C2).

\subsection{Evasion and the levels of guarantee}
The last constraint concerns the guarantee levels of Section 2 when an adversary influences the inputs. In one study, mimicry attacks evaded each of the evaluated graph-based provenance detectors without changing the attack's effect~\cite{goyaletal2023}, and an analysis of public Sigma rules found many that small changes to commands or process attributes evade~\cite{uetzetal2024}, so in the evaluated settings neither learned nor rule-based detection was robust by construction. Defenses exist on both sides: R-CAID ties embeddings to root causes~\cite{goyaletal2024}, and KnowHow reports robustness to several mimicry attacks~\cite{mengetal2026}. Behavioral mimicry can evade detection while the records remain faithful. Deleting or falsifying records, or corrupting their extraction into facts, can instead break the correspondence between input facts and host behavior and violate the preconditions of an evaluation-level guarantee. Events missed upstream cannot be derived from their own records, although other observations, domain rules or abductive hypotheses can sometimes recover them. Mis-extracted facts yield soundly derived but false conclusions, and deleted or back-dated records can invalidate temporal conclusions whose derivations use them or assume their completeness~\cite{paccagnellaetal2020b}. Which correspondence guarantee is achievable depends on the attacker's capability over the evidence, which most provenance-era detectors state only as an exclusion (Section 4.3).

\subsection{Takeaways}
\begin{itemize}
\item \textbf{T5.} Rule-language expressiveness is a major determinant of the worst-case cost of symbolic reasoning, alongside workload, implementation, retained state and output volume, so symbolic reasoning is not slow as such, and simple tag-propagation rules and complex event automata run in real time.
\item \textbf{T6.} Logical characterizations bound what specific GNN architectures, including temporal message passing, can distinguish on stated tasks, yet no surveyed detector says which patterns its full pipeline can tell apart.
\item \textbf{T7.} For the tasks reviewed here, uncertainty, unrestricted generation of entities and stable-model negation raise worst-case reasoning complexity, while restricted fragments, such as stratified negation in DatalogMTL, avoid some of these increases, so the choice of formalism and fragment is a security design decision.
\item \textbf{T8.} Learned and rule-based detectors can both be evaded by manipulated behavior, while manipulated evidence can break the correspondence between records and host behavior, which no evaluation-level guarantee repairs.
\end{itemize}

\section{Temporal KRR for security reasoning problems}
We organize the state of the art in KRR around six reasoning problems that recur across the families of Section 4. For each problem we state what it requires, which KRR domains address it and how they have advanced, what has reached security practice, and what remains open. Section 7.7 assesses the formalisms on nine attack-reasoning patterns, and Appendix B analyzes each KRR domain in depth.

\subsection{R1: Recognizing ordered multi-step patterns}
\noindent \textbf{Requirement.} Detection needs patterns that combine order, deadlines, joins across entities and counts within windows, such as a process that reads a credential file and opens an outbound connection within ten minutes.

\smallskip\noindent \textbf{State of the art.} Several established approaches address R1. Temporal-logic monitors and complex event processing, described as families S2 and S3 in Section 4.2, give such patterns a formal semantics with efficient evaluation: MonPoly adds aggregation over monitorable fragments~\cite{basinetal2015}, DejaVu monitors first-order past-time logic with binary decision diagrams~\cite{havelundetal2017}, and CEL compiles fixed queries with unary filters to automata with constant update cost per event, with result enumeration counted separately~\cite{grezetal2019}. Time-aware graph queries express such patterns over stored provenance~\cite{gaoetal2018a}.

\smallskip\noindent \textbf{In security.} Families S2, S3 and S5 address R1 (Section 4.2), and recent LLM pipelines generate Sigma rules from threat reports~\cite{caietal2026}.

\smallskip\noindent \textbf{Gap.} Sigma and the vendor correlation languages we examined specify their syntax and describe their matching in prose, but none has a published formal semantics or a translation into a formal CEP or monitoring language. CEP engines have been evaluated on host-monitoring workloads: SAQL builds on the Siddhi engine and compares against it on audit-derived queries~\cite{gaoetal2018b}. What is missing is a comparison of the surveyed temporal reasoning formalisms on common audit tasks with aligned pattern semantics and explicit clock and delivery assumptions. Expert-written rules can be evaded, as an evaluation of public Sigma rules shows~\cite{uetzetal2024}.

\subsection{R2: Composing campaigns across windows and hosts}
\noindent \textbf{Requirement.} Campaigns span days and hosts. Composing them needs temporal context over long horizons, joins across hosts, ordering taken from a specific attack pattern and, for unbounded ancestry or reach, recursion or an equivalent path construct.

\smallskip\noindent \textbf{State of the art.} Allen's interval algebra orders stages qualitatively and can absorb imprecise timestamps when the admissible relations account for their uncertainty~\cite{allen1983}. The Event Calculus (EC), as implemented in RTEC, recognizes composite events with interval-valued fluents over windows~\cite{artikisetal2015}, including events whose effects begin only after a delay~\cite{mantenoglouetal2025}. Stream reasoning, for example in the LARS framework, gives declarative semantics to reasoning over unbounded streams~\cite{dellaglioetal2017,becketal2018}. Temporal rule languages such as DatalogMTL combine recursion with metric time and have engines with stated guarantees~\cite{wangetal2024,bellomarinietal2025a} (Appendix B.3), complex event recognition has a relational automaton model in symbolic register transducers~\cite{alevizosetal2024}, and temporal ASP has a metric extension over timed traces with an implementation~\cite{beckeretal2024,hahnetal2026}.

\smallskip\noindent \textbf{In security.} The selected provenance-based systems of Table 8 compose campaigns through system-specific procedures or objectives, learned state, causal graph queries or LLM agents, and temporal ASP has composed SIEM alerts against attack patterns in a proof of concept (S1).

\smallskip\noindent \textbf{Gap.} Beyond CEP-based query systems such as SAQL, no temporal rule language, Event Calculus engine or stream reasoner has, to our knowledge, been applied to host audit data. RTEC has been evaluated with delayed and out-of-order events outside security~\cite{artikisetal2015}, but not on host audit streams under the clock, delivery and attacker conditions considered here.

\subsection{R3: Reasoning under uncertainty}
\noindent \textbf{Requirement.} Learned detectors and text matchers produce uncertain technique hypotheses that must be combined over time, including when several hypotheses rest on shared evidence.

\smallskip\noindent \textbf{State of the art.} Probabilistic and weighted logics attach probabilities or weights to rules, as in Markov logic~\cite{richardsonanddomingos2006} and probabilistic logic programming~\cite{deraedtetal2007}. Temporal variants followed: annotated probabilistic temporal logic bounds the probability that one action follows another within a time window~\cite{shakarianetal2011}, a probabilistic Event Calculus handles uncertain input events~\cite{skarlatidisetal2015}, Alevizos et al. survey probabilistic complex event recognition~\cite{alevizosetal2017}, and metric temporal logic has been evaluated approximately over uncertain streams~\cite{delengandheintz2019}. Learning has since been combined with these logics: DeepProbLog grounds probabilistic rules in neural predicates~\cite{manhaeveetal2018}, the robustness of signal temporal logic (STL) serves as a differentiable loss and STL formulas can be learned by gradient descent~\cite{frondaandabbas2022,leungetal2023,licaietal2024}, differentiable Datalog and embedding-based probabilistic logic extend neuro-symbolic frameworks~\cite{lihuangandnaik2023,maeneandderaedt2023}, PyReason implements annotated logic over graphs with temporal extensions~\cite{adityaetal2023}, DeepProbCEP couples a neural event classifier to complex-event rules and is evaluated under adversarial conditions~\cite{roigvilamalaetal2023}, and recent work gives neuro-symbolic models temporal structure through automata, relational Markov models and relational temporal knowledge~\cite{manginasetal2025,desmetetal2025,lorelloetal2025}.

\smallskip\noindent \textbf{In security.} Alert correlation combined alerts probabilistically, by feature similarity and causal networks~\cite{valdesandskinner2001,qinandlee2004}, and assigned probabilities to attack occurrences under time constraints~\cite{albaneseetal2011}. Annotated probabilistic temporal logic has been used to allocate monitoring resources over alerts~\cite{jajodiaetal2016}, R2U2 combines temporal observers with Bayesian diagnosis~\cite{schumannetal2015}, and Event Calculus security monitoring has been combined with Bayesian estimation of threat likelihood~\cite{amalioandspanoudakis2008}. On host data, temporal dependence networks infer security states over object histories derived from system calls with a Bayesian model~\cite{yangetal2018}. Bayesian attack graphs (S6) and knowledge-guided GNNs (L4) handle uncertainty without metric temporal operators, and CAPTAIN learns thresholds rather than reasoning about uncertainty explicitly~\cite{wangetal2025b}.

\smallskip\noindent \textbf{Gap.} None of the surveyed learned provenance pipelines documents a calibration and interface procedure that turns detector outputs into probabilistic temporal hypotheses while specifying their shared evidence. An anomaly score is not automatically a probability, and calibrated marginals alone do not determine the joint probability of a campaign. The confidences that fuzzy or t-norm semantics propagate are truth degrees, not probabilities, and do not specify the joint distribution needed when hypotheses share evidence (AP7). The security uses above work on alerts or on dependency models of their own, and the declarative probabilistic temporal logics surveyed here have not been applied to the outputs of learned provenance detectors or to whole-system provenance data. The temporal neuro-symbolic models above have been evaluated on perception and sequence tasks rather than on audit data.

\subsection{R4: Missing, conflicting and manipulated evidence}
\noindent \textbf{Requirement.} Reasoning must cope with deleted events, contradicting records and manipulated times, and must report a claim as undetermined, rather than negative, when the evidence under the stated completeness assumptions does not settle it.

\smallskip\noindent \textbf{State of the art.} Belief revision formalizes minimal change~\cite{alchourronetal1985}, and answer set programming encodes defaults and, once the admissible hypotheses and their selection are specified, abduction of missing facts~\cite{gelfondandlifschitz1988,gebseretal2019}, with temporal extensions~\cite{cabalaretal2019}. Abstract and structured argumentation adjudicate conflicting evidence~\cite{dung1995,garciaandsimari2004,modgilandprakken2014}, and interval-algebra consistency checking detects contradictory temporal constraints~\cite{nebelandburckert1995}. Three-valued monitoring semantics returns an inconclusive verdict when a finite prefix of the trace does not determine a property~\cite{baueretal2011}, which addresses uncertainty about how a trace continues, whereas missing historical records additionally need a model of the admissible completions of the observed trace. Revising conclusions when input is retracted or expires has stream-reasoning treatments in retractable complex event processing~\cite{anicicetal2011} and in answer update for LARS programs by truth maintenance~\cite{becketal2015}, and out-of-order monitoring handles late events~\cite{basinetal2020}. Inconsistency-tolerant query answering over temporal data has been defined for ontology-mediated queries~\cite{bourgauxetal2019}, and conflicts and repairs over time-annotated facts have been defined for DatalogMTL~\cite{bienvenuetal2025}.

\smallskip\noindent \textbf{In security.} Secure logging and tamper-evident auditing address manipulation at the logging layer (Section 4.3). Formal digital forensics reasons explicitly about anti-forensic attacks that delete or alter evidence~\cite{rekhisandboudriga2012}, answer set programming has generated investigation hypotheses from evidence~\cite{costantinietal2015}, and argumentation has been used for attribution (S7).

\smallskip\noindent \textbf{Gap.} Formal forensics accounts for anti-forensic attacks after the fact, but no reasoning layer yet takes integrity signals from the logging layer, such as authenticated completeness or sequence-gap indicators, as input to its conclusions, and no detection system yet reasons formally about the completeness of its evidence or the consistency of its timestamps under a stated attacker model.

\subsection{R5: Explanation and forensic evidence}
\noindent \textbf{Requirement.} Analysts and forensic reviewers need explanations they can check, ideally derivations that link an alert to the events and rules that produced it, whereas identifying a counterfactual root cause is a separate task that needs a causal model.

\smallskip\noindent \textbf{State of the art.} Actual causation, with responsibility and blame, formalizes root cause~\cite{chocklerandhalpern2004,halpernandpearl2005,halpern2016} at high computational cost~\cite{eiterandlukasiewicz2002}, and argumentation produces explanations as accepted arguments. Datalog engines can record derivation provenance at scale~\cite{zhaoetal2020}. For temporal monitoring, verdicts can be explained as proof trees in a sound and complete proof system, for MTL~\cite{limaetal2023} and for MFOTL, where WhyMon's proof trees are certified by a checker formally verified in Isabelle~\cite{limaetal2024}. Actual causation has recently been brought to event traces, with temporal causal reasoning over structural equation models~\cite{gladyshevetal2025} and causal explanations of safety-property violations in discrete event systems~\cite{goleretal2026}.

\smallskip\noindent \textbf{In security.} Scenario graphs~\cite{milajerdietal2019a}, tactical provenance graphs~\cite{hassanetal2020}, root-cause-aware embeddings~\cite{goyaletal2024} and LLM-generated attack stories~\cite{alyetal2025} explain alerts (S4, L1 and L3 in Section 4).

\smallskip\noindent \textbf{Gap.} Apart from WhyMon, we found no temporal reasoner that documents a derivation output checkable outside the reasoner, and no use of certified temporal explanations in intrusion analysis. LLM narratives are not checked against a derivation in the systems we found, although structured claims in a narrative could be linked to checked derivations (C1), and the root causes reported by provenance systems are dependency origins, not counterfactual causes under a stated causal model.

\subsection{R6: Acquiring and maintaining knowledge}
\noindent \textbf{Requirement.} Rules and patterns must be obtained from threat intelligence and data, validated, and kept current as techniques drift.

\smallskip\noindent \textbf{State of the art.} Ontology-based data access provides a shared ontology and mappings for querying heterogeneous sources with logical entailment~\cite{calvaneseetal2007,artaleetal2017}. Temporal knowledge can be learned in several formalisms: temporal logic point processes place propositional temporal rules inside the intensity function of a continuous-time point process, so that rules and weights are learned by gradient~\cite{lietal2020,yanetal2023,yangetal2024}, probabilistic Event Calculus theories are learned online from streams~\cite{katzourisetal2021}, STL formulas and their interval parameters by gradient descent~\cite{frondaandabbas2022,licaietal2024}, DatalogMTL rules are mined window by window with a Datalog rule learner and scored by time-sensitive confidence~\cite{wangwaegaandcuencagrau2024}, and LTLf formulas are learned by inductive logic programming~\cite{ieloetal2025}.

\smallskip\noindent \textbf{In security.} Families S8 and S9 (Section 4.2) cover the acquisition of rules and threat-intelligence knowledge. Among them, ThreatRaptor and SIGMERGE synthesize queries and rules from reports~\cite{gaoetal2021,caietal2026}, KnowHow applies report knowledge through a general indicator representation~\cite{mengetal2026}, CAPTAIN adapts rule parameters to each environment~\cite{wangetal2025b}, and Eckhoff et al. translate threat-report sentences into multi-step LTLf attack patterns, using one pairwise elicitation template whose outputs are composed into longer formulas~\cite{eckhoffetal2025}.

\smallskip\noindent \textbf{Gap.} In host security, the synthesized multi-step temporal patterns we found are proofs of concept built from one pairwise elicitation template and evaluated on short illustrative traces, while most synthesized rules capture single events or steps and are validated beyond syntax only ad hoc. Apart from the early learning of specifications for single programs (S8), we found no application of inductive logic programming to whole-system provenance or to metric, multi-stage campaign patterns.

\subsection{A catalog of attack-reasoning patterns}
We turn the requirements of Sections 7.1 to 7.6 into nine concrete attack-reasoning patterns: AP1 to AP3 instantiate pattern recognition (R1), AP4, AP5 and AP9 composition across windows and hosts (R2), AP7 uncertainty (R3), and AP6 and AP8 missing and conflicting evidence (R4). AP3 is the credential reuse of the running example (Section 3). We then assess whether eleven symbolic formalisms and a learned baseline provide constructs for them (Table 9). Appendix D states the conventions that fix the intended matches, such as the treatment of equal timestamps, strict temporal order and the interval endpoints of "within", and gives the encodings and constructions behind the symbolic ratings. The patterns isolate reasoning requirements from attack scenarios: a match establishes the specified behavior, while interpreting it as malicious requires a security policy and contextual evidence, since several patterns also describe legitimate administration. Table 9 assesses expressibility over a fixed input trace under the conventions of Appendix D. It does not assess clock-error assumptions, delivery guarantees or the deployment policies for issuing, revising and finalizing verdicts.

\begin{itemize}
\item \textbf{AP1 Ordered chain with deadline.} The same process reads a credential file and opens an outbound connection within ten minutes of the read.
\item \textbf{AP2 Fan-out with fixed k.} One process writes to at least k distinct controller endpoints within one minute, where k is a constant of the specification, as in the Triton case. If k is an input value, the pattern needs general counting.
\item \textbf{AP3 Cross-entity join.} A credential c read by process p is later used to authenticate by a different process q (p $\neq$ q). The join is on c, which must be represented as an entity or attribute.
\item \textbf{AP4 Time-respecting ancestry.} The exfiltrating process descends from a process started from a downloaded file, through a chain of spawn edges each later than its predecessor, and the whole chain lies within 72 hours of the download.
\item \textbf{AP5 Ordering from a given attack flow.} The steps of a specific attack flow, for example from Attack Flow or ChronoCTI, occur in that flow's order on one host over days, with unbounded gaps. We encode the totally ordered case, since a partial order with incomparable branches additionally needs shared step instances, which we do not assess. ATT\&CK tactics themselves impose no linear order~\cite{cisa2023}.
\item \textbf{AP6 Absence with exceptions.} (a) Over complete data: a privileged process starts with no logon in the preceding five minutes, unless it belongs to approved maintenance whose binary is signed. (b) As a default: the conclusion is retracted when updated data contain a matching logon or exception.
\item \textbf{AP7 Uncertain composition.} The probability of a campaign hypothesis is computed from uncertain technique hypotheses that may share evidence, such as two hypotheses derived from the same alert.
\item \textbf{AP8 Temporal inconsistency.} (a) As data comparison: a child process carries an earlier event time than its parent's creation, which violates a process-creation invariant but is not by itself evidence of manipulation. (b) As constraint consistency: qualitative temporal constraints from different sources, such as host logs and network sensors, are jointly unsatisfiable.
\item \textbf{AP9 Time-respecting reach.} Influence from a compromised workstation reaches a controller's network zone over a changing connectivity graph, using only connections that exist at increasing times, within a time bound $\Delta$.
\end{itemize}

AP4 is the pattern whose formalization is least obvious, because its deadline counts from the download and not from the previous step. An encoding must therefore preserve the originating download and its deadline throughout the match, and the encodings in Appendix D do so by carrying the download identifier d through the recursion: each spawn must be strictly later than the one before, and every step and the exfiltration must lie within 72 hours of the download. Restarting a 72-hour window at each step instead would wrongly accept a chain with steps at hours 70 and 140. Appendix D gives AP4 encodings or constructions for the positively assessed symbolic rows, except for the Event Calculus, whose support rests on the cited cyclic-dependency extension without a constructed encoding, and graph-query support remains unassessed. The formalisms differ in whether recursion over metric operators is part of the language, as in DatalogMTL and in MFOTL with the recursive let of VeriMon, or whether the deadline needs explicit time variables with comparisons, as in LARS, temporal ASP and Datalog. AP9 is formalized analogously, with connection edges and the compromise episode e.

\begin{table*}[!tp]
\caption{Constructs for the attack-reasoning patterns, by formalism.}
\label{tab:9}
\centering \scriptsize
\begin{tabular}{>{\raggedright\arraybackslash}p{0.115\textwidth} >{\raggedright\arraybackslash}p{0.056\textwidth} >{\raggedright\arraybackslash}p{0.056\textwidth} >{\raggedright\arraybackslash}p{0.056\textwidth} >{\raggedright\arraybackslash}p{0.056\textwidth} >{\raggedright\arraybackslash}p{0.056\textwidth} >{\raggedright\arraybackslash}p{0.056\textwidth} >{\raggedright\arraybackslash}p{0.056\textwidth} >{\raggedright\arraybackslash}p{0.056\textwidth} >{\raggedright\arraybackslash}p{0.056\textwidth} >{\raggedright\arraybackslash}p{0.165\textwidth}}
\toprule
\textbf{Formalism} & \textbf{AP1} & \textbf{AP2} & \textbf{AP3} & \textbf{AP4} & \textbf{AP5} & \textbf{AP6} & \textbf{AP7} & \textbf{AP8} & \textbf{AP9} & \textbf{Applied to security data} \\
\midrule
Temporal message-passing GNN & L & L & L & L & L & a L, b L & L & a L, b -- & L & Yes, provenance detectors (L1, L2) \\
MFOTL (MonPoly, VeriMon) & \checkmark{} & \checkmark{} & \checkmark{} & \checkmark{} recursive let (VeriMon) & \checkmark{} & a \checkmark{}, b \checkmark{} & -- & a \checkmark{}, b -- & \checkmark{} recursive let (VeriMon) & Mainly compliance logs; ORCHIDS uses its own temporal logic \\
CEP (chronicles) & \checkmark{} & \checkmark{} & \checkmark{} & -- & \checkmark{} & a \checkmark{}, b \checkmark{} & -- & a \halfcirc{}, b -- & -- & Chronicles over intrusion detection alerts \\
Event Calculus (RTEC) & \halfcirc{} & \halfcirc{} & \checkmark{} & \halfcirc{} & \checkmark{} & a \halfcirc{}, b \halfcirc{} & \halfcirc{} probabilistic EC & a \checkmark{}, b -- & \halfcirc{} & Event Calculus for security monitoring~\cite{amalioandspanoudakis2008}; none for RTEC \\
LARS & \checkmark{} & \checkmark{} & \checkmark{} & \halfcirc{} & \checkmark{} & a \checkmark{}, b \checkmark{} & -- & a \checkmark{}, b \halfcirc{} & \halfcirc{} & We found none \\
Temporal ASP (telingo) & \halfcirc{} metric temporal ASP & \halfcirc{} & \checkmark{} & \halfcirc{} & \checkmark{} & a \halfcirc{}, b \halfcirc{} & ? probabilistic ASP & a \halfcirc{}, b \halfcirc{} & \halfcirc{} & Short SIEM alert traces; alert-cluster labeling with ordering preferences~\cite{eckhoffetal2025} \\
Graph queries (AIQL; RPQ for paths) & \checkmark{} & \checkmark{} & \checkmark{} & ? & \checkmark{} & a \halfcirc{}, b \halfcirc{} & -- & a \checkmark{}, b -- & ? & Audit data (AIQL, SAQL, POIROT) \\
Datalog with comparisons and stratified negation & \halfcirc{} & \halfcirc{} & \halfcirc{} & \halfcirc{} & \halfcirc{} & a \halfcirc{}, b \halfcirc{} & -- & a \halfcirc{}, b -- & \halfcirc{} & Attack graphs (MulVAL); static analysis \\
DatalogMTL & \checkmark{} & \halfcirc{} needs $\neq$ & \halfcirc{} needs $\neq$ & \checkmark{} & \checkmark{} & a \halfcirc{}, b \halfcirc{} stratified negation over $\mathbb{Q}$ & -- & a \checkmark{}, b -- & \checkmark{} & We found none \\
Probabilistic logic programming (ProbLog) & \halfcirc{} & \halfcirc{} & \halfcirc{} & \halfcirc{} & \halfcirc{} & a \halfcirc{}, b \halfcirc{} & \checkmark{} & a \halfcirc{}, b -- & \halfcirc{} & We found none on audit data \\
Allen interval algebra & -- & -- & -- & -- & \halfcirc{} & -- & -- & a --, b \checkmark{} & -- & Rarely, in alert correlation \\
STREL & \halfcirc{} & -- & -- & -- & \halfcirc{} & a \halfcirc{}, b \halfcirc{} & -- & -- & -- & We found none \\
\bottomrule
\end{tabular}
\par\vspace{3pt}\begin{minipage}{0.96\textwidth}\scriptsize A \checkmark{} marks direct support in the named language or fragment, \halfcirc{} marks support via a stated encoding, such as timestamps as data values, or via a named extension with source support, ? marks a candidate extension whose language and semantics have not been selected and assessed and which is not counted as support, -- marks the absence of a native construct in the base language, which is not a non-expressibility result, and L marks that a learned approach applies, without establishing expressibility or learnability of the particular pattern. Appendix D marks the extension cells for which no encoding is constructed. Where a pattern has two variants, the cell gives a and b. We found no proved non-expressibility result for these patterns, so no cell claims one. Appendix D gives the encodings and constructions behind the symbolic rows, with the conventions and the rule for metric bounds that the ratings follow. Where a row names several extensions, the cells do not show that they combine in one implementation. The cells were coded once from those formalizations (Section 9). RPQ denotes regular path queries.\end{minipage}
\end{table*}
Each row of Table 9 rates one language, or one named language with its stated extensions, as Appendix D encodes it. Recognition (AP1 to AP3) is covered by every symbolic formalism except interval algebra and STREL, with differences in detail, such as whether a time bound is a native window or arithmetic on timestamps, and whether inequality is available, which base DatalogMTL lacks for AP3. DatalogMTL's native recursion over metric operators also comes at a cost, since fact entailment is PSPACE-complete in data complexity over the rationals and its stratified negation is not straightforward to maintain incrementally on a live stream (Appendix B.3). Composition across windows (AP4, AP5, AP9) separates the formalisms. Ordered flows (AP5) are native in seven of them, but among the fragments assessed, the native constructions for time-respecting ancestry and reach (AP4, AP9) use recursion over metric operators: in MFOTL with the recursive let of VeriMon, whose recursion must be past-guarded~\cite{zinggetal2022}, and in DatalogMTL. LARS, temporal ASP, Datalog and ProbLog encode them with explicit timestamps and comparisons, and the Event Calculus rests on a cited extension. Graph-query extensions for AP4 and AP9 remain unassessed, and chronicles, interval algebra and STREL have no construction in the assessed fragments. Snapshot reachability, such as the reach operator of the spatio-temporal reach and escape logic STREL, which quantifies over paths in the graph S(t) at a single time point~\cite{nenzietal2022}, does not suffice: if A--B exists only at time 1 and B--C only at time 2, a time-respecting path from A to C exists, but no snapshot contains one. Uncertain composition (AP7) is native in the assessed ProbLog configuration, while Event Calculus support rests on the cited probabilistic extension and probabilistic ASP remains an unassessed candidate. For missing and conflicting evidence, absence with exceptions within a time window (AP6) is native in three formalisms, while the consistency of qualitative constraints from different sources (AP8b) is native only in interval algebra.

No assessed base language provides native constructs for the whole catalog. Positive encoded cells rely on the constructions or extensions in Appendix D, and unassessed candidates (?) are not counted as coverage. Temporal ASP has been applied to short SIEM alert traces~\cite{eckhoffetal2025}, an earlier Event Calculus to security monitoring~\cite{amalioandspanoudakis2008}, chronicles to intrusion alerts, graph queries to audit data and Datalog to attack-graph derivation, whereas LARS, RTEC and the assessed probabilistic logic-programming configuration have not been applied to host audit streams. The table is an inventory of constructs, not a ranking of end-to-end systems.

\subsection{Takeaways}
\begin{itemize}
\item \textbf{T9.} Each reasoning problem has established KRR methods for specified subproblems, but uptake in security concentrates on pattern recognition (R1) and explanation (R5), and for uncertainty (R3) and manipulated evidence (R4) we found few security applications, although inconsistency-tolerant semantics and certified temporal explanations exist.
\item \textbf{T10.} Every assessed language leaves part of the catalog to encodings or extensions, and the rows differ in temporal composition, uncertain inference and qualitative consistency checking. Their positive cells must be read with the language fragments, input mappings and extension status recorded in Appendix D, and support across separate extensions does not establish a combined implementation or a deployed host-audit system.
\end{itemize}

\section{Open challenges}
Each challenge below states a research question and how progress on it could be measured.

\smallskip\noindent \textbf{C1. Typed interfaces between learned recognition and symbolic composition (T1, T2).} Can learned detectors emit technique hypotheses, with entities, intervals and confidences, that a temporal formalism composes into campaigns? Success can be measured by how well the emitted facts match labeled entities and intervals (precision and recall), how far their interval endpoints deviate under controlled timing jitter, and how well their confidences are calibrated on held-out data, and at campaign level by reconstruction accuracy, fewer joins of unrelated steps and the way errors of recognition propagate into composition. The baseline composes the same typed facts, with the same background knowledge, by an LLM investigator or by a system-specific procedure such as HOLMES, with any adapter between the detector's output and the procedure's input documented. Such an interface also lets fast filters handle raw event rates while the temporal reasoner composes the far smaller set of typed hypotheses.

\smallskip\noindent \textbf{C2. Comparative evidence across temporal formalisms (T4, T10).} Which formalism, or combination, covers realistic attack patterns, at what cost in throughput, latency and memory? A comparison would encode a shared pattern catalog such as Table 9 in each candidate language under common conventions, check the encodings against an independently specified reference relation on positive and negative boundary traces, restricted to the patterns each language supports, and then measure throughput, latency and memory as event rate and horizon grow, on the same audit facts replayed in and out of order. Candidates include MonPoly or VeriMon, Flink CEP, RTEC, telingo and MeTeoR.

\smallskip\noindent \textbf{C3. Reasoning over incomplete and manipulated evidence (T3, T8, T9).} How can reasoning use integrity signals from the logging layer, abduce missing steps and detect inconsistent times under a stated attacker model? An evaluation would degrade multi-host audit streams with labeled campaigns by deleting, delaying and reordering records, skewing clocks across hosts and timestomping files, each separately and at stated rates, so that the ground truth of each change is known. A reference model states which records may be added, corrected or rejected under its integrity assumptions, and its verdicts range over the resulting admissible histories, with evidence that admits none reported as inconsistent. The evaluation would measure detection and reconstruction accuracy, and, against the reference verdicts defined by this model, how often claims are determinate and correct, wrongly determinate, or unnecessarily undetermined, and how verdicts recover as late evidence arrives. The baseline is the same reasoning under the closed-world assumption.

\smallskip\noindent \textbf{C4. Validated acquisition of temporal knowledge (T9).} How can varied multi-step temporal rules be acquired from threat intelligence or host evidence and validated with stated guarantees before deployment? Useful measures are the false-positive rate on benign data, precision and recall on labeled attacks at stated base rates, ATT\&CK technique coverage, and agreement with analyst-written rules for the same reports, used as a reference rather than as ground truth. This needs threat reports paired with host audit data that contain the techniques they describe. The baselines are analyst-written rules and single-step synthesized rules such as those of SIGMERGE.

\smallskip\noindent \textbf{C5. Benchmarks with formal ground truth and adversarial evaluation (T6, T8).} Can benchmarks define campaigns as temporal specifications and include evasive variants? Such a benchmark would release each campaign with its specification, node-level labels and the evidence each step leaves, executed on instrumented hosts by adversary-emulation tooling. Generating fresh instances from the specification would reduce exact-instance overlap with the training data of LLM-based systems, while overlap in templates and attack knowledge needs separate controls such as held-out templates. Systems would be scored on reconstruction, ordering against the specification and explanation fidelity, including under mimicry and rule evasion, with labels for conformance to the specification kept separate from labels for attack status, including benign and ambiguous cases. The point of comparison is current evaluation on DARPA-derived datasets such as Transparent Computing and OpTC, as re-examined by Bilot et al.~\cite{bilotetal2025}.

\smallskip\noindent \textbf{Further directions.} Derivation provenance for metric temporal operators, decidable fragments for inventing unobserved entities, interventional causality over provenance and spatio-temporal models of hosts and networks need further study (Appendix B).

\section{Threats to validity}
The main limitation of this paper is that the comparative ratings (Table 6) and the pattern assessment (Table 9) rest on a reading of the literature and on formalizations checked against each language's documentation rather than on experiments, whereas only Table 5 rests on a systematic check of full texts.

\begin{itemize}
\item \textbf{Status of the central results.} The findings of Section 4.3 rest on a full-text check of 160 papers selected by the rule of Appendix A.4 and documented in Table C1 and the evidence ledger. Thirteen selected papers could not be obtained, mostly recent temporal-graph detectors from the claim check, and two were coded from an abstract or a description elsewhere, for Q1 and Q2 only. Citation rank favors older and influential work within each period, by design. Empty cells follow the candidate populations of the screening roles (Appendix A.4). A rule-based sample bounds, but does not eliminate, the chance that an unsampled system states an assumption we report as absent, so our negative statements concern the sampled systems. None of these codings was repeated by an independent second coder, although an agreement measure alone would not replace checking the encodings.
\item \textbf{Absence claims.} Every statement in this paper that something is absent from the literature, however phrased, is bounded by our search. A rarely cited KRR paper outside the listed venues can be missed, since the open and recent KRR sets were only checked for highly cited work (Appendix A).
\item \textbf{Selection.} Screening of both sets, by title and where unclear by abstract, was performed once, with AI assistance (see Use of Generative AI) and without an independent second screening, but the roles and decisions per record are released.
\item \textbf{Verification.} Every reference was matched to a bibliographic record, and for most the full text was opened, while for the remainder claims were checked against the abstract or publisher page. Whether each claim is supported as stated was checked once, with AI assistance.
\item \textbf{Reported results.} We quote figures as each paper reports them, on different datasets, splits and metrics. Such figures can be optimistic: re-evaluations of learned detectors show this~\cite{bilotetal2025}, synthetic background activity and conspicuous attacks inflate detection rates~\cite{liuetal2025}, the choice of benchmark and protocol shapes the conclusions drawn~\cite{guerraetal2027}, and the gap between evaluation and operation was raised long before provenance graphs~\cite{sommerandpaxson2010}.
\end{itemize}

\section{Conclusion}
The question this systematization raises is not whether host-based detectors use time, but what they assume when they do. Clock bounds, delivery and the integrity of records decide which campaign conclusions hold, and the systems that draw these conclusions rarely state what they assume about them.

On RQ1, no sampled system states a numeric bound on clock offset. Among the 52 audited host-detector papers that consume host records, four describe an input-ordering condition or convention, and none specifies conditions under which its conclusions remain valid across a stated class of delayed or missing records. In 28 of 38 provenance-era detector papers, capture or stored evidence is trusted or its manipulation is placed out of scope, again without saying which conclusions survive it. These assumptions are not unknown: some alert correlators state them, and digital forensics measures clock error. On RQ2, temporal KRR offers formal treatments of late, imprecise and incomplete records as well as certified explanations, which the sampled systems use only for incomplete logs. No assessed base language covers all nine attack patterns natively, and support spread over separate extensions does not add up to one implementation. Security applications exist in alert correlation, monitoring, attack-graph derivation and audit querying, so what remains open are particular combinations of language, input and task. On RQ3, a campaign claim becomes open to independent checking when it supplies a formally specified derivation, names its input records and rules and states its correspondence assumptions. A checker can then establish the inference relative to those inputs, while whether the assumptions about capture, identity and attacker influence hold remains a separate matter. No provenance-based detector we audited documents this complete evidence interface.

These findings are bounded by our search and by a single coding (Section 9), and Section 8 turns them into five research questions with ways to measure progress.

\FloatBarrier
\section*{Ethics Considerations}
This paper systematizes published work. It involves no human subjects, no personal data and no experiments on live systems, and it develops no attack and discloses no vulnerability. The full texts were obtained from open-access copies, author and venue pages, and copies supplied by the authors. We considered whether describing the unstated assumptions of detectors could help attackers. The capabilities over the evidence discussed in Section 2 are documented attack techniques, such as timestomping~\cite{mitreattckt1070006}, and we judge that making the dependence of detectors on them explicit helps defenders more than attackers.

\FloatBarrier
\section*{Open Science}
All artifacts are archived at doi:10.5281/zenodo.23530002: the search scripts, every OpenAlex request with its retrieval date, the returned records, the screening decisions with an argument for each record, the ranked candidates of the citation chase, and the evidence ledger behind Table 5 and Table C1 (Appendix A). Appendix D provides schematic formulas and construction descriptions for Table 9, identifies reused constructions, and distinguishes assessed support from unconstructed or unassessed extensions.

\FloatBarrier
\section*{Competing Interests}
Both authors are founders of datAInsights GmbH, which develops the temporal reasoning engine Tyla. During his doctoral work at TU Wien, M. Nissl contributed to the Temporal Vadalog system~\cite{bellomarinietal2025a} and to DatalogMTL with existential rules~\cite{lanzingeretal2023}, both discussed in this paper. That work started from limits of the available temporal formalisms, standard DatalogMTL included, which Appendix B.3 and Section 7.7 record from the published literature. The paper does not describe, evaluate or rate Tyla, and none of its findings depends on it. Table 9 rates DatalogMTL under the same rubric as every other formalism and ranks none, and Appendix D states the evidence status of every symbolic cell. Whether any engine, ours included, closes the gaps documented here is the comparative question of challenge C2.

\FloatBarrier
\section*{Use of Generative AI}
The literature work and the manuscript were produced with generative AI systems (Google Gemini, Anthropic Claude and OpenAI models) under the authors' direction. The systems ran the searches, screened the records, searched full texts for the evidence ledger, drafted ratings, encodings and text, and checked bibliographic metadata. The authors defined the research questions, the inclusion rules, the rubric and the conventions, directed every revision and reviewed the results, and take responsibility for the manuscript's claims and citations. Prompts and intermediate outputs were not archived, so the AI steps are not reproducible, but their results are released in the artifact, where each decision can be checked against its source.

\FloatBarrier

\appendices
\makeatletter\@addtoreset{table}{section}\makeatother
\renewcommand{\thetable}{\Alph{section}\arabic{table}}

\FloatBarrier
\section{Selection method}
We searched the security and the KRR literature with the same design on OpenAlex, in three rounds. The main round, on 23 October 2026, covered 2017 to 2026 for security and 2015 to 2026 for KRR. An extension on 24 October 2026 ran the same queries over the same venues for the earlier years from 2000. A final round on the same day added the queries and venues that the citation chase (Appendix A.3) showed to be missing, over 2000 to 2026. Each query of the main round was evaluated as four sets: an open set over all venues in the Computer Science field and a recent set over 2024 to 2026, each a sample of the 100 most cited records, a venue set restricted to a list of sources and retrieved in full, and its counterpart for the earlier years. The final-round queries used open and recent sets and a single venue set over 2000 to 2026. The venue sets were screened in full. The capped open and recent samples were screened in full for security and used only to check for highly cited work in KRR. In each literature, titles and, where unclear, abstracts were screened with AI assistance under the authors' direction against one stated rule, and each screened record has an inclusion or exclusion decision with a written reason in the released decision sheets. Writing and checking these reasons led us to revisit earlier exclusions under the same final rule. Table A1 summarizes the records screened and retained per round.

OpenAlex does not index several LNCS-published conferences as sources of their own (ESORICS, DIMVA, RAID before 2019, CAV, TACAS, RV, FORMATS, LPNMR, JELIA, RuleML+RR, ESWC), so both venue sets include the Lecture Notes in Computer Science series as one source and let the query do the narrowing.

\begin{table*}[!tp]
\caption{Records screened and retained, by literature and round.}
\label{tab:A1}
\centering \scriptsize
\begin{tabular}{>{\raggedright\arraybackslash}p{0.25\textwidth} r r r r}
\toprule
\textbf{Literature} & \textbf{Main round} & \textbf{Earlier years} & \textbf{Final round} & \textbf{Total} \\
\midrule
Security, screened & 3,045 & 1,064 & 760 & 4,869 \\
Security, retained & 613 & 282 & 108 & 1,003 \\
KRR, screened & 1,216 & 608 & 726 & 2,550 \\
KRR, retained & 661 & 259 & 89 & 1,009 \\
\bottomrule
\end{tabular}
\end{table*}

\subsection{Security literature}
\noindent \textbf{Sources.} 21 venues (22 OpenAlex source identifiers): IEEE S\&P, USENIX Security, ACM CCS, NDSS, ACSAC, RAID (ACM, 2019 onward), IEEE EuroS\&P, ACM AsiaCCS, DSN, IEEE CSF, IEEE TDSC, IEEE TIFS, ACM TOPS and its predecessor ACM TISSEC, ACM DTRAP, the Journal of Computer Security, Computers \& Security, the Journal of Information Security and Applications, Forensic Science International: Digital Investigation, ACM Computing Surveys, and LNCS (ESORICS, DIMVA, RAID before 2019 and others). The Journal of Computer Security and ACM TISSEC entered in the final round.

\smallskip\noindent \textbf{Queries.} Nine title-and-abstract queries, one per line of work (Table A2). Q1 to Q7 ran in the main round and its extension to earlier years. Q8, for attack languages, misuse detection and audit trails, and Q9, for the time and integrity of security evidence, ran in the final round, which also ran Q1 to Q7 over the two added journals. Q8 and Q9 retained 38 and 58 records without overlap. The runs over the added journals retained 12 more (from Q2, Q3, Q4, Q5 and Q6, partly overlapping), which gives the 108 of Tables A3 and A1.

\begin{table*}[!tp]
\caption{Security queries and hits (OpenAlex, 23 and 24 October 2026).}
\label{tab:A2}
\centering \scriptsize
\begin{tabular}{l >{\raggedright\arraybackslash}p{0.10\textwidth} >{\raggedright\arraybackslash}p{0.43\textwidth} r r r r r r}
\toprule
\textbf{Query} & \textbf{Line of work} & \textbf{Search expression} & \textbf{Open} & \textbf{Recent} & \textbf{Venue} & \textbf{Included} & \textbf{Early} & \textbf{Included early} \\
\midrule
Q1 & Provenance-based detection and investigation & (provenance OR "causal graph" OR "dependency graph" OR "audit graph") AND (intrusion OR attack OR attacks OR threat OR threats OR APT OR "advanced persistent" OR forensic OR forensics OR investigation OR malware OR "lateral movement") & 4101 & 3037 & 320 & 174 & 98 & 10 \\
Q2 & Audit-log and host-based detection & ("audit log" OR "audit logs" OR "system audit" OR "kernel audit" OR "system logs" OR "security logs" OR "event logs" OR "system call" OR "system calls" OR "host-based" OR "host intrusion detection" OR endpoint OR "endpoint detection and response" OR EDR OR "Windows event" OR Sysmon OR eBPF OR ETW) AND (attack OR attacks OR intrusion OR "advanced persistent" OR APT OR "threat hunting" OR forensic OR investigation OR "lateral movement") & 8001 & 4467 & 571 & 216 & 423 & 123 \\
Q3 & Multi-step attacks and campaign reconstruction & ("alert correlation" OR "multi-step attack" OR "multi-step attacks" OR "multistage attack" OR "multi-stage attack" OR "multi-stage attacks" OR "attack scenario" OR "attack scenarios" OR "kill chain" OR "attack reconstruction" OR "attack story" OR "attack stories" OR "attack campaign" OR "attack campaigns" OR "attack graph" OR "attack graphs" OR SIEM OR "security information and event management" OR "event reconstruction" OR "timeline reconstruction" OR "lateral movement") & 12251 & 7729 & 905 & 209 & 369 & 137 \\
Q4 & Symbolic and knowledge-based security reasoning & ("temporal logic" OR datalog OR "logic programming" OR "answer set programming" OR "rule-based" OR "rule based" OR ontology OR "knowledge graph" OR "event calculus" OR "complex event processing" OR "runtime verification" OR "query language") AND (intrusion OR "attack detection" OR "threat hunting" OR "attack investigation" OR "threat detection" OR forensics OR "security monitoring" OR "audit logs" OR "audit data") & 3314 & 2355 & 120 & 52 & 81 & 32 \\
Q5 & Learned detection over host data & ("graph neural network" OR "graph neural networks" OR transformer OR "language model" OR "language models" OR "anomaly detection" OR "representation learning") AND (provenance OR "audit log" OR "audit logs" OR "system logs" OR "host-based" OR endpoint OR "system call") AND (attack OR attacks OR intrusion OR APT OR threat) & 1641 & 1342 & 99 & 119 & 40 & 29 \\
Q6 & Threat-intelligence knowledge and ATT\&CK & ("threat intelligence" OR CTI OR "ATT\&CK" OR TTP OR TTPs OR "attack technique" OR "attack techniques") AND (extraction OR "knowledge graph" OR ontology OR mapping OR detection OR provenance OR "audit" OR "attack graph" OR correlation) & 6107 & 4677 & 336 & 162 & 21 & 2 \\
Q7 & Detection-rule and signature synthesis & ("detection rules" OR "detection rule" OR "Sigma rule" OR "Sigma rules" OR "signature generation" OR "rule generation" OR "rule synthesis" OR "query synthesis" OR "specification mining") AND (attack OR attacks OR intrusion OR threat OR malware OR "threat hunting" OR "security") & 1289 & 694 & 113 & 21 & 114 & 6 \\
Q8 & Attack languages, misuse detection and audit trails & ("misuse detection" OR "attack language" OR "attack languages" OR "attack specification" OR "signature language" OR "audit trail" OR "audit trails" OR "state transition analysis") AND (intrusion OR intrusions OR attack OR attacks OR misuse) & 1577 & 477 & 116 & 38 & -- & -- \\
Q9 & Time and integrity of security evidence & (timestamp OR timestamps OR "clock skew" OR "clock synchronization" OR "time synchronization" OR timestomping OR "time manipulation" OR "log integrity" OR "tamper-evident" OR "secure logging" OR "log tampering" OR "event ordering") AND (log OR logs OR audit OR forensic OR forensics OR provenance OR intrusion OR attack) & 5555 & 3093 & 356 & 58 & -- & -- \\
\bottomrule
\end{tabular}
\par\vspace{3pt}\begin{minipage}{0.96\textwidth}\scriptsize Open = Computer Science, 2017--2026. Recent = 2024--2026. Venue = the sources above, 2017--2026 (2000--2026 for Q8 and Q9). Included = distinct records returned by the query in these three sets and retained after screening, for Q8 and Q9 only records not seen in earlier rounds. Early = the sources above, 2000--2016, and Included early = early records retained. Open and Recent give total hits, of which the 100 most cited were retrieved. Included counts overlap across queries and must not be added.\end{minipage}
\end{table*}
\smallskip\noindent \textbf{Screening.} The rule: the paper detects, investigates or reconstructs attacks from host audit data, provenance graphs, security logs or alert streams, or it provides the symbolic machinery compared in Section 4 (alert correlation, attack graphs, threat-intelligence knowledge, detection-rule synthesis), or it concerns the integrity or reliability of audit evidence, or it evaluates, benchmarks or surveys such work. Across the three rounds, 1,003 of 4,869 records were retained (Table A1). The main round excluded 2,425 records: 907 outside the rule with no reasoning over host event history, 545 on networks, protocols, IoT or cyber-physical systems, 435 on machine-learning security, web attacks, media forensics or malware classification, 201 on cryptography, authentication, privacy or blockchains, 145 on risk assessment, game-theoretic defense or mitigation, 118 on side channels, hardware or physical attacks, 65 surveys outside the rule, and 9 audit capture systems without analysis. Seven further records were second indexings of retained papers. The earlier years and the final round were excluded under the same codes. Writing the arguments for the exclusions moved 64 records of the main round and 10 of the earlier years from excluded to included. Each retained paper has one primary role, the closest match (Table A3). Host-based detection over system calls and audit records, a distinct earlier line, appears mainly in the earlier years. The retained sets, together with snowballing, are the population from which Section 4 describes the families and Section 4.3 samples systems by the rule in Appendix A.4. Argumentation-based attribution (S7) has no separate role in Table A3, and its representatives were reached by snowballing.

\begin{table*}[!tp]
\caption{Retained security papers by role and round.}
\label{tab:A3}
\centering \scriptsize
\begin{tabular}{>{\raggedright\arraybackslash}p{0.40\textwidth} r r r r}
\toprule
\textbf{Role} & \textbf{Main round} & \textbf{Earlier years} & \textbf{Final round} & \textbf{Total} \\
\midrule
Learned detection (L1, L2) & 186 & 0 & 4 & 190 \\
LLM investigation (L3) & 14 & 0 & 0 & 14 \\
Knowledge-guided learning (L4) & 8 & 0 & 0 & 8 \\
Alert correlation and plan recognition (S1) & 61 & 69 & 3 & 133 \\
Temporal-logic and event-pattern detection (S2, S3) & 0 & 11 & 16 & 27 \\
Rule-based analysis and queries (S4, S5) & 26 & 0 & 0 & 26 \\
Attack graphs (S6) & 56 & 55 & 8 & 119 \\
Rule and specification learning (S8) & 12 & 4 & 1 & 17 \\
Threat-intelligence knowledge (S9) & 85 & 9 & 2 & 96 \\
Host behavior anomaly detection over system calls or audit data (L5) & 0 & 88 & 19 & 107 \\
Provenance tracing, attack investigation and data reduction (S10) & 68 & 23 & 0 & 91 \\
Audit integrity and the time of evidence (context) & 32 & 10 & 55 & 97 \\
Evaluation, evasion, datasets and surveys (context) & 65 & 13 & 0 & 78 \\
Total & 613 & 282 & 108 & 1,003 \\
\bottomrule
\end{tabular}
\end{table*}
\smallskip\noindent \textbf{Snowballing.} From the retained papers and from the surveys among them, we followed references backward and citations forward for alert correlation before 2017, USENIX Security systems that OpenAlex does not index, and the foundational systems of each family. Works cited in Sections 4 and 5 that are not in the retained set entered this way.

\subsection{KRR literature}
\noindent \textbf{Sources.} 39 venues (42 OpenAlex source identifiers): KR, IJCAI, AAAI, ECAI, JAIR, Artificial Intelligence, AAMAS, TPLP, ICLP, Argument \& Computation, LICS, RV (reached through the LNCS series, see A.1), FMCAD, TIME, Journal of Logic and Computation, Formal Methods in System Design, ACM TOCL, Journal of the ACM, LMCS, JAR, Information and Computation, LNCS, NeurIPS, ICML, ICLR, Neurosymbolic Artificial Intelligence, PVLDB, VLDB Journal, SIGMOD, PODS, ICDT, TODS, ICDE, IEEE TKDE, DEBS, Journal of Web Semantics, Semantic Web, ISWC and ACM Computing Surveys, which entered in the final round.

\smallskip\noindent \textbf{Queries.} Twelve queries: one for each of the ten KRR domains of Appendix B, and two supplemental queries on event stream processing and its time model (Table A4). B1 to B10 ran in the main round and its extension. B2r, which widens the vocabulary of B2 to event stream processing and out-of-order or imprecise input, and B11, for the time model of stream processing (event and processing time, watermarks, heartbeats and punctuations), ran in the final round.

\begin{table*}[!tp]
\caption{KRR queries and hits (OpenAlex, 23 and 24 October 2026).}
\label{tab:A4}
\centering \scriptsize
\begin{tabular}{>{\raggedright\arraybackslash}p{0.15\textwidth} >{\raggedright\arraybackslash}p{0.44\textwidth} r r r r r r}
\toprule
\textbf{Domain} & \textbf{Search expression} & \textbf{Open} & \textbf{Recent} & \textbf{Venue} & \textbf{Included} & \textbf{Early} & \textbf{Included early} \\
\midrule
B1 Temporal logics and runtime verification & ("runtime verification" OR "runtime monitoring" OR "online monitoring" OR "metric temporal logic" OR "metric first-order temporal logic" OR MFOTL OR "first-order temporal logic" OR "linear temporal logic" OR LTL OR LTLf OR "temporal logic") AND (monitor OR monitoring OR monitors OR monitorability) AND (log OR logs OR trace OR traces OR stream OR streams OR timestamp OR timestamps OR "out-of-order") & 748 & 303 & 235 & 149 & 63 & 39 \\
B2 Event recognition, complex event processing and stream reasoning & ("complex event processing" OR "complex event recognition" OR "composite event" OR "event pattern" OR "event patterns" OR "event calculus" OR "chronicle recognition" OR "stream reasoning") AND (semantics OR formal OR logic OR "temporal constraints" OR intervals OR "interval-based" OR recognition) & 566 & 131 & 131 & 97 & 129 & 64 \\
B3 Datalog-based temporal rule languages & (DatalogMTL OR "metric temporal datalog" OR "temporal datalog" OR vadalog OR (datalog AND ("metric temporal" OR "temporal operators" OR "temporal reasoning" OR streams))) & 121 & 33 & 59 & 45 & 6 & 3 \\
B4 Non-monotonic reasoning, argumentation and belief revision & ("temporal answer set programming" OR "temporal equilibrium logic" OR telingo OR ("answer set programming" AND (temporal OR "time-dependent" OR stream OR streams OR "dynamic domains")) OR ("belief revision" AND temporal) OR ("argumentation framework" AND temporal) OR ("argumentation frameworks" AND temporal) OR ("defeasible logic" AND temporal) OR ("defeasible reasoning" AND temporal) OR ("abductive reasoning" AND (temporal OR "event"))) & 287 & 121 & 104 & 79 & 76 & 33 \\
B5 Uncertainty and neuro-symbolic reasoning & ("probabilistic logic programming" OR "probabilistic logic" OR "probabilistic temporal logic" OR "temporal logic point process" OR "Markov logic" OR "annotated logic" OR "neuro-symbolic" OR neurosymbolic OR "differentiable logic" OR "logic tensor networks" OR "probabilistic event calculus") AND (temporal OR time OR sequence OR sequences OR events OR streams) & 1655 & 1192 & 207 & 63 & 75 & 20 \\
B6 Description logics, ontologies and ontology-based data access & ("description logic" OR "description logics" OR "ontology-based data access" OR OBDA OR "ontology-mediated") AND (temporal OR "time-stamped" OR "temporal data" OR "temporal query" OR "temporal queries") & 193 & 35 & 47 & 38 & 58 & 27 \\
B7 Graph pattern and query languages & ("regular path queries" OR "regular path query" OR "graph query language" OR "graph query languages" OR "property graph" OR "property graphs" OR "temporal graph query" OR "temporal graph queries" OR "time-respecting" OR "temporal paths" OR "temporal reachability") AND (query OR queries OR reachability OR temporal OR semantics) & 1338 & 617 & 261 & 110 & 60 & 34 \\
B8 Causal reasoning & ("actual causation" OR "actual causality" OR "causal explanation" OR "causal explanations" OR "structural causal model" OR "structural causal models") AND (temporal OR trace OR traces OR "event log" OR "event logs" OR "discrete event" OR "event sequence" OR "event sequences" OR "events") & 300 & 217 & 32 & 15 & 3 & 1 \\
B9 Qualitative and spatio-temporal reasoning & ("interval algebra" OR "qualitative temporal reasoning" OR "qualitative spatial reasoning" OR "spatio-temporal logic" OR "spatio-temporal reach" OR STREL OR "spatial temporal logic" OR "spatio-temporal logics" OR "signal temporal logic") AND (reasoning OR consistency OR monitoring OR logic OR specification OR robustness OR semantics) & 1248 & 515 & 188 & 111 & 138 & 45 \\
B10 Rule and specification learning, and LLMs with formal reasoning & ("specification mining" OR "temporal logic inference" OR "learning temporal logic" OR "learning LTL" OR "signal temporal logic inference" OR "temporal rule mining" OR "temporal rule learning" OR "mining temporal rules" OR "learning temporal rules" OR ("inductive logic programming" AND (temporal OR events)) OR ("natural language" AND "temporal logic" AND (LLM OR "language model" OR "language models"))) & 379 & 190 & 52 & 38 & 22 & 10 \\
B2r Event and stream processing, revised vocabulary & ("complex event processing" OR "complex event recognition" OR "event stream processing" OR "event streams" OR "stream processing" OR "data stream" OR "data streams" OR "composite event" OR "event pattern" OR "event patterns" OR "sequence pattern" OR "event calculus" OR "chronicle recognition" OR "stream reasoning") AND ("out-of-order" OR disorder OR disordered OR "late arrival" OR "late data" OR "imprecise timestamps" OR uncertain OR uncertainty OR semantics OR formal OR logic) & 4303 & 1001 & 688 & 75 & -- & -- \\
B11 Time model of stream processing & ("event time" OR "processing time" OR "application time" OR watermark OR watermarks OR heartbeat OR heartbeats OR punctuation OR punctuations OR "out-of-order" OR "late data" OR "late arrival" OR "clock synchronization") AND (stream OR streams OR "event processing" OR "stream processing" OR "continuous queries" OR "continuous query") & 3463 & 707 & 263 & 34 & -- & -- \\
\bottomrule
\end{tabular}
\par\vspace{3pt}\begin{minipage}{0.96\textwidth}\scriptsize Open = Computer Science, 2015--2026. Recent = 2024--2026. Venue = the sources above, 2015--2026 (2000--2026 for B2r and B11). Included = venue-set records retained after screening, for B2r and B11 only records not seen in earlier rounds. Early = the sources above, 2000--2014, and Included early = early records retained. Open and Recent give total hits, of which the 100 most cited were retrieved. Included counts overlap across queries and must not be added.\end{minipage}
\end{table*}
\smallskip\noindent \textbf{Screening.} The rule: the work defines or implements reasoning over timestamped or ordered events with a stated semantics, or learns such specifications. Domain applications without a reasoning contribution, graph data management without a temporal or path-query contribution, learning methods without a stated logical semantics, planning and control synthesis, model checking of system models, and overviews or editorials were excluded. Across the three rounds, 1,009 of 2,550 venue-set records were retained (Table A1). The earlier years excluded 60 records as spatial reasoning without a temporal component, and the final round retained 56 records on event and stream processing semantics, 30 on out-of-order, uncertain or imprecise input and 3 others. Writing the arguments for the exclusions moved 44 records of the main round and two of the earlier years from excluded to included. The open and recent sets were retrieved and are released, but were not screened record by record. Works in the retained set that bear on a claim in Section 7 or Appendix B are cited where the claim is made.

\subsection{Citation chase}
As a bounded supplement to the queries, which favors established work close to our references, we took every reference of this paper that had an OpenAlex identifier (107 at the time of the chase; the frozen seed list and the ranked candidates are released in the artifact), retrieved the works it cites and the 50 most cited works that cite it, and ranked the works linked to at least two of our references that were neither cited nor retained. We inspected the 130 most strongly linked of the 650 ranked candidates by title. They revealed eleven lines: the time model of stream processing, the foundations of metric temporal logic, temporal databases and logical clocks, attack specification languages from before 2000, provenance capture and audit-data reduction, learned detection over system logs and security events, secure logging, early alert correlation, event and stream processing surveys, temporal ontology-mediated queries and temporal equilibrium logic, and critiques of machine learning for intrusion detection. Their key works are cited in the main text and Appendix B, and the final round of the search added queries and venues for them.

\subsection{Sampling and full-text check of systems}
\noindent \textbf{Sampling rule.} Each retained security record has one primary role (Table A3). Roles that describe detection or investigation define the sampling strata. Three of them cover two families each (L1 and L2, S2 and S3, S4 and S5) and were ranked jointly, and each sampled paper was assigned to one family when it was coded, so a paper retained under learned detection that models log sequences rather than provenance graphs, such as LogBERT, is coded under L5. Argumentation (S7) has no retained role and is represented only by the systems that Section 4 names. Audit integrity, evaluation, datasets and surveys are context, not families. The citation rate is the OpenAlex citation count stored in the artifact divided by the number of calendar years from publication to 2026, both included. For each eligible screening role and each of the periods 2000 to 2006, 2007 to 2016 and 2017 to 2026, we ranked the retained system papers by this citation rate and took the first three. A survey, position paper or dataset was replaced by the next in rank. Ties were broken by the order of the records in the decision sheets, which is incidental, not a deliberate secondary key. Index records were consolidated by normalized title before ranking. To this rule sample we added every system that Sections 4.1 and 4.2 name, an author-curated list of landmark systems per family and period, chosen as the first or most influential system of a mechanism within a family, together with five digital-forensics papers on time and event reconstruction, and, as a check of the findings, every retained record whose title concerns clocks, timestamps, out-of-order or incomplete data, tampering or anti-forensics and that describes a detection, investigation or reasoning system. Retained papers on the security of time synchronization, secure logging and forensic timestamp artifacts are cited as context and not coded. The selection comprises 175 papers: 160 checked in full text, 2 coded from an abstract or a description in another paper, and 13 that could not be obtained. Full texts were obtained from open-access copies (Unpaywall, arXiv, OpenAlex), author and venue pages, and copies supplied by the authors. They are not redistributed.

\smallskip\noindent \textbf{Coding.} For each checked system, we searched the full text, without its reference list, for keywords per checklist question and read every matched passage, together with the threat-model and limitations sections. The keywords were timestamp, time stamp, event time, wall clock and time encoding for Q1, time window, sliding window, temporal order, chronological, across windows and order of events for Q2, clock, synchronization, NTP, skew and drift for Q3, out-of-order, late arrival, delayed, lost, missing or dropped events, records or logs, and reorder for Q4, PID, process id, UUID, unique identifier, identifier reuse and across hosts for Q5, and threat model, tamper, trusted computing, trusted kernel, integrity of logs or provenance, assumptions of trust, mimicry and evasion for Q6. A negative code therefore means that no statement was located by this procedure, not that a system makes no such assumption. For papers coded from an abstract or a description elsewhere, only positive statements on Q1 and Q2 were recorded. Two full texts were available only as scans and were searched after text recognition. The coding per system, the passages behind it, the sampling source (rule, named, landmark or claim check) and the period form the evidence ledger, which also lists the papers we could not obtain.

\smallskip\noindent \textbf{Delivery follow-up.} For the 52 papers eligible for Q4, two further searches followed the first coding. The first covered arrival, out-of-order and reorder, lost or dropped records, partial or causal order, buffering, completeness and lossless capture, and the second monotonic or non-decreasing timestamps, sorting by time, chronological or timestamp order, sequence numbers, and incomplete or missing data. We read the input, streaming and evaluation sections around every match and recorded separately input-ordering conditions or conventions, observations about missing records, and statements that link delivery failures to the validity or completeness of specified outputs. Sorting stored or batched events by timestamp was not counted as an input-ordering condition, and an ordering convention or a report of no dropped records was not counted as an account of output validity under delivery failure. We distinguish an observed success or failure on incomplete data from an account that specifies conditions under which conclusions remain valid across a stated class of delivery failures, which need not take the form of a formal theorem.

\subsection{Artifacts}
The artifacts are archived at Zenodo (doi:10.5281/zenodo.23530002). They contain the query scripts with the source identifiers, the API request and hit count for each query and set, all retrieved records with OpenAlex identifier, DOI, venue, citation count and the sets in which each appears, the decision sheets with the argument for every record, one per search and round, the seed list and ranked candidates of the citation chase, and the evidence ledger of Appendix A.4. The archive preserves the requests, returned records and hit counts of each retrieval date. Re-running the scripts against the live index can return different results, because the index changes. The supplemental roster from snowballing can be inspected, but the traversal cannot be replayed, because its seed-to-paper links were not retained.

\FloatBarrier
\section{In-depth analysis of KRR domains}
Sections B.1 to B.10 correspond to the queries B1 to B10 of Table A4 and carry their names. The final-round queries B2r and B11 widened B2 and feed Section B.2. Each section is analyzed with the same template: formalisms and reasoning tasks, expressiveness for intrusion detection, complexity, engines, and open questions. Table B1 summarizes complexity across the classes. It separates design-time tasks, such as checking whether a specification is satisfiable, from detection-time tasks, such as monitoring a trace or answering a query over data, because the latter run repeatedly over audit data.

\begin{table*}[!tp]
\caption{Complexity across KRR domains, by task.}
\label{tab:B1}
\centering \scriptsize
\begin{tabular}{>{\raggedright\arraybackslash}p{0.132\textwidth} >{\raggedright\arraybackslash}p{0.263\textwidth} >{\raggedright\arraybackslash}p{0.282\textwidth} >{\raggedright\arraybackslash}p{0.188\textwidth}}
\toprule
\textbf{Domain} & \textbf{Design-time task} & \textbf{Detection-time task} & \textbf{Sources} \\
\midrule
Propositional LTL & Satisfiability: PSPACE-complete & Past-time monitoring of a fixed propositional formula with memory independent of trace length & ~\cite{sistlaandclarke1985,havelundandrosu2002} \\
Metric temporal logic & Satisfiability: EXPSPACE-complete over integer time; undecidable over the reals in the pointwise and continuous semantics of infinite traces, with MITL EXPSPACE-complete & Online monitoring of monitorable first-order fragments & ~\cite{alurandhenzinger1993,aluretal1996,basinetal2015} \\
Complex event logic & --- & Fixed queries in the unary-filter fragment: constant update cost per event, then constant-delay enumeration of results & ~\cite{grezetal2019} \\
Datalog & --- & Query evaluation: P-complete in data, EXPTIME-complete combined & ~\cite{dantsinetal2001} \\
DatalogMTL & --- & Fact entailment: PSPACE-complete in data, EXPSPACE-complete combined; tractable fragments in Appendix B.3 & ~\cite{brandtetal2018,waegaetal2019} \\
Answer set programming & --- & Stable model existence for finite propositional programs: NP-complete for normal, $\Sigma_2^P$-complete for disjunctive programs & ~\cite{dantsinetal2001} \\
Temporal ASP & Existence of a temporal equilibrium model: EXPSPACE-complete over infinite traces & Depends on trace length and grounding; telingo uses the finite-trace variant & ~\cite{bozzelliandpearce2015,cabalaretal2018} \\
Abstract argumentation & --- & Grounded acceptance polynomial; credulous preferred NP-complete; skeptical preferred $\Pi_2^P$-complete & ~\cite{dvorakanddunne2018} \\
Probabilistic logic programming & --- & Exact success probability: \#P-hard & ~\cite{deraedtetal2007} \\
Description logics & ALC knowledge base satisfiability: EXPTIME-complete; subsumption in EL and its tractable extension EL++: polynomial & DL-Lite query answering: AC\textsuperscript{0} in data complexity & ~\cite{baaderetal2003,baaderetal2005,calvaneseetal2007} \\
Graph queries & --- & Conjunctive queries NP-complete combined; regular path queries NL-complete in data under walk semantics & ~\cite{chandraandmerlin1977,barcelo2013} \\
Allen's interval algebra & --- & Network consistency: NP-complete; ORD-Horn subclass polynomial & ~\cite{vilainandkautz1986,nebelandburckert1995} \\
Actual causation & --- & Deciding causality under the original Halpern--Pearl definition: $\Sigma_2^P$-complete in general, NP-complete for binary variables; later definitions differ & ~\cite{eiterandlukasiewicz2002} \\
\bottomrule
\end{tabular}
\end{table*}

\subsection{Temporal logics and runtime verification}
\noindent \textbf{Formalisms and tasks.} Linear temporal logic (LTL) describes orderings over infinite traces~\cite{pnueli1977}. LTLf interprets it over finite traces, which suits logs~\cite{degiacomoandvardi2013}. Metric temporal logic (MTL)~\cite{koymans1990} annotates operators with time intervals, and metric first-order temporal logic (MFOTL) adds quantification over data values such as process and file identifiers. Monitoring algorithms for MTL~\cite{thatiandrosu2005}, three-valued semantics for finite prefixes~\cite{baueretal2011} and parametric monitoring frameworks such as MOP~\cite{meredithetal2012} form the basis of the field. The tasks are satisfiability, model checking and, most relevant here, monitoring: computing online whether the trace seen so far satisfies or violates a formula, or does not yet determine either, under the chosen trace semantics.

\smallskip\noindent \textbf{Expressiveness for intrusion detection.} MFOTL expresses ordered, time-bounded patterns over entities, such as a read of a credential file followed within ten minutes by a connection from the same process. Past-time operators look backwards from the current event, which matches detection. Standard MFOTL provides no recursive definition construct. Our AP4 and AP9 encodings use VeriMon's past-recursive let extension, and we do not establish a non-expressibility result for base MFOTL. The recursive let extension adds recursive rules interleaved with temporal formulas, with recursive occurrences guarded to lie strictly in the past, so it supports past-guarded recursion rather than arbitrary fixpoints within one time point~\cite{zinggetal2022}.

\smallskip\noindent \textbf{Complexity.} LTL satisfiability is PSPACE-complete~\cite{sistlaandclarke1985}. MTL satisfiability over integer time is EXPSPACE-complete~\cite{alurandhenzinger1993}. Over the reals and infinite traces, in the pointwise and continuous semantics, it is undecidable, while the MITL fragment without punctual intervals is EXPSPACE-complete~\cite{aluretal1996}. These results concern infinite traces and do not carry over unchanged to every finite timed-word variant. First-order temporal satisfiability is undecidable, but monitoring is effective for monitorable fragments. Propositional past-time formulas can be monitored with memory independent of trace length~\cite{havelundandrosu2002}.

\smallskip\noindent \textbf{Engines.} MonPoly monitors a monitorable fragment of MFOTL, including aggregation~\cite{basinetal2015}. VeriMon is a formally verified MFOTL monitor~\cite{schneideretal2019} extended with a past-guarded recursive let~\cite{zinggetal2022}. DejaVu monitors first-order past-time logic using binary decision diagrams~\cite{havelundetal2017}. WhyMon outputs proof trees that explain each verdict and are certified by a checker verified in Isabelle~\cite{limaetal2024}, following earlier proof-tree explanations for MTL~\cite{limaetal2023}. Monitoring over out-of-order streams has been studied explicitly~\cite{basinetal2020}.

\smallskip\noindent \textbf{Limitations and open questions.} Whether bounded past-time monitoring suffices for realistic attack patterns, or whether recursive rules are needed and affordable, is open. Monitoring under incomplete observation has a formal treatment, for LTL over lossy traces~\cite{joshietal2017} and as assumption-based runtime verification under partial observability~\cite{cimattietal2019}, but neither has been evaluated on manipulated audit data.

\subsection{Event recognition, complex event processing and stream reasoning}
\noindent \textbf{Formalisms and tasks.} The Event Calculus represents how events initiate and terminate fluents over time~\cite{kowalskiandsergot1986}. Composite event recognition derives high-level situations, such as an attack stage, from streams of low-level events. Chronicles specify sets of events with metric constraints between them. Complex event processing (CEP) languages specify sequence patterns with Kleene closure, negation and time windows. Complex event logic gives such a language a denotational and compositional semantics~\cite{grezetal2019}. Data-stream systems give continuous queries a semantics over time-varying relations~\cite{arasuetal2006}, SASE+ adds Kleene closure to pattern matching over streams~\cite{agrawaletal2008}, and Cugola and Margara survey event and stream processing~\cite{cugolaandmargara2012}. Stream reasoning studies declarative semantics for continuous queries, as in LARS~\cite{becketal2018}.

\smallskip\noindent \textbf{Expressiveness for intrusion detection.} These formalisms compose recognized events into higher-level situations over time, the campaign composition stage for which the surveyed learned detectors provide no declarative temporal specification. Interval-based fluents represent ongoing states such as an active foothold, and inertia keeps them until a terminating event, so a missing termination record can make a foothold persist, and the completeness of termination evidence must therefore be stated.

\smallskip\noindent \textbf{Complexity.} For a fixed query of complex event logic with unary filters, compiled to complex event automata, each event costs constant update time, followed by constant-delay enumeration of results~\cite{grezetal2019}. Many industrial CEP engines define semantics operationally, so their complexity is usually stated per operator. RTEC's windowed evaluation bounds the input retained per query time by the window contents~\cite{artikisetal2015}. The work also depends on the rules and the number of candidate intervals. LARS inherits answer-set complexity for its general fragment~\cite{becketal2018}.

\smallskip\noindent \textbf{Engines.} RTEC implements the Event Calculus for streams with windowing and caching~\cite{artikisetal2015}. It has been extended to events whose effects begin only after a delay, with correctness and complexity results~\cite{mantenoglouetal2025}. LARS models can be updated incrementally when input arrives or expires, by extending truth maintenance to streams~\cite{becketal2015}, and complex event processing with event revocation has been formalized~\cite{anicicetal2011}. Complex event recognition has a relational automaton model in symbolic register transducers~\cite{alevizosetal2024}. A probabilistic Event Calculus handles uncertain input events~\cite{skarlatidisetal2015}. SASE introduced high-performance sequence matching over streams~\cite{wuetal2006}, and Esper and Flink CEP are production event-processing implementations. Forward-propagating DatalogMTL was proposed for stream reasoning~\cite{waegakaminskiandcuencagrau2019}.

\smallskip\noindent \textbf{Limitations and open questions.} A translation of SIEM correlation rules into complex event logic or a stream reasoning formalism would give them a formal semantics and make them comparable, but we found none.

\subsection{Datalog-based temporal rule languages}
\noindent \textbf{Formalisms and tasks.} Datalog is function-free Horn logic under least-model semantics. The main task is query answering. Temporal deductive databases were an early form of temporal rule languages~\cite{chomickiandimielinski1988}. Datalog$\pm$ adds existential quantification in rule heads, and its warded fragment keeps query answering tractable in data complexity~\cite{bergeretal2022}. DatalogMTL adds interval-indexed past and future box and diamond operators and Since and Until, interpreted over the rational or integer timeline~\cite{brandtetal2018,waegaetal2019}. Window-based stream reasoning in temporal Datalog, with bounds on the history that must be retained, was studied earlier for the integer timeline~\cite{roncaetal2018}. Query answering returns, for an answer predicate, the tuples and intervals entailed by the program and data~\cite{bellomarinietal2025a}. Extensions add stratified negation~\cite{tenacucalaetal2021}, stable-model negation~\cite{waegaetal2021}, existential rules~\cite{lanzingeretal2023}, and inconsistency-tolerant semantics based on conflicts and repairs of time-annotated facts~\cite{bienvenuetal2025}.

\smallskip\noindent \textbf{Expressiveness for intrusion detection.} Datalog expresses recursive reachability and joins over provenance graphs. DatalogMTL combines this with metric constraints. Appendix D gives a program for AP4 that binds the 72-hour deadline to the originating download and requires strictly increasing spawn times. Existential rules can entail the existence of entities not named in the input, such as a session that a rule requires. Selecting such entities as explanatory hypotheses needs an abductive semantics.

\smallskip\noindent \textbf{Complexity.} Datalog query evaluation is P-complete in data complexity~\cite{dantsinetal2001}. For DatalogMTL over the rationals, fact entailment is PSPACE-complete in data complexity and EXPSPACE-complete in combined complexity~\cite{brandtetal2018,waegaetal2019}, also with stratified negation~\cite{tenacucalaetal2021}, and the core and linear fragments, restricted further in their temporal operators, are tractable in data complexity (TC\textsuperscript{0} and NL)~\cite{waegaetal2020}. We have not classified the programs of Appendix D against these fragments. Stable-model negation makes reasoning undecidable over the rationals but not over the integers~\cite{waegaetal2021}. Existential rules make it undecidable under the natural semantics. Under a uniform semantics, consistency checking for weakly acyclic programs is 2-EXPSPACE-complete in combined complexity under the open-world assumption, and EXPSPACE-complete for arbitrary programs under the closed-world assumption~\cite{lanzingeretal2023}. Without time, warded Datalog$\pm$ keeps query answering polynomial in data complexity~\cite{bergeretal2022}. For inconsistent DatalogMTL knowledge bases, the complexity depends on the repair notion, the timeline and the inference task. Where repairs delete whole facts, recognizing or generating a single repair is tractable for the tractable fragments, and so is query entailment under brave and intersection semantics for the core fragment~\cite{bienvenuetal2025}.

\smallskip\noindent \textbf{Engines.} Soufflé compiles Datalog to parallel C++ for large program analyses~\cite{jordanetal2016} and can record derivation provenance~\cite{zhaoetal2020}. MeTeoR combines materialization with automata-based reasoning. It is the first practical algorithm for full DatalogMTL over the rationals that is sound, complete and terminating, and its authors report scaling to tens of millions of temporal facts~\cite{wangetal2022,wangetal2024}. Related work identifies when pure materialization terminates~\cite{waegaetal2023a} and gives a terminating procedure for bounded intervals~\cite{waegaetal2023b}. Incremental maintenance of such materializations has been proposed~\cite{zhaoetal2026}. Temporal Vadalog, by the Bank of Italy and TU Wien, answers DatalogMTL queries in a data-pipeline architecture with interval merging and time-series support. Its authors report memory-efficient pipeline execution and use in financial applications~\cite{bellomarinietal2025a,bellomarinietal2025b}. Magic-sets rewriting speeds up ground DatalogMTL queries against MeTeoR on several benchmarks~\cite{wangetal2025a}. DatalogMTL programs can also be embedded into Datalog with arithmetic, which preserves finite materializability and lets standard Datalog engines evaluate them~\cite{vannoortandwaega2026}.

\smallskip\noindent \textbf{Limitations and open questions.} Pure materialization terminates only under conditions such as bounded intervals, whereas MeTeoR's hybrid procedure terminates in general. Neither guarantee by itself bounds memory or latency for continuous stream maintenance, and goal-driven optimizations have been evaluated on ground queries only. The engines assume a single timeline of trusted event times, have no native uncertainty, and, as far as we found, document no derivation output.

\subsection{Non-monotonic reasoning, argumentation and belief revision}
\noindent \textbf{Formalisms and tasks.} Answer set programming (ASP) defines stable models of logic programs with default negation~\cite{gelfondandlifschitz1988}, supporting defaults, choice and constraints, and encodings of abduction once the admissible hypotheses and their selection are specified. Temporal equilibrium logic gives ASP a temporal semantics~\cite{cabalarandperezvega2007,aguadoetal2013}, and temporal ASP extends it with temporal operators over finite traces~\cite{cabalaretal2019}, and metric temporal equilibrium logic over timed traces adds metric bounds, with an implementation on clingo~\cite{beckeretal2024,hahnetal2026}. Abstract argumentation evaluates sets of arguments and attacks under semantics such as grounded, preferred and stable~\cite{dung1995}. Structured argumentation builds arguments from strict and defeasible rules, as in ASPIC+ and defeasible logic programming~\cite{modgilandprakken2014,garciaandsimari2004}. Belief revision specifies how a knowledge base changes minimally when new, conflicting information arrives~\cite{alchourronetal1985}.

\smallskip\noindent \textbf{Expressiveness for intrusion detection.} Detection rules often need exceptions, for example: a pattern is malicious unless it is the patch job, unless the patch job's binary is unsigned. Default negation and argumentation represent such chains directly, and argumentation also adjudicates conflicting threat reports. Abduction hypothesizes unobserved attack steps that would explain the evidence, which addresses missing or deleted logs and labels such steps as hypotheses instead of treating missing observations as proof that they did not occur.

\smallskip\noindent \textbf{Complexity.} Stable model existence is NP-complete for normal and $\Sigma_2^P$-complete for disjunctive propositional programs~\cite{dantsinetal2001}. In abstract argumentation, grounded acceptance is polynomial, credulous acceptance under preferred semantics NP-complete and skeptical acceptance $\Pi_2^P$-complete~\cite{dvorakanddunne2018}.

\smallskip\noindent \textbf{Engines.} Clingo is a mature ASP system with multi-shot solving for incremental and reactive reasoning~\cite{gebseretal2019}, and telingo builds temporal ASP on it~\cite{cabalaretal2019}. ASPARTIX encodes argumentation semantics in ASP~\cite{eglyetal2010}.

\smallskip\noindent \textbf{Limitations and open questions.} The cost of incremental non-monotonic reasoning with metric time depends on the semantics, the grounding and the retained history, and the evaluations we found do not establish its performance on audit streams.

\subsection{Uncertainty and neuro-symbolic reasoning}
\noindent \textbf{Formalisms and tasks.} Probabilistic logic programming assigns probabilities to facts under the distribution semantics, as in ProbLog~\cite{deraedtetal2007}. Markov logic attaches weights to first-order formulas~\cite{richardsonanddomingos2006}. Annotated probabilistic temporal logic bounds the probability that one action follows another within a time window~\cite{shakarianetal2011}, and generalized annotated logic, implemented in PyReason, propagates interval-valued annotations over graphs with temporal extensions~\cite{adityaetal2023}. Metric temporal logic has been evaluated approximately over uncertain streams~\cite{delengandheintz2019}. Neuro-symbolic frameworks connect logic to learning: DeepProbLog uses neural predicates inside probabilistic logic programs~\cite{manhaeveetal2018}, DeepSoftLog adds soft unification over embeddings to it~\cite{maeneandderaedt2023}, Scallop makes Datalog differentiable through provenance semirings~\cite{lihuangandnaik2023}, and logic tensor networks map first-order formulas to differentiable truth degrees~\cite{badreddineetal2022}. On the temporal side, signal temporal logic robustness can be used as a differentiable loss~\cite{leungetal2023}, STL formulas with their interval parameters can be inferred by gradient descent~\cite{frondaandabbas2022,licaietal2024}, neuro-symbolic automata, relational neuro-symbolic Markov models and sequence classifiers with relational and temporal knowledge give learned models temporal structure~\cite{manginasetal2025,desmetetal2025,lorelloetal2025}, and DeepProbCEP trains an event classifier through probabilistic complex-event rules and reports robustness under adversarial training conditions~\cite{roigvilamalaetal2023}. Truth degrees under t-norm semantics and probabilities under the distribution semantics are different objects: the former do not by themselves specify the joint probabilistic model needed when hypotheses share evidence (AP7). The latter do when the shared probabilistic facts are modeled explicitly, with exact inference \#P-hard in general.

\smallskip\noindent \textbf{Expressiveness for intrusion detection.} Depending on the formalism, these can carry the score of a learned detector into symbolic reasoning as a probability, weight or truth degree, account for hypotheses that share evidence (AP7), express that a technique usually follows another within a time bound, and let rules shape learning.

\smallskip\noindent \textbf{Complexity.} Exact inference in probabilistic logic programs is \#P-hard~\cite{deraedtetal2007}, and practical systems rely on knowledge compilation or approximation.

\smallskip\noindent \textbf{Engines.} ProbLog, DeepProbLog and PyReason are publicly available, and PyReason emphasizes explainable inference traces over graphs.

\smallskip\noindent \textbf{Limitations and open questions.} Among the systems reviewed here, we did not identify a documented pipeline that combines learned hypotheses, relational recursion and metric temporal constraints with an independently checkable derivation of the resulting campaign claim. Such a combination needs an explicit input and uncertainty model as well as a checking procedure.

\subsection{Description logics, ontologies and ontology-based data access}
\noindent \textbf{Formalisms and tasks.} Description logics (DLs) represent classes, relations and constraints with decidable reasoning tasks such as consistency checking, classification and query answering~\cite{baaderetal2003}. Temporal ontology-mediated queries have been studied for first-order rewritability~\cite{artaleetal2015} and with MTL operators over timed words~\cite{kikotetal2018}. Ontology-based data access (OBDA) answers queries over heterogeneous sources through an ontology and mappings~\cite{calvaneseetal2007}. Temporal OBDA extends this to timestamped data, using temporal DLs or temporal query languages~\cite{artaleetal2017}, including metric temporal queries over log data~\cite{brandtetal2018}.

\smallskip\noindent \textbf{Expressiveness for intrusion detection.} DLs type entities and relations, check consistency of extracted facts, and organize technique taxonomies. OBDA could give heterogeneous audit sources, such as Linux Audit, Windows event logs and network sensors, a single queryable vocabulary.

\smallskip\noindent \textbf{Complexity.} ALC knowledge base satisfiability is EXPTIME-complete~\cite{baaderetal2003}. Subsumption in EL and its tractable extension EL++ is polynomial~\cite{baaderetal2005}. DL-Lite conjunctive query answering is first-order rewritable, which gives AC\textsuperscript{0} data complexity and allows implementations in SQL~\cite{calvaneseetal2007}. Combining DLs with temporal operators ranges from tractable to undecidable depending on what may change over time~\cite{artaleetal2017}.

\smallskip\noindent \textbf{Engines.} Ontop implements OBDA by query rewriting over relational sources~\cite{calvaneseetal2017}, and standard OWL reasoners cover the expressive profiles.

\smallskip\noindent \textbf{Limitations and open questions.} OWL has no temporal operators, and we found no application of temporal OBDA to audit data. Inconsistency-tolerant ontology-mediated query answering over temporal data~\cite{bourgauxetal2019} has not been evaluated on conflicting security evidence.

\subsection{Graph pattern and query languages}
\noindent \textbf{Formalisms and tasks.} Conjunctive queries match structural patterns with shared variables. Regular path queries add constraints on paths of unbounded length~\cite{barcelo2013}. Practical languages include Cypher, SPARQL property paths and GQL, an ISO standard since 2024~\cite{isoiec390752024}. Security-specific languages add time: AIQL expresses multi-event patterns with temporal relations~\cite{gaoetal2018a}, SAQL runs anomaly queries over streams with sliding windows~\cite{gaoetal2018b}, and TBQL describes threat behaviors extracted from reports~\cite{gaoetal2021}.

\smallskip\noindent \textbf{Expressiveness for intrusion detection.} Graph queries express structural signatures with joins across non-adjacent nodes and path constraints along provenance dependency paths. Whether a path's edges occur in increasing time order is not expressible in plain regular path queries and needs path predicates or a temporal extension.

\smallskip\noindent \textbf{Complexity.} Conjunctive query evaluation is NP-complete in combined complexity~\cite{chandraandmerlin1977}, and regular path queries are evaluable in nondeterministic logarithmic space in data complexity~\cite{barcelo2013}.

\smallskip\noindent \textbf{Engines.} General graph databases implement Cypher or GQL. Security query systems use storage and execution optimizations specific to provenance data~\cite{gaoetal2018a}.

\smallskip\noindent \textbf{Limitations and open questions.} Temporal constructs are defined per system rather than by a shared formal semantics, so comparing their expressiveness with temporal logics or rule languages requires an explicit semantics or translation first. Appendix D illustrates selected patterns but gives no general semantics-preserving translation. The query interfaces we found return matches rather than derivations that can be checked independently.

\subsection{Causal reasoning}
\noindent \textbf{Formalisms and tasks.} Structural causal models represent mechanisms as equations over variables and support interventions and counterfactuals~\cite{pearl2009}. Actual causation defines when a particular event caused a particular outcome in a specific scenario~\cite{halpernandpearl2005,halpern2016}, and extensions quantify responsibility and blame~\cite{chocklerandhalpern2004}.

\smallskip\noindent \textbf{Expressiveness for intrusion detection.} Given a specified structural model, causal analysis distinguishes candidate causes of a compromise from the many events it merely depends on, and counterfactual queries ask which countermeasure would have prevented an outcome.

\smallskip\noindent \textbf{Complexity.} Under the original Halpern--Pearl definition, deciding causality is $\Sigma_2^P$-complete in general and NP-complete for binary variables~\cite{eiterandlukasiewicz2002}. The later definitions cited above have different bounds.

\smallskip\noindent \textbf{Engines.} General causal inference libraries exist, but we found none applied to system provenance. Actual causation over event traces has been defined for safety-property violations in discrete event systems~\cite{goleretal2026} and for structural equation models that evolve over time~\cite{gladyshevetal2025}, and neither has been applied to provenance graphs.

\smallskip\noindent \textbf{Limitations and open questions.} We found no validated structural causal model for host audit data, and mapping interventions to concrete countermeasures, for example D3FEND techniques, is open.

\subsection{Qualitative and spatio-temporal reasoning}
\noindent \textbf{Formalisms and tasks.} Allen's interval algebra relates intervals by thirteen base relations such as before, overlaps and during~\cite{allen1983}. The main task is consistency checking of interval networks. Signal temporal logic specifies properties of real-valued signals and gives a quantitative robustness degree~\cite{malerandnickovic2004}. Spatio-temporal logics such as STREL add reach and escape operators over graphs that evolve in time~\cite{bartoccietal2017}. Qualitative spatial calculi such as RCC relate regions~\cite{randelletal1992}.

\smallskip\noindent \textbf{Expressiveness for intrusion detection.} Interval relations express overlapping attack stages without exact timestamps, which helps under clock skew (step 3 of the running example). Signal temporal logic captures physical-process constraints in industrial systems. STREL can express whether a controller's network zone is reachable from a compromised workstation in the graph at a given time and whether that holds within a time bound. Propagation along edges that exist at increasing times is not expressed by its reach operator (Appendix D).

\smallskip\noindent \textbf{Complexity.} Consistency in the full interval algebra is NP-complete~\cite{vilainandkautz1986}, and the ORD-Horn subclass is the maximal tractable subclass containing all base relations~\cite{nebelandburckert1995}. Signal temporal logic and STREL admit monitoring algorithms with stated complexity bounds~\cite{malerandnickovic2004,bartoccietal2017}.

\smallskip\noindent \textbf{Engines.} MoonLight monitors STREL properties~\cite{bartoccietal2020}.

\smallskip\noindent \textbf{Limitations and open questions.} We found no application of STREL to host-based intrusion detection, and host provenance, network zones and physical invariants are reasoned about separately rather than in one spatio-temporal model.

\subsection{Rule and specification learning, and LLMs with formal reasoning}
\noindent \textbf{Formalisms and tasks.} Specification mining learns temporal-logic formulas or invariants from traces. Inductive logic programming learns logic rules from examples, and LLMs translate natural language into formal artifacts such as temporal-logic formulas, detection rules, queries or knowledge graphs.

\smallskip\noindent \textbf{Expressiveness for intrusion detection.} These methods address the specification acquisition and maintenance that many symbolic approaches depend on: they can turn threat reports and normal behavior into rules that the families in Section 4.2 execute.

\smallskip\noindent \textbf{Complexity.} The learning problems are typically intractable in general, and evaluating a learned or generated artifact costs at most what its target language costs in the worst case. A particular rule may lie in an easier fragment.

\smallskip\noindent \textbf{Engines and tools.} The learning methods differ in what they learn. Template-free methods learn both the structure and the parameters of formulas, as STL inference by gradient descent~\cite{frondaandabbas2022,licaietal2024} and LTLf learning by inductive logic programming~\cite{ieloetal2025} do, and template-based methods fit only the parameters of a given formula. Temporal logic point processes learn weights for propositional rules together with the rules~\cite{lietal2020}, and DatalogMTL rule mining scores candidate rules by time-sensitive confidence over windows~\cite{wangwaegaandcuencagrau2024}. The template-free methods cited here learn from examples labeled as normal and anomalous, whereas Indri et al. learn ensembles of formulas in one shot from normal data alone~\cite{indrietal2022}. LLM front ends such as nl2spec translate natural language into temporal logic interactively, leaving the check of the result to the user~\cite{cosleretal2023}.

\FloatBarrier
\onecolumn
\section{Evidence for Tables 5 and 6}

\subsection{Evidence for Table 5}
{\scriptsize
\begin{longtable}{>{\raggedright\arraybackslash}p{0.125\textwidth} >{\raggedright\arraybackslash}p{0.035\textwidth} >{\raggedright\arraybackslash}p{0.185\textwidth} >{\raggedright\arraybackslash}p{0.16\textwidth} >{\raggedright\arraybackslash}p{0.165\textwidth} >{\raggedright\arraybackslash}p{0.20\textwidth}}
\caption{Statements about order, clocks, delivery and trust in the evidence in the sampled systems.}\label{tab:C1}\\
\multicolumn{6}{p{0.96\textwidth}}{\scriptsize Q1 and Q5 are summarized per family in Table 5 and coded per system in the evidence ledger, which gives the passage behind every cell and also records the input layer and the eligibility of each paper for the Q4 and Q6 counts of Section 4.3. --- means that no statement was located, and n/a means that the question does not apply. † marks systems coded from an abstract or a description in another paper, for which Q3 to Q6 could not be assessed. Family F lists the digital-forensics papers on time and event reconstruction, which are not detection systems. The ledger distinguishes delivery conditions and observations from statements about output validity under delivery failures.}\\[2pt]
\toprule
\textbf{System} & \textbf{Family} & \textbf{Order (Q2)} & \textbf{Clocks (Q3)} & \textbf{Delivery (Q4)} & \textbf{Trust in the evidence (Q6)} \\
\midrule
\endfirsthead
\multicolumn{6}{l}{\textit{Table C1, continued}}\\
\toprule
\textbf{System} & \textbf{Family} & \textbf{Order (Q2)} & \textbf{Clocks (Q3)} & \textbf{Delivery (Q4)} & \textbf{Trust in the evidence (Q6)} \\
\midrule
\endhead
\bottomrule
\endfoot
Liu et al. 2019 (Log2vec) & L1 & Chronological edges between log entries of a user & --- & --- & --- \\
Wang et al. 2020 (ProvDetector) & L1 & Causal paths in temporal order & --- & --- & Kernel, audit system or logs trusted (TCB) \\
ATLAS & L1 & Temporal order in an entity's neighborhood & --- & --- & OS and audit trusted; causal graph tamper-proof \\
KAIROS & L1 & Windows of wall-clock time; queue across windows & --- & Edges ingested chronologically as they appear in the stream & Kernel, audit framework and analysis code trusted \\
King and Huang 2022 (Euler) & L1 & Temporal link prediction across windows & --- & --- & --- \\
NODLINK & L1 & Not stated; an in-memory cache retains state instead of fixed windows & --- & --- & None found; uncaptured channels out of scope \\
R-CAID & L1 & Causal paths ordered by edge timestamps & --- & --- & Logs cannot be altered; tamper-evident logging out of scope \\
SIGL & L1 & Sliding window over installation sequences & --- & --- & OS and audit framework trusted \\
TFLAG & L1 & Time windows; encoded time differences & --- & --- & Tamper-proof log storage assumed \\
Unicorn & L1 & Order of graph sketches across the stream & --- & Streaming updates exploit partial arrival order (incoming edges before outgoing at each vertex), with extra updates when it does not hold; capture races discussed; no guarantee for late or lost records & Relies on capture guarantees; discusses capture races \\
Wang et al. 2022b (ThreaTrace) & L1 & Node-level graph learning & --- & --- & --- \\
Zeng et al. 2022 (ShadeWatcher) & L1 & Interaction sequences ordered by timestamps & --- & --- & Kernel, audit system or logs trusted (TCB); tamper-evident auditing named as future integration \\
King et al. 2023 (EdgeTorrent) & L1 & Temporal order preserved in message passing over an edge stream & --- & Edges streamed in the order in which they were observed & --- \\
Jia et al. 2024 (MAGIC) & L1 & Masked graph learning & --- & --- & Kernel, audit system or logs trusted (TCB) \\
Khoury et al. 2024 (Jbeil) & L1 & Temporal graph over evolving edges & --- & --- & --- \\
Jiang et al. 2025b (Orthrus) & L1 & Temporal graph learning; dependency analysis for attribution & --- & --- & Trusted: follows the threat model of prior provenance detectors (kernel and audit trusted) \\
Landauer et al. 2025 & L1 & Delta times between consecutive probes; chronological sorting alone insufficient & --- & --- & --- \\
Sang et al. 2025 (STGAN) & L1 & Temporal encoding of dynamic provenance graphs & --- & --- & Kernel, audit system or logs trusted (TCB) \\
EagleEye & L2 & Graph traversal in temporal order & --- & Missing behavior events can change the classification & Audit framework and logs not compromised \\
FLASH & L2 & Events sorted by timestamp & --- & --- & Tamper-resistant log storage assumed \\
ANCHOR & L3 & Across windows through a tracking cache & --- & --- & Capture mechanism and OS trusted \\
OCR-APT & L3 & Timestamp-ordered subgraphs; LLM composes across them & Coarsened to seconds; no clock bound & --- & Audit framework trusted \\
ProvAgent & L3 & Event sequences and idle periods & --- & --- & Kernel and auditing trusted; logs immutable once collected \\
ProvSEEK & L3 & Chronological order of interactions & --- & --- & Collection and audit pipeline trusted \\
Hmimou et al. 2025 & L3 & Agents correlate events & --- & --- & --- \\
Karunanayake et al. 2025 & L3 & Not addressed & --- & --- & --- \\
Loumachi et al. 2025 & L3 & Chronological timeline reconstruction & --- & --- & Anti-forensics explicitly excluded from the evaluated scenarios \\
CAPTAIN & L4 & Event order in the stream & --- & --- & Kernel and auditing tools trusted \\
Chen et al. 2022 (APT-KGL)† & L4 & Local subgraphs around key entities; offline embedding, online detection & not assessable & not assessable & not assessable \\
KnowGraph & L4 & Windows of 1,800 s & --- & --- & None found \\
Yue et al. 2024 & L4 & Attack steps in event order; tracing in reverse timestamp order & --- & --- & Audit data assumed tamper-proof \\
Denning 1986 & L5 & Statistical profiles over time intervals & Clock used for periodic updates; no synchronization statement & --- & --- \\
Forrest et al. 1996 & L5 & Short sequences within a trace & n/a & --- & --- \\
Ko et al. 1997 & L5 & Trace policies over program executions & Synchronization only as program behavior; clocks not addressed & --- & --- \\
Hofmeyr et al. 1998 & L5 & Short sequences within a trace & n/a & --- & --- \\
Lee and Stolfo 1998 & L5 & Frequent episodes over records & --- & --- & --- \\
Warrender et al. 1999 & L5 & Sequences within a trace & --- & --- & --- \\
Sekar et al. 2001 & L5 & Automaton over call sequences & n/a & --- & --- \\
Wagner and Dean 2001 & L5 & Model derived from program code & n/a & --- & --- \\
Liao and Vemuri 2002 & L5 & None (bag of calls per process) & n/a & --- & --- \\
Kruegel et al. 2003 & L5 & Within one system call & --- & No audit records dropped in the evaluation & --- \\
Lanzi et al. 2010 (AccessMiner) & L5 & System-centric access models & --- & --- & --- \\
Maggi et al. 2010 & L5 & Sequences and arguments & --- & --- & --- \\
Shu et al. 2015 & L5 & Long call sequences & --- & --- & --- \\
Guo et al. 2021 (LogBERT) & L5 & Sequences of log keys & --- & --- & --- \\
Cuppens and Ortalo 2000 & S1 & Event algebra of sequence, choice and synchronized execution & --- & --- & None stated \\
Debar and Wespi 2001 & S1 & Alert aggregation and correlation; arrival out of time order allowed & Synchronized probes required; no bound & Out-of-order arrival allowed; a missing expected alert indicates a failed probe & None stated; spoofed source addresses modeled \\
Geib and Goldman 2001 & S1 & Partial order of plan steps; unobserved actions inferred from state changes & --- & Unobserved actions inferred from reported state changes (incomplete record of actions) & None stated \\
Templeton and Levitt 2001 & S1 & Requires/provides capabilities between attacks & --- & --- & --- \\
Valdes and Skinner 2001 & S1 & Alerts fused by feature similarity & Clock drift tolerated by a minimum time similarity; no numeric bound & Late reports considered & Spoofed source addresses considered in similarity \\
Cuppens and Miège 2002 & S1 & Post-condition before pre-condition & --- & Missing facts abduced to complete a diagnosis (virtual alerts) & None stated; missing facts abduced \\
Dain and Cunningham 2002 & S1 & Alerts assigned to scenarios by probability & --- & Long attack latencies tolerated & Forged IP addresses considered \\
Morin et al. 2002 (M2D2) & S1 & Events compared by close timestamps & Clock synchronization impossible; constant close to the maximum gap between synchronized clocks & --- & --- \\
Ning et al. 2002 & S1 & Prerequisite before consequence by interval timestamps & --- & Partial matches may indicate attacks missed by IDSs & None stated; missed attacks indicated by partial matches \\
Porras et al. 2002 (M-Correlator) & S1 & Alerts ranked and clustered by mission impact & --- & Incomplete fact base or topology tolerated & --- \\
Cheung et al. 2003 (CAML) & S1 & Module pre- and post-conditions with temporal constraints & --- & Missing sensor coverage acknowledged; policy for missing information & --- \\
Julisch 2003 & S1 & Clusters of alarms; no scenario order & --- & --- & Spoofed source addresses noted for correlation systems \\
Ning and Xu 2003 & S1 & Partial temporal order of attack steps & --- & Missed detections identified via strategy similarity & None stated \\
Qin and Lee 2003 & S1 & Time-series causality with lags & --- & Missing alerts acknowledged; future work & None stated \\
Ning et al. 2004 & S1 & Prerequisite before consequence by interval timestamps & --- & Missed attacks hypothesized from indirect causal relations and validated against raw audit data & None stated; missed attacks hypothesized \\
Qin and Lee 2004 & S1 & Partial order in causal networks; earlier alert before later & --- & Partial order and unobserved steps handled; missing alert hypothesized & None stated \\
Valeur et al. 2004 & S1 & Alert fusion in a sliding time window; scenario reconstruction & Sensor clocks synchronized by NTP; timestamp heuristics may introduce errors; no bound & --- & --- \\
King et al. 2005 & S1 & Across hosts through recorded packets & --- & --- & None stated \\
Wang et al. 2006 & S1 & Queue graph; neither timestamp order nor arrival order trusted & Sensor clocks loosely synchronized by NTP; no bound & Missing alerts hypothesized; attacker may delay steps to escape sliding window & Delayed steps and injected alerts considered; clock tampering left open \\
Zhai et al. 2006 & S1 & Alert times matched to lifetimes of host objects & Window widened for clock discrepancy; no bound & Hypotheses of missed attacks via OS-level dependencies & Host event logs trusted over alerts \\
Fava et al. 2008 & S1 & Variable-length Markov models over alert sequences & --- & --- & --- \\
Albanese et al. 2011 & S1 & Temporal windows on activity sequences in attack models & --- & --- & --- \\
Roschke et al. 2011 & S1 & Alerts mapped to attack-graph paths in time order & --- & --- & --- \\
HERCULE & S1 & Across logs by a timestamp threshold & Cross-log comparison without a clock bound & --- & None stated \\
Ramaki et al. 2015 (RTECA) & S1 & Episodes of alerts within time windows & --- & --- & --- \\
Ghafir et al. 2018 (MLAPT) & S1 & Stage alerts correlated within a time window & --- & Missing a stage acknowledged as failure mode & --- \\
Tian et al. 2019 & S1 & Alarms added as evidence in chronological order & --- & --- & --- \\
Zimba et al. 2020 & S1 & Correlation of APT stages within time windows & --- & --- & --- \\
Eckhoff et al. 2025 & S1 & Over the trace of SIEM alerts & --- & --- & None stated \\
Kumar and Spafford 1994 & S2 & Patterns over tagged events with sequence, duration and interval conditions & Interval conditions hold up to clock accuracy; no bound & --- & None stated \\
Ilgun et al. 1995 & S2 & State transitions in order within a scenario & --- & --- & None stated \\
Porras and Neumann 1997 (EMERALD)† & S2 & Recursive coordination of analyses across monitors and layers & not assessable & not assessable & not assessable \\
Lindqvist and Porras 1999 & S2 & Time windows and timestamp differences within rules & --- & --- & None stated \\
Vigna and Kemmerer 1999 (NetSTAT) & S2 & Within a scenario; across hosts after merging audit trails & Merging trails from several hosts with synchronization named; no bound & --- & --- \\
Pouzol and Ducassé 2001 & S2 & Order within a declarative signature & --- & --- & None stated \\
Morin and Debar 2003 & S2 & Chronicles with time constraints over alert timestamps & Synchronized sensor clocks with a gap coherent with the chronicle time resolution; delayed alerts integrated under this condition & Delays from sensor processing or transport handled when the clock condition holds & Attacker may vary time gaps to evade detection \\
STATL & S2 & Within a scenario, with timers & --- & --- & None stated \\
Meier et al. 2005 & S2 & Order within a signature (EDL) & --- & --- & --- \\
Roger and Goubault-Larrecq 2001 & S3 & Temporal-logic constraints on the order of events in a log & --- & --- & --- \\
ORCHIDS & S3 & Within a temporal-logic signature & --- & --- & None stated \\
Naldurg et al. 2004 & S3 & Across sources after merging by timestamp & Cross-source merge without a clock bound & --- & None stated \\
Rekhis and Boudriga 2005 & S3 & Temporal logic of actions for forensic reconstruction & --- & Hypotheses formulated where details are missing & --- \\
Garg et al. 2011 & S3 & Total order of integer event times & --- & Incomplete logs as three-valued partial structures; isFinal: table complete up to last update time; residual policy & None stated; incomplete logs as three-valued structures \\
Chowdhury et al. 2014 & S3 & Temporal mode-checking of first-order policies & --- & Potentially incomplete logs (three-valued) & --- \\
Basin et al. 2015 (MonPoly) & S3 & Metric first-order temporal logic over a timestamped trace & Limited clock precision acknowledged; no bound on offset between sources & Time stamps must be monotonically increasing (in-order trace assumed) & --- \\
R2U2 & S3 & Bounded temporal-logic intervals over ticks of one system clock & --- & --- & Spoofed sensor and GPS data are among the diagnosed attacks \\
Jajodia et al. 2016 & S3 & Probabilistic temporal rules over security events & --- & --- & --- \\
HOLMES & S4 & Kill-chain stage along information flows & --- & One scenario flagged despite missing events & Kernel and auditing engine trusted \\
KnowHow & S4 & Order of lifecycle stages, partly unconstrained & --- & --- & Prior threat model of trusted monitoring \\
RapSheet & S4 & Temporal order of alerts & --- & --- & EDR not compromised; logs not tampered \\
SLEUTH & S4 & Per-subject order of events & Millisecond resolution & --- & Audit data trustworthy \\
Satvat et al. 2021 (Extractor) & S5 & Within one report & n/a & n/a & --- \\
AIQL & S5 & Temporal relations between events, per host & NTP synchronization with server-side check; no bound & Timestamps of collected data assumed to increase monotonically per host & Prior threat model of trusted monitoring \\
KRYSTAL & S5 & Causal paths over provenance relations; no temporal constraints in its reconstruction queries & Time synchronization named as an open issue for distributed online analysis & --- & None stated \\
POIROT & S5 & Alignment of a query graph & --- & --- & Kernel logs trusted; separate secure log server \\
SAQL & S5 & Sliding windows over the stream & --- & --- & Prior threat model of trusted monitoring \\
ThreatRaptor & S5 & Order constraints from edge sequence numbers & --- & --- & Prior threat model; kernel trusted \\
Ammann et al. 2002 & S6 & n/a & n/a & n/a & --- \\
Sheyner et al. 2002 & S6 & Atemporal & n/a & n/a & None stated \\
MulVAL & S6 & Atemporal & n/a & n/a & None stated \\
Noel et al. 2004 & S6 & Event paths by attack-graph distance, in arrival order & --- & Missed detections bridged by attack-graph distance & None stated \\
Ingols et al. 2006 (NetSPA) & S6 & n/a & n/a & n/a & --- \\
Ou et al. 2006 & S6 & Atemporal & n/a & n/a & None stated \\
Wang et al. 2008 & S6 & n/a & n/a & n/a & --- \\
Ingols et al. 2009 & S6 & n/a & n/a & n/a & --- \\
Poolsappasit et al. 2012 & S6 & Atemporal; posterior update as attack evidence is observed & n/a & --- & None stated \\
Sun et al. 2018 & S6 & Bayesian network over object instances with timestamps & --- & --- & --- \\
Özdemir Sönmez et al. 2022 & S6 & n/a & n/a & n/a & --- \\
Zhang et al. 2025 (AttacKG+) & S6 & Within one report & n/a & n/a & --- \\
Nunes et al. 2016 & S7 & Not applicable & --- & --- & Attacker may plant false evidence \\
Karafili et al. 2020 & S7 & Not applicable & --- & Missing evidence abduced and suggested to the analyst & Attacker may forge sources; missing evidence abduced \\
Ko 2000 & S8 & Order of accesses ignored by design & n/a & --- & Training traces from a securely configured host \\
Nanda and Chiueh 2008 & S8 & Signatures from execution traces & --- & --- & --- \\
Li et al. 2009 & S8 & Signatures from execution traces & --- & --- & --- \\
Godefroy et al. 2015 & S8 & Generated correlation rules & Knowledge base may be out of sync with the environment (not clocks) & --- & --- \\
Feng et al. 2019 & S8 & Invariants over sensor readings at discrete time points & --- & --- & None stated \\
Raff et al. 2020 & S8 & Within one generated rule & n/a & n/a & --- \\
Indri et al. 2022 & S8 & Within one learned formula & --- & --- & Not applicable \\
SIGMERGE & S8 & Within one generated rule & --- & --- & Threat reports assumed untampered \\
Schwartz et al. 2025 (LLMCloudHunter) & S8 & Within one generated rule & --- & --- & --- \\
Martin and Plaza 2003 (SOID) & S9 & Within one report or ontology & n/a & n/a & --- \\
Undercoffer et al. 2003 & S9 & Not addressed; timestamps only as recorded data & n/a & --- & None stated \\
He et al. 2004 & S9 & Within one report or ontology & n/a & n/a & --- \\
Coppolino et al. 2009 & S9 & Attack start and end from lowest and highest timestamps & --- & --- & --- \\
Balduccini et al. 2015 & S9 & Not addressed & n/a & Incomplete knowledge of the analyst noted & --- \\
Liao et al. 2016 (iACE) & S9 & Not addressed & n/a & n/a & --- \\
Husari et al. 2017 (TTPDrill) & S9 & Temporal order of kill-chain phases from text & n/a & n/a & --- \\
CTINexus & S9 & Not addressed & n/a & n/a & None stated \\
ChronoCTI & S9 & Relative order of techniques from text & n/a & n/a & None stated \\
Li et al. 2022 (AttacKG) & S9 & Within one report & n/a & n/a & --- \\
Ren et al. 2023 (CSKG4APT) & S9 & Within one report or ontology & n/a & n/a & --- \\
Hu et al. 2024 (LLM-TIKG) & S9 & Within one report or ontology & n/a & n/a & --- \\
Goel et al. 2005 (Taser) & S10 & Dependencies over logged file-system operations & --- & --- & Logging system assumed not corrupted by the attack \\
Joshi et al. 2005 (IntroVirt) & S10 & Vulnerability predicates over replayed past execution & Hardware clock known in replay only where it was read & --- & Virtual machine monitor trusted \\
Purohit et al. 2004 & S10 & Tracing process ancestry to its root & --- & --- & Root attacker may tamper with a user-level daemon; motivates kernel tracing \\
Arnes et al. 2006 (ViSe) & S10 & Order of events reconstructed in a virtual testbed & Distributed events not necessarily synchronized; order may be non-deterministic; no bound & --- & Anti-forensic tools to be considered in reconstruction \\
Martignoni et al. 2010 & S10 & Live forensic analysis & --- & --- & OS assumed compromised; hypervisor-based analysis cannot be tampered with \\
Mokhov et al. 2010 (Forensic Lucid) & S10 & Event reconstruction in an intensional logic & --- & --- & --- \\
Rekhis and Boudriga 2012 & S10 & Formal investigation over action sequences & --- & --- & Anti-forensic attacks modeled \\
Rekhis and Boudriga 2012b & S10 & Hypotheses over action sequences & Clock adjustment hypotheses discussed (after Willassen) & --- & Anti-forensic attacks that alter evidence modeled \\
Lee et al. 2013a (BEEP) & S10 & Execution units partition long-running processes & --- & Delays in user interaction and asynchronous tasks make timestamps ineffective for causality & Attacker without privileges to tamper with instrumented binaries assumed \\
Ma et al. 2016 (ProTracer) & S10 & Units and taint propagation over event order & --- & Daemon processes events concurrently and out of order across applications & Log assumed untampered by the administrator \\
Xu et al. 2016 & S10 & Causality-preserving reduction over event time windows & --- & --- & --- \\
Ji et al. 2017 (RAIN) & S10 & Record and replay; refinable information-flow tracking & --- & --- & Kernel, audit system or logs trusted (TCB) \\
Liu et al. 2018 (PrioTracker) & S10 & Backward and forward tracing in temporal order within a time limit & --- & --- & Kernel, audit system or logs trusted (TCB) \\
Yang et al. 2018 & S10 & Temporal dependence network over objects & --- & --- & --- \\
Hassan et al. 2019 (NoDoze) & S10 & Dependency graph paths; frequencies in a time window & --- & --- & Kernel, audit system or logs trusted (TCB) \\
Hossain et al. 2020 (MORSE) & S10 & Tag propagation along event order & --- & Incomplete DARPA TC logs noted in evaluation & --- \\
Pai Kasturi et al. 2020 (TARDIS) & S10 & Sliding-window ranking of temporal snapshots & --- & Robust to out-of-order attack payloads & --- \\
Fang et al. 2022 (DepImpact) & S10 & Dependency impact back-propagated with time-based weights & --- & --- & Kernel, audit system or logs trusted (TCB) \\
Yang et al. 2022 (RATScope) & S10 & Recorded RAT behaviors reconstructed & --- & Missing RAT semantics recorded and reconstructed & Kernel, audit system or logs trusted (TCB) \\
Neale et al. 2026 & S10 & Reasoning about which artifacts were tampered & --- & --- & Tampering of artifacts is the subject of reasoning \\
Gladyshev and Patel 2004 & F & Event reconstruction by finite state machines & --- & Partial knowledge of the event sequence & --- \\
Schatz et al. 2006 & F & Correlation of timestamps across sources & Clock skew and drift estimated by correlating with reference sources & --- & Clock tampering named as a factor \\
Buchholz and Tjaden 2007 & F & n/a & Measured clock offsets: about 74\% of observed hosts within 10 s of reference & Network delay in synchronization noted & --- \\
Willassen 2008 & F & Clock hypotheses checked against observed evidence & Clock adjustments modeled as hypotheses; manipulated or erroneous clocks & n/a & Timestamps may be manipulated; hypotheses test consistency \\
Hargreaves and Patterson 2012 & F & Automated timeline reconstruction & Assumes the clock of the investigated system is correct; no generic clock offset supported & --- & --- \\
\end{longtable}
}
\twocolumn

\subsection{Rubric and evidence for Table 6}
Table C2 states what each dimension of Table 6 assesses and what full support requires. The evidence behind each family's ratings follows.

\begin{table*}[!tp]
\caption{Assessment criteria for the dimensions of Table 6.}
\label{tab:C2}
\centering \scriptsize
\begin{tabular}{>{\raggedright\arraybackslash}p{0.150\textwidth} >{\raggedright\arraybackslash}p{0.282\textwidth} >{\raggedright\arraybackslash}p{0.226\textwidth} >{\raggedright\arraybackslash}p{0.207\textwidth}}
\toprule
\textbf{Dimension} & \textbf{What is assessed} & \textbf{++ requires} & \textbf{$-$ or $--$ indicates} \\
\midrule
Correct evaluation of a declarative attack specification & Whether outputs are correct with respect to a stated declarative specification of attack behavior (Section 2, evaluation level); it concerns an explicit campaign or attack-pattern specification, and specifications of permitted program behavior (L5) are distinguished in the evidence and not counted & Sound, and where claimed complete, evaluation of such a specification & No declarative attack specification is evaluated ($--$), or generation is unconstrained ($-$) \\
Explanation form & The kind of justification an alert comes with & Not rated; the form is named & --- \\
Independently checkable & Whether a single output can be verified by a checker separate from the producing system & A separate checker for the explanation format exists; whether that checker is itself formally verified is noted in the evidence per family below & No checkable format \\
Order and metric time & Native support for event order and time bounds & Metric temporal operators or constraints & Order ignored or only within windows \\
Long horizons & Whether reasoning relates evidence across windows, hours and days without a horizon-dependent cut-off & No cut-off on how far back evidence can be related; memory use is reported separately and may grow with the number of entities & Context capped by window or sequence length \\
Recursion and joins & Unbounded paths and joins across non-adjacent entities & Recursive rules or path queries, together with joins on shared variables & Bounded by message-passing depth, or no join construct \\
Counting & Multiplicities and aggregates & Native aggregates & Averaging that loses multiplicity \\
Uncertainty & Whether the published pipeline gives the combination of uncertain evidence a stated semantics & An explicit probabilistic or annotated semantics; calibration of the inputs and treatment of shared evidence are reported separately (Section 7.3) and not inferred from this rating & + a combination mechanism whose semantic commitments are partly specified; $-$ scores without a combination semantics \\
Defaults, incompleteness, conflicts & Exceptions, missing evidence and conflicting evidence, each with a stated semantics & A stated semantics for all three, such as argumentation, or abduction with defaults; + means a stated semantics for at least one, named in the evidence per family below & Closed-world assumption without exceptions, or handling without a stated semantics \\
Semantic context & Use of command lines, paths and attributes; the codes describe representation mechanisms, not validated interpretation & Attribute values interpreted by a learned or ontology-based representation, such as attribute encoders, language models or ontologies & + attribute values used through explicit tests, parsing or deterministic encoding; $-$ structure or event types only \\
Demonstrated raw-stream operation & Whether the cited evaluation demonstrates operation on raw host audit streams; it does not compare throughput across systems or imply a shared real-time threshold & Demonstrated operation on raw audit streams at the event rates of the cited evaluations & Requires curated evidence ($--$) \\
Late or out-of-order data & Correct results when records arrive late or reordered & A stated correctness result for a named verdict or answer relation under a specified disorder model, with its ordering constraints, lateness or retention bounds, and whether outputs are provisional, retractable or final & Results depend on arrival order \\
Adversarial manipulation & Resistance to evasion, deletion and time manipulation under a stated attacker capability (Section 2) & A stated guarantee that named conclusions remain valid for all manipulations the attacker model permits (empirical resistance to evaluated manipulations: +) & Documented evasion \\
Knowledge update & Adding or changing detection knowledge without retraining; the evidence names the object changed (rules, ontology or mappings, prompts or retrieval content, program-derived model, learned parameters) and whether the change needs retraining, reanalysis or only re-evaluation; event ingestion does not count & Declarative rule changes & Retraining required ($--$) \\
\bottomrule
\end{tabular}
\end{table*}
\smallskip\noindent \textbf{Evidence per family.} We list the evidence behind each family's ratings. A cell rated dep depends on an input that the family does not produce, such as an extractor, n/a means that the dimension does not apply to the family's task, and n/s means that the capability was not evaluated in the sources we inspected.

\begin{itemize}
\item \textbf{L1.} The pipelines output scores and summary graphs~\cite{hanetal2020,chengetal2024}. Only NODLINK learns embeddings of command lines, paths and addresses (\textsuperscript{1})~\cite{lietal2024}, whereas KAIROS encodes path and network attributes by feature hashing, which uses attributes without learning their semantics~\cite{chengetal2024}. The pipelines expose no recursive-query or join construct and their reconstruction steps are fixed procedures, while logical characterizations of aggregate-combine and temporal message passing bound their learned backbones on stated tasks~\cite{barceloetal2020,xuetal2019,waegaandrawson2025}. They operate in real time with windowed context~\cite{chengetal2024,lietal2024} and produce anomaly scores without a combination semantics. Mimicry evades them~\cite{goyaletal2023}, and resistance under a stated mimicry attacker is evaluated for R-CAID only (\textsuperscript{1})~\cite{goyaletal2024}. New detection knowledge requires adapting or retraining the model, unlike the ingestion of new events.
\item \textbf{L2.} The family rates as L1, plus attribute and command-line semantics~\cite{rehmanetal2024,gyseletal2024}, and CrossGuard adds window limits and an assumption of intact logs~\cite{yoonetal2025}. FLASH reports resistance to the evaluated mimicry attack, which inserts benign structures into attack graphs (+\textsuperscript{1}), without establishing resistance to deletion, timestamp manipulation or stronger adaptive attacks~\cite{rehmanetal2024}, and EagleEye reports real-time resource measurements on endpoint event streams (++ for raw-stream operation).
\item \textbf{L3.} The agents produce narratives without derivations and rely on cross-window caches and evidence curation~\cite{alyetal2025,zhengetal2026}. The pipelines ingest audit logs and select the evidence passed to the LLM, which their authors motivate by cost and context limits, but they report no operation on raw streams at stated event rates (n/s). Missing and conflicting evidence are handled in narratives without a stated semantics ($-$), and prompts and baselines update knowledge without adherence guarantees. Command lines, paths and other attributes are interpreted as text by the language model (++ for semantic context under the criterion of Table C2).
\item \textbf{L4.} Rules carry learned weights or parameters~\cite{zhouetal2024,wangetal2025b}. CAPTAIN runs in real time with persistent tags, and the ASP labeling of alert-embedding clusters uses weak constraints that align timestamp order with kill-chain order, which makes it the only representative with explicit order (\textsuperscript{1}) and defaults (\textsuperscript{1})~\cite{eckhoffetal2025}, while the other rules are atemporal. KnowGraph's reasoning component is a Markov logic network over weighted first-order formulas, a probabilistic semantics (++ for uncertainty)~\cite{zhouetal2024}. Its knowledge models predict node attributes and semantic entities with learned models (++ for semantic context). CAPTAIN evaluates streaming audit-log processing (++\textsuperscript{1} for raw-stream operation) and reports resistance in bounded mimicry and contaminated-training-data experiments (+\textsuperscript{1}), without establishing resilience to corruption of the deployed audit stream~\cite{wangetal2025b}.
\item \textbf{L5.} The models describe normal system-call sequences, arguments or access behavior, learned from training traces, specified as policies or derived from program code~\cite{forrestetal1996,sekaretal2001,kruegeletal2003,lanzietal2010}. They capture order within one process trace and frequencies as counts, operate online on system-call streams and evaluate no declarative attack specification, and mimicry attacks evade them~\cite{wagnerandsoto2002}. Learned models are relearned, whereas specification-based monitors and models derived from program code change by editing the policy or reanalyzing the program~\cite{koetal1997,wagneranddean2001}. Learned models of system-call arguments interpret attribute values~\cite{kruegeletal2003,maggietal2010} (++ for semantic context), while sequence-only models use event types alone.
\item \textbf{S1.} Alerts are chained logically into scenario graphs~\cite{ningetal2002,cuppensandortalo2000}, plan recognition~\cite{geibandgoldman2001} and temporal ASP over SIEM alerts~\cite{eckhoffetal2025} use abduction, and attack steps are modeled declaratively. Metric time constraints and probabilities of temporally constrained attack occurrences~\cite{albaneseetal2011} and causal networks over alerts~\cite{qinandlee2004} come without calibration of their inputs. Out-of-order arrival and late reports are tolerated without a formal treatment~\cite{debarandwespi2001,valdesandskinner2001}. Injected alerts and delayed steps~\cite{wangetal2006} and forged source addresses~\cite{dainandcunningham2002} are considered, but without a guarantee for host evidence.
\item \textbf{S2.} Chronicles and CEP engines support metric constraints and windowed counts~\cite{morinanddebar2003,wuetal2006}, and out-of-order sequence-pattern processing comes with correctness guarantees for its fragment~\cite{liuetal2009,chandramoulietal2010}, rated ++ under the same fragment rule as S3. CEL gives CEP a formal core with constant-time-per-event evaluation of a fragment~\cite{grezetal2019}, whereas the SIEM languages have no formal semantics, and evaluated public Sigma rules are evaded by small changes to commands or process attributes~\cite{uetzetal2024}. Liu et al. process sequence patterns either conservatively, emitting final results once a bound on disorder guarantees no earlier event remains, or aggressively, emitting provisional results that are later compensated~\cite{liuetal2009}. CEL and the pattern evaluators of Zhang et al. have correctness results within their supported fragments~\cite{grezetal2019,zhangetal2010}, which meets the evaluation criterion within those fragments (++) without establishing correctness for other CEP languages or their security deployments. Zhang et al. also combine event-time distributions into probabilities of pattern matches, which supports ++ for timing uncertainty without establishing calibrated technique probabilities or the shared-evidence campaign inference of AP7.
\item \textbf{S3.} Monitors give exact verdicts with metric first-order operators and aggregation~\cite{basinetal2015,havelundandrosu2002}, formally verified evaluation and recursive rules~\cite{schneideretal2019,zinggetal2022}, and certified proof-tree explanations, so far for one monitor, WhyMon (\textsuperscript{1})~\cite{limaetal2024}. Out-of-order monitoring handles late events~\cite{basinetal2020}. Policy auditing treats incomplete logs as three-valued structures~\cite{gargetal2011,chowdhuryetal2014}, and incomplete and disagreeing logs receive verdicts that mark what the logs do not determine~\cite{basinetal2012}. The family supports explicit exception conditions, as in the AP6 encoding of Appendix D, but no general defeasible-default semantics and no repair semantics for conflicting evidence. Annotated probabilistic temporal logic~\cite{jajodiaetal2016} and Bayesian diagnosis on top of temporal observers~\cite{schumannetal2015} give uncertainty a semantics, and R2U2 diagnoses spoofed sensor and GPS data under a bounded attacker model (\textsuperscript{1}), which does not extend to corrupted host logs. The certified checker of WhyMon is itself formally verified. The out-of-order monitor of Basin et al. outputs a verdict only once it is determined by the events seen so far, so its verdicts are final under arbitrary reordering~\cite{basinetal2020}. Past-guarded recursive rules with first-order joins (\textsuperscript{1}, VeriMon) and native aggregation meet the full criteria for recursion and joins and for counting, within the stated language restrictions.
\item \textbf{S4.} The systems build scenario graphs ordered by kill-chain stage~\cite{milajerdietal2019a,hassanetal2020} and propagate persistent tags in real time~\cite{hossainetal2017}, and only KnowHow adds natural-language lifting, lifecycle ordering and reported robustness to mimicry (\textsuperscript{1})~\cite{mengetal2026}.
\item \textbf{S5.} Queries are declarative over paths and time windows, with bounded temporal relations between events such as before[1-2 minutes] in AIQL~\cite{gaoetal2018a,gaoetal2018b}, and further systems align graphs and synthesize queries~\cite{milajerdietal2019b,gaoetal2021} or use RDF knowledge graphs~\cite{kurniawanetal2022}. SAQL demonstrates operation on host-event streams at its evaluated workload (++\textsuperscript{1} for raw-stream operation), which is not a cross-system throughput ranking~\cite{gaoetal2018b}. AIQL and SAQL provide native aggregation, including counts and statistical aggregates, which meets the full criterion for counting.
\item \textbf{S6.} Declarative attack-graph derivation~\cite{ouetal2005}, probabilistic assessment with Bayesian attack graphs~\cite{poolsappasitetal2012}, and applications that correlate observed alerts or infer attack paths from system-call evidence~\cite{noeletal2004,sunetal2018} make up the family. The core graph models have no metric event-time semantics. Delivery and order assumptions nevertheless apply to the evidence-consuming applications, and since none documents a treatment of late input, the late-input rating is n/s rather than n/a, which remains for prospective configuration-only models.
\item \textbf{S7.} The systems derive defeasible conclusions from conflicting evidence~\cite{nunesetal2016,karafilietal2020} and report no streaming evaluation (n/s for demonstrated raw-stream operation). Defeasible conflict resolution is not a semantics for combining uncertain evidence, and none was assessed for the cited systems (n/s for uncertainty).
\item \textbf{S8.} Learned or generated rules are readable~\cite{fengetal2019,indrietal2022,caietal2026}, and multi-step LTLf patterns are derived from reports with one pairwise elicitation template~\cite{eckhoffetal2025}. Validation ranges from detection evaluation on testbed data for mined invariants and learned formulas to the acceptance of generated Sigma rules into a public repository, without formal checks of the rules against their source reports.
\item \textbf{S9.} The family includes vocabulary resources and report-derived relations~\cite{rahmanetal2024,mitreattackflow,chengetal2025}, whose cited uses are retrieval and lookup, as well as active ontology-based detection. Undercoffer et al. use rules derived from the ontology in detection~\cite{undercofferetal2003}, while Coppolino et al. generate event-processing queries that join symptoms by temporal proximity and by source or target and combine symptom scores with weights for the trustworthiness of probes~\cite{coppolinoetal2009}, a partial combination mechanism (+) that establishes neither calibrated probabilities nor a joint model of shared evidence. Delivery, raw-stream operation and output checking apply to these active systems even where the sources do not evaluate them (n/s), and n/a remains for the resource-only members.
\item \textbf{S10.} Investigation traces dependencies backward and forward over audit records in timestamp order~\cite{kingandchen2003,liuetal2018}, with unbounded paths and joins over entities, and dependency graphs can be re-traced from the stored records. ProTracer processes online with out-of-order handling across applications (\textsuperscript{1})~\cite{maetal2016}, NoDoze prioritizes probabilistically without a combination semantics~\cite{hassanetal2019}, and one system analyzes under a compromised operating system~\cite{martignonietal2010}. Tracing systems encode no attack knowledge, while vulnerability predicates and forensic attack libraries are added without retraining~\cite{joshietal2005,rekhisandboudriga2012}, and temporal dependence networks infer security states with a Bayesian model~\cite{yangetal2018}. ProTracer lets processing proceed out of order across applications while keeping it in order within an application and resolving the resulting dependencies during analysis, which supports the bounded partial rating but not tolerance of arbitrary lateness or final verdicts under missing input.
\end{itemize}

\FloatBarrier
\section{Formalizations behind Table 9}
The formalizations below document the symbolic rows of Table 9 with schematic formulas and construction descriptions, identify reused constructions and distinguish assessed support from extensions. Cells that rest on a named extension without a constructed encoding are the Event Calculus on AP4, AP7 and AP9 and temporal ASP on AP1. Temporal ASP on AP7 and graph queries on AP4 and AP9 are marked ? because no extension has been selected and assessed for them. The non-probabilistic cells of ProbLog reuse the Datalog constructions, since ProbLog extends Datalog with probabilistic facts. They follow the conventions stated below, were checked against each language's documented operators but have not been executed. Programs in tool syntax are sketches of the relevant rules, not complete runnable programs. Formulas use each language's operators in mathematical notation rather than a specific tool's concrete syntax. $\blacklozenge$ is MFOTL's "once", $\Diamond^{-}$ and $\Diamond^{+}$ are DatalogMTL's past and future diamonds, and times are in seconds unless marked.

\smallskip\noindent \textbf{Conventions.} The patterns of Section 7.7 fix the intended matches together with the following conventions, which every encoding below follows or for which it names the deviation. Events are punctual and carry their event time in integer seconds. Where a language works on trace states, all events with the same timestamp form one state, and state timestamps strictly increase, so that a strictly earlier state is a strictly earlier time. Two events with equal timestamps are not ordered, so "later", "after" and "before" are strict, and "within n seconds of x" means a delay in the left-open interval (0, n]. For absence (AP6) and counting (AP2) the window [t $-$ n, t] ending at the evaluation time t is closed, so that a logon with the same timestamp suppresses the alert. Descent (AP4) and reach (AP9) are strict: they need at least one spawn or connection edge. Every Exec(d, x) belongs to the unique download episode d and is no earlier than Download(d), and the exfiltration in AP4 is no earlier than the last spawn of the chain. Identifiers of downloads (d) and compromise episodes (e) denote single episodes, and process identifiers are not reused within a pattern's horizon. Static relations, such as membership of a network zone or of an approved maintenance list, are persistent. Connections are instantaneous, and influence may wait at a host between connections. AP2 counts distinct endpoints e with ControllerEndpoint(e) per process, under set semantics. Stratified negation suffices for AP6b, and whether an engine retracts incrementally is an implementation question. AP5 is encoded for the totally ordered case of three steps. A partial order with incomparable branches additionally needs shared step instances. AP8a assumes stable process identities and comparable clocks. Each encoding derives an alert for every match. Encodings may differ in the variables they project and in the time at which they derive the alert, so the ratings compare whether a match is found, not the bindings returned.

\smallskip\noindent \textbf{Rating rule for metric bounds.} One rule is applied uniformly across the rows: a metric bound expressed through a native window, metric operator or deadline construct counts as \checkmark{}, and one expressed through arithmetic on timestamps stored as data counts as \halfcirc{}. This is why Datalog and temporal ASP are \halfcirc{} on AP2 (the 60-second window is arithmetic on data), and LARS is \halfcirc{} on AP4 (the 72-hour deadline counts from the download, which needs time variables with comparisons rather than a window).

\smallskip\noindent \textbf{Temporal message-passing GNNs.} Cells marked L indicate that a learned approach may be applied to the pattern. They establish neither expressibility nor learnability, and no formalization applies. The expressive bounds of temporal message passing are characterized separately~\cite{waegaandrawson2025}.

\smallskip\noindent \textbf{MFOTL.} Free variables are output, and negation is applied only to safely bounded subformulas.

\begin{progblock}
AP1\hspace*{1.1em}Connect(p, a) $\wedge$ $\blacklozenge$(0,600] ReadCred(p, f)\\
AP2\hspace*{1.1em}(n $\leftarrow$ CNT e; p. $\blacklozenge$[0,60] (Write(p, e) $\wedge$ ControllerEndpoint(e))) $\wedge$ n $\geq$ k\hspace*{1.7em}(MonPoly aggregation over distinct e, grouped by p)\\
AP3\hspace*{1.1em}Auth(q, c) $\wedge$ $\blacklozenge$(0,$\infty$) ReadCred(p, c) $\wedge$ $\neg$(p = q)\\
AP4\hspace*{1.1em}letpast Desc(d, y) = $\exists$x. Spawn(x, y) $\wedge$ $\blacklozenge$(0,259200] (Exec(d, x) $\vee$ Desc(d, x))\\
\hspace*{2.8em}in\hspace*{1.7em}Exfil(z) $\wedge$ $\blacklozenge$[0,259200] Desc(d, z) $\wedge$ $\blacklozenge$[0,259200] Download(d)\\
AP5\hspace*{1.1em}S3(h) $\wedge$ $\blacklozenge$(0,$\infty$)(S2(h) $\wedge$ $\blacklozenge$(0,$\infty$) S1(h))\hspace*{15.4em}(steps of one flow, unbounded gaps)\\
AP6a PrivStart(p, u) $\wedge$ $\neg$$\blacklozenge$[0,300] Logon(u) $\wedge$ $\neg$MaintSigned(p)\\
AP8a Start(par) $\wedge$ $\blacklozenge$(0,$\infty$) (Start(c) $\wedge$ ChildOf(c, par))\\
\end{progblock}
In AP4 every recursive occurrence of Desc sits under $\blacklozenge$ with a left-open interval, so it lies strictly in the past, which is the past-guarded form VeriMon admits~\cite{zinggetal2022}. AP9 follows the same construction over connection edges, with the window bound to the compromise episode. Because the row names VeriMon, whose language includes this recursive let, AP4 and AP9 are rated \checkmark{}. These two positive ratings therefore depend on VeriMon's recursive let, and no separate construction for MonPoly is assessed here. The final conjuncts check the deadline relative to the download at the time of the exfiltration. AP6b holds by re-evaluation on the updated trace, as for the other rows. Revising a verdict that was already emitted is the separate capability of out-of-order monitoring~\cite{basinetal2020}. AP1 and AP3 use left-open intervals, so a read with the same timestamp as the connection or the authentication does not match.

\smallskip\noindent \textbf{CEP (chronicles).} Chronicles are sets of events with variables and bounded temporal constraints between their occurrence times, with forbidden events for absence~\cite{morinanddebar2003}. The row covers chronicles only. CEL has different constructs for joins, iteration and negation and is not encoded here.

\begin{progblock}
AP1\hspace*{1.1em}chronicle cred\_then\_connect(p):\\
\hspace*{3.9em}event ReadCred(p, f) at t1;\hspace*{1.1em}event Connect(p, a) at t2;\hspace*{1.1em}t1 < t2 $\leq$ t1 + 600\\
AP3\hspace*{1.1em}chronicle cred\_reuse(p, q, c):\\
\hspace*{3.9em}event ReadCred(p, c) at t1;\hspace*{1.1em}event Auth(q, c) at t2;\hspace*{1.1em}t1 < t2;\hspace*{1.1em}p $\neq$ q\\
AP5\hspace*{1.1em}chronicle flow(h):\\
\hspace*{3.9em}event S1(h) at t1;\hspace*{1.1em}event S2(h) at t2;\hspace*{1.1em}event S3(h) at t3;\hspace*{1.1em}t1 < t2 < t3\\
AP6a chronicle unexplained\_priv(p, u):\\
\hspace*{3.9em}event PrivStart(p, u) at t;\hspace*{1.1em}forbidden Logon(u) in [t - 300, t];\hspace*{1.1em}not MaintSigned(p)\\
\end{progblock}
Chronicles have no recursive path construct, and we provide no variable-length construction for AP4 or AP9 in the assessed fragment (-- for both). These cells do not claim non-expressibility, including by finite unrolling under a fixed integer time bound. AP2, with k fixed by the specification, is written as k events with pairwise distinct variables, as in LARS (\checkmark{}). AP6b holds by re-evaluation on the updated trace, as for the other rows (\checkmark{}).

\smallskip\noindent \textbf{Event Calculus (RTEC).} Fluents hold over maximal intervals computed from initiating and terminating events, and rules can also compare the timestamps of input events retained in the working memory~\cite{artikisetal2015}.

\begin{progblock}
AP1\hspace*{1.1em}happensAt(alert(P), T) :- happensAt(connect(P,\_), T),\\
\hspace*{17.1em}happensAt(readCred(P,\_), T1), T1 < T, T - T1 =< 600.\\
AP5\hspace*{1.1em}initiatedAt(seen1(H)=true, T)\hspace*{1.1em}:- happensAt(s1(H), T).\\
\hspace*{2.8em}initiatedAt(seen12(H)=true, T) :- happensAt(s2(H), T), holdsAt(seen1(H)=true, T).\\
\hspace*{2.8em}happensAt(alert(H), T)\hspace*{5.0em}:- happensAt(s3(H), T), holdsAt(seen12(H)=true, T).\\
AP6\hspace*{1.1em}recentLogon(U, T)\hspace*{3.3em}:- happensAt(privStart(\_,U), T),\\
\hspace*{17.1em}happensAt(logon(U), T1), T1 =< T, T - T1 =< 300.\\
\hspace*{2.8em}happensAt(alert(P), T) :- happensAt(privStart(P,U), T),\\
\hspace*{17.1em}\textbackslash{}+ recentLogon(U, T), \textbackslash{}+ maintSigned(P).\\
AP8a incons(C) :- childOf(C,P), happensAt(start(C),TC), happensAt(start(P),TP), TC < TP.\\
\end{progblock}
AP1 and AP6 compare timestamps over retained input events: the read in AP1 must occur in [T $-$ 600, T), and AP6 tests the closed interval [T $-$ 300, T], so a read exactly 600 seconds earlier matches, a simultaneous read does not, and a logon exactly 300 seconds earlier or at the same time suppresses the alert. Repeated reads or logons add witnesses and cannot remove one. The events covering these windows must still be in RTEC's working memory when the rules are evaluated, and AP6 uses negation over the data assumed complete for variant (a), while variant (b) needs re-evaluation when the facts change. Both patterns are therefore rated encoded (\halfcirc{}). A fluent initiated at T holds after T, so AP5 is strict, and it uses separate fluents for the prefixes seen so far, so that a repeated first step does not reset a partial match. AP4 and AP9 need a fluent whose definition depends on itself, such as desc(D, Y) initiated by a spawn while desc(D, X) holds. The original RTEC compilation assumes a hierarchy of fluent definitions~\cite{artikisetal2015}. A later extension handles cyclic dependencies~\cite{mantenoglouetal2022}. The \halfcirc{} for AP4 and AP9 rests on that named extension, for which we have not constructed an encoding. Uncertain composition requires a probabilistic Event Calculus (AP7, \halfcirc{})~\cite{skarlatidisetal2015}. AP3 uses a persistent fluent seenRead(P,C), initiated by readCred(P,C), and raises a match at an authentication by Q with the same credential when P $\neq$ Q. Since initiation takes effect after the initiating time, the read is strictly earlier. AP2 keeps timestamped write events and applies the distinct-endpoint count over the closed 60-second window, an explicit timestamp and count encoding (\halfcirc{}).

\smallskip\noindent \textbf{LARS.} Windows are native: $\boxplus$\textsuperscript{n} restricts the stream to the last n seconds, including the current time point, $\boxplus$\textasciicircum{}$\infty$ to the whole stream so far, $\Diamond$ and $\Box$ quantify over it, and @T names an explicit time point~\cite{becketal2018}. Strict order is written with @T and a comparison.

\begin{progblock}
AP1\hspace*{1.1em}alert(P) @T\hspace*{2.2em}$\leftarrow$ connect(P) @T,\hspace*{1.1em}$\boxplus$\textasciicircum{}600 @T1 readCred(P),\hspace*{1.1em}T1 < T.\\
AP2\hspace*{1.1em}fan(P)\hspace*{5.0em}$\leftarrow$ $\boxplus$\textasciicircum{}60 $\Diamond$ write(P,E1), $\boxplus$\textasciicircum{}60 $\Diamond$ write(P,E2), $\boxplus$\textasciicircum{}60 $\Diamond$ write(P,E3), ctl(E1), ctl(E2), ctl(E3), E1$\neq$E2, E1$\neq$E3, E2$\neq$E3.\\
AP3\hspace*{1.1em}alert(P,Q,C) @T $\leftarrow$ auth(Q,C) @T,\hspace*{1.1em}$\boxplus$\textasciicircum{}$\infty$ @T1 readCred(P,C),\hspace*{1.1em}T1 < T,\hspace*{1.1em}P $\neq$ Q.\\
AP4\hspace*{1.1em}desc(D,Y) @T\hspace*{1.7em}$\leftarrow$ spawn(X,Y) @T,\hspace*{1.1em}exec(D,X) @T1,\hspace*{1.1em}T1 < T,\hspace*{1.1em}download(D) @T0,\hspace*{1.1em}T $\leq$ T0 + 259200.\\
\hspace*{2.8em}desc(D,Y) @T\hspace*{1.7em}$\leftarrow$ spawn(X,Y) @T,\hspace*{1.1em}desc(D,X) @T1,\hspace*{1.1em}T1 < T,\hspace*{1.1em}download(D) @T0,\hspace*{1.1em}T $\leq$ T0 + 259200.\\
\hspace*{2.8em}alert(D,Z) @T\hspace*{1.1em}$\leftarrow$ exfil(Z) @T,\hspace*{1.1em}desc(D,Z) @T1,\hspace*{1.1em}T1 $\leq$ T,\hspace*{1.1em}download(D) @T0,\hspace*{1.1em}T $\leq$ T0 + 259200.\\
AP6\hspace*{1.1em}alert(P)\hspace*{3.9em}$\leftarrow$ privStart(P,U),\hspace*{1.1em}not $\boxplus$\textasciicircum{}300 $\Diamond$ logon(U),\hspace*{1.1em}not maintSigned(P).\\
\end{progblock}
AP4 and AP9 need time variables with comparisons to bind the deadline to the origin (\halfcirc{}). Strict order, as in AP3 and AP5, uses the native @ operator with a comparison (\checkmark{}). Because $\boxplus$\textsuperscript{n} includes the current time point, AP1 and AP3 use @T with a comparison for strict order, and AP6 uses the closed window as the conventions require. Qualitative constraint consistency (AP8b) requires encoding the composition table of an interval algebra (\halfcirc{}).

\smallskip\noindent \textbf{Temporal ASP (telingo).} telingo provides past and future operators over the steps of a trace, such as $\blacklozenge$ (once) and $\bullet$ (previous). Clock time is a data term~\cite{cabalaretal2019}. Metric temporal ASP adds native metric bounds over timed traces~\cite{beckeretal2024,hahnetal2026}, which is the extension the AP1 cell names.

\begin{progblock}
AP3\hspace*{1.1em}alert(P,Q,C) :- auth(Q,C), \&tel\{ < <? readCred(P,C) \}, P != Q.\hspace*{2.8em}\% readCred at some strictly earlier step\\
AP5\hspace*{1.1em}done3(H)\hspace*{2.8em}:- s3(H), \&tel\{ < <? (s2(H) \& < <? s1(H)) \}.\\
AP6\hspace*{1.1em}alert(P)\hspace*{2.8em}:- privStart(P,U,T), not recent\_logon(U,T), not maint\_signed(P).\\
\hspace*{2.8em}recent\_logon(U,T) :- privStart(\_,U,T), logon(U,T1), T1 <= T, T - T1 <= 300.\\
AP2\hspace*{1.1em}fan(P,T)\hspace*{2.8em}:- time(T), proc(P), \#count\{ E : write(P,E,T1), ctl(E), T - 60 <= T1, T1 <= T \} >= 3.\hspace*{1.7em}\% write/3 kept as timestamped historical facts\\
\end{progblock}
Order is native (AP3, AP5 \checkmark{}) under the grouping of equal timestamps into one state. Metric bounds are arithmetic on timestamps as data (AP1, AP2, AP6 \halfcirc{}), and so are deadlines relative to an origin (AP4 \halfcirc{}). Facts of earlier states are not visible in the current state unless they are kept as historical facts with their timestamp, as the AP2 and AP6 rules assume. The AP4, AP8a and AP9 cells follow the timestamp constructions of the Datalog subsection below, which an ASP program states with the same rules and timestamps as data. For AP8b, choice rules guess one atomic relation per pair of intervals and constraints from the composition table reject inconsistent guesses, while the assessed stratified Datalog fragment has no corresponding choice construct and we provide no complete AP8b encoding for it, so its -- records that assessment rather than a proved non-expressibility result. Probabilistic ASP frameworks could supply AP7, but this row selects no probability semantics and constructs no shared-evidence model, so AP7 is left unassessed (?).

\smallskip\noindent \textbf{Graph queries.} AIQL expresses multi-event patterns with shared entities, temporal relations between events and sliding time windows with aggregation~\cite{gaoetal2018a}, and regular path queries add unbounded paths.

\begin{progblock}
AP2\hspace*{1.1em}proc p write file f[ctl = true] as evt\hspace*{1.1em}\#time(60 s)\hspace*{1.1em}group by p\hspace*{1.1em}having count(distinct f) >= 3\\
AP1\hspace*{1.1em}proc p read file f[cred = true] as e1, proc p connect ip i as e2\hspace*{1.1em}with e1 before[0-10 minutes] e2\\
AP3\hspace*{1.1em}proc p read file f as e1, proc q auth with f as e2\hspace*{1.1em}with e1 before e2, p != q\\
AP5\hspace*{1.1em}proc p1[agentid = h] s1 as e1, proc p2[agentid = h] s2 as e2, proc p3[agentid = h] s3 as e3\hspace*{1.1em}with e1 before e2, e2 before e3\\
\end{progblock}
An unbounded path whose successive edges carry increasing timestamps (AP4, AP9) is not a regular path query over a single graph. A recursive or temporal-path language is a candidate, but none is selected and assessed here, and the AIQL and plain RPQ constructions do not establish it (?). AP1 uses AIQL's native bounded temporal relation (\checkmark{}). We assume that before is strict and its bounds inclusive, which matches the interval (0, 600] of the conventions. AP5 is a fixed chain of three event patterns on one host with two before relations (\checkmark{}). Its gaps are unbounded but its length is not, so it needs no path query. In AP1, f ranges over the credential files supplied by the input mapping. The AP3 query is schematic: authentication is not an AIQL event type and must be supplied as an event relation.

\smallskip\noindent \textbf{Datalog with comparisons and stratified negation.} Times are data values, so both order and metric bounds need arithmetic comparisons, which is why AP1 to AP6 and AP8a are \halfcirc{} rather than \checkmark{}.

\begin{progblock}
AP4\hspace*{1.1em}desc(D,Y,T)\hspace*{1.7em}:- exec(D,X,T1), spawn(X,Y,T), T1 < T, download(D,T0), T <= T0 + 259200.\\
\hspace*{2.8em}desc(D,Y,T)\hspace*{1.7em}:- desc(D,X,T1), spawn(X,Y,T), T1 < T, download(D,T0), T <= T0 + 259200.\\
\hspace*{2.8em}alert(D,Z)\hspace*{2.2em}:- desc(D,Z,T1), exfil(Z,T), T1 <= T, download(D,T0), T <= T0 + 259200.\\
AP6\hspace*{1.1em}alert(P,T)\hspace*{2.2em}:- priv\_start(P,U,T), not recent\_logon(U,T), not approved\_signed\_maintenance(P,T).\\
\hspace*{2.8em}recent\_logon(U,T) :- priv\_start(\_,U,T), logon(U,T1), T1 <= T, T - T1 <= 300.\\
AP8a incons(C)\hspace*{2.8em}:- child\_of(C,P), start(C,TC), start(P,TP), TC < TP.\\
\end{progblock}
\smallskip\noindent \textbf{DatalogMTL.} $\Diamond^{-}$I A holds at time t if A held at some time t $-$ d with d $\in$ I, $\Diamond^{+}$I is its future counterpart, and a head $\boxplus$[a,b] A makes A hold throughout [t + a, t + b]. Negated metric atoms require the stratified extension~\cite{tenacucalaetal2021}, which keeps the timeline $\mathbb{Q}$. In AP4, Download(d) holds at the time of download d and Exec(d, x) when process x is started from its file. Win(d) holds from the download until 72 hours later.

\begin{progblock}
AP1\hspace*{1.1em}Alert(p)\hspace*{2.2em}$\leftarrow$ Connect(p) $\wedge$ $\Diamond^{-}$(0,600] ReadCred(p)\\
AP2\hspace*{1.1em}Fan(p)\hspace*{3.3em}$\leftarrow$ $\Diamond^{-}$[0,60] Write(p,e1) $\wedge$ $\Diamond^{-}$[0,60] Write(p,e2) $\wedge$ $\Diamond^{-}$[0,60] Write(p,e3) $\wedge$ Ctl(e1) $\wedge$ Ctl(e2) $\wedge$ Ctl(e3) $\wedge$ e1$\neq$e2 $\wedge$ e1$\neq$e3 $\wedge$ e2$\neq$e3\hspace*{1.7em}(k = 3; needs $\neq$)\\
AP3\hspace*{1.1em}Alert(p,q)\hspace*{1.1em}$\leftarrow$ Auth(q,c) $\wedge$ $\Diamond^{-}$(0,$\infty$) ReadCred(p,c) $\wedge$ p$\neq$q\hspace*{22.6em}(needs $\neq$)\\
AP4\hspace*{1.1em}$\boxplus$[0,72h] Win(d)\hspace*{1.1em}$\leftarrow$ Download(d)\\
\hspace*{2.8em}Desc(d,y)\hspace*{4.4em}$\leftarrow$ $\Diamond^{-}$(0,72h] Exec(d,x) $\wedge$ Spawn(x,y) $\wedge$ Win(d)\\
\hspace*{2.8em}Desc(d,y)\hspace*{4.4em}$\leftarrow$ $\Diamond^{-}$(0,72h] Desc(d,x) $\wedge$ Spawn(x,y) $\wedge$ Win(d)\\
\hspace*{2.8em}Alert(d,z)\hspace*{3.9em}$\leftarrow$ $\Diamond^{-}$[0,72h] Desc(d,z) $\wedge$ Exfil(z) $\wedge$ Win(d)\\
AP5\hspace*{1.1em}Done2(h)\hspace*{2.2em}$\leftarrow$ S2(h) $\wedge$ $\Diamond^{-}$(0,$\infty$) S1(h)\\
\hspace*{2.8em}Done3(h)\hspace*{2.2em}$\leftarrow$ S3(h) $\wedge$ $\Diamond^{-}$(0,$\infty$) Done2(h)\\
AP6\hspace*{1.1em}Alert(p)\hspace*{2.2em}$\leftarrow$ PrivStart(p,u) $\wedge$ not $\Diamond^{-}$[0,300] Logon(u) $\wedge$ not MaintSigned(p)\hspace*{11.6em}(stratified negation)\\
AP8a Incons(c)\hspace*{1.7em}$\leftarrow$ Start(par) $\wedge$ ChildOf(c,par) $\wedge$ $\Diamond^{-}$(0,$\infty$) Start(c)\\
AP9\hspace*{1.1em}$\boxplus$[0,$\Delta$] Win(e)\hspace*{1.1em}$\leftarrow$ Compromised(e,w)\\
\hspace*{2.8em}Reach(e,z)\hspace*{2.8em}$\leftarrow$ $\Diamond^{-}$(0,$\Delta$] Compromised(e,x) $\wedge$ Conn(x,z) $\wedge$ Win(e)\\
\hspace*{2.8em}Reach(e,z)\hspace*{2.8em}$\leftarrow$ $\Diamond^{-}$(0,$\Delta$] Reach(e,x) $\wedge$ Conn(x,z) $\wedge$ Win(e)\\
\hspace*{2.8em}Alert(e,z)\hspace*{2.8em}$\leftarrow$ Reach(e,z) $\wedge$ InZone(z, controller)\\
\end{progblock}
AP9 follows the same pattern as AP4 over connection edges, with the window bound to the compromise episode e, so that a later compromise of the same workstation opens a new window instead of extending an old one. Ctl abbreviates ControllerEndpoint, and InZone and Ctl are static relations. AP6 covers both variants: on re-evaluation with a later logon or exception record, the conclusion is no longer derived.

\smallskip\noindent \textbf{Probabilistic logic programming (ProbLog).} Shared evidence is represented once as a probabilistic fact, so the distribution semantics does not double-count it (AP7). The campaign probability depends on this generative model: the technique marginals alone do not determine it.

\begin{progblock}
0.9::alert\_genuine(a17).\\
0.7::maps\_to(a17, credential\_access).\\
0.6::maps\_to(a17, lateral\_movement).\\
tech(T)\hspace*{1.1em}:- alert\_genuine(a17), maps\_to(a17, T).\\
campaign :- tech(credential\_access), tech(lateral\_movement).\\
query(campaign).\\
\end{progblock}
Here P(campaign) = 0.9 $\times$ 0.7 $\times$ 0.6 = 0.378, whereas multiplying the two technique confidences, 0.63 and 0.54, as if they were independent gives about 0.34. The difference is the shared evidence a17 counted twice. The computation assumes that the three base facts are independent, and further correlations would need an extended model. Where the confidences come from, and whether they are calibrated, is the open problem of R3.

\smallskip\noindent \textbf{Allen interval algebra.} Only qualitative relations between intervals are expressible. AP8b is its native task: for example, the network \{session S during process P, process P before logon L, logon L during session S\} is inconsistent, which path consistency detects~\cite{allen1983,nebelandburckert1995}. Path consistency is not complete for arbitrary interval-algebra networks. A qualitative order of attack stages (AP5) can be stated as a chain of before relations, but without metric bounds or entities (\halfcirc{}).

\smallskip\noindent \textbf{STREL.} Temporal operators apply to signals at the locations of a spatial model. The reach operator quantifies over paths in the graph at one time point~\cite{nenzietal2022}. A location-local pattern such as AP1 can be written as a temporal formula at each location, \halfcirc{} because entities and their attributes are not first-order data. AP4 and AP9 need time-respecting paths across changing graphs, which per-snapshot reach does not provide (--). For AP5, the three step predicates at a fixed host location are composed with nested strict-past temporal operators (Since). AP6 needs a signal at each location that already associates a privileged start with the user's logon history and maintenance status, after which the temporal condition excludes a logon in the closed preceding 300-second window. Both are encoded (\halfcirc{}), because the entity binding and the input mapping are supplied outside STREL's location-based operators.

\end{document}